\documentclass[a4,12pt]{article}
\usepackage{soul}
\usepackage{graphicx}   
\usepackage{subcaption} 
\usepackage{multirow}  
\usepackage{hyperref}
\hypersetup{colorlinks=true,citecolor=blue,linkcolor=blue} 
\usepackage[usenames,dvipsnames,table]{xcolor}
\usepackage{dcolumn,rotating,lscape,longtable}
\usepackage[super,comma,numbers,sort&compress]{natbib}
\usepackage[normalem]{ulem}
\usepackage{mathptmx,mathtools,newtxtext,newtxmath,xfrac,amsmath}  
\usepackage{siunitx}  
\usepackage{pgffor}
\usepackage{booktabs} 
\usepackage{enumitem} 
\usepackage{pifont}
\usepackage{tikz} 
\usepackage{fontawesome} 
\usepackage{float}
\usepackage{colortbl} 
\usepackage{xcolor}   
\usepackage[pagewise]{lineno}
\usepackage{caption}
\usepackage{arydshln}
\usepackage[percent]{overpic}  
  
\graphicspath{{./FIG.s/}}
  
\usepackage{caption}
\begin{document}
	\title{\textbf{\textit{Ab Initio} Adiabatic Potential Energy Surfaces and Non-adiabatic Couplings for O$_3$: Construction of Four State Diabatic Hamiltonian}}
	
	\author{
		Avik Guchait $^{1}$, 
		Gourhari Jana $^{1}$, 
		Satyam Ravi $^{2}$,
		Koushik Naskar $^{1}$, 
		Satrajit Adhikari $^{1,}$\footnote{Corresponding author; e-mail: pcsa@iacs.res.in}
		\vspace{.5cm}
		\and {$^1$School of Chemical Sciences, Indian Association for the Cultivation of Science}
		\and {2A \& 2B Raja S. C. Mullick Road, Jadavpur, Kolkata 700032}
		\vspace{.3cm}
		\and {$^2$ School of Advance Science and Languages, VIT Bhopal University, Bhopal, India}}
	
	\date{}
	\maketitle
	\clearpage
	
	\section*{ABSTRACT}
	We compute highly accurate first principle based \textit{ab initio} adiabatic potential energy surfaces (PESs) using State-Averaged Multi-Configurational Self-Consistent Field (SA-MCSCF) followed by internally contracted Multi-Reference Configuration Interaction method incorporating fixed-reference Davidson corrections [ic-MRCI(Q)], where a full valence active space of 18 electrons in 12 orbitals and aug-cc-pVQZ basis set are employed for the low-lying four singlet electronic states of ozone ($\tilde{X}^1A'$, $1~^1A''$, $1~^1A'$ and $2^1A''$). It accurately reproduces the dissociation energies of ozone (1.101 eV) as well as the molecular oxygen (5.106 eV) along with vibrational frequencies of O$_3$ in comparison with experimental data. To ensure appropriate accuracy and proper convergence in the interaction as well as asymptotic regions, we (a) extend the number of electronic states in SA-MCSCF calculation (singlet as well as triplet and quintet); (b) systematically expand the active space [(12e,9o) $\rightarrow$ (18e,12o) $\rightarrow$ (24e,15o)] and basis set size (AVDZ $\rightarrow$ AV6Z $\rightarrow$ Complete Basis Set limit); (c) incorporate multi-reference character along with Davidson correction. Conical intersections between the adjacent electronic states (1-2, 2-3 and 3-4) are located at \textit{C}$_{2v}$, \textit{D}$_{3h}$ as well as \textit{C}$_{s}$ geometries through the four-state adiabatic-to-diabatic transformation of non-adiabatic coupling terms (NACTs) computed at Coupled-Perturbed Multi-Configurational Self-Consistent Field (CP-MCSCF) method along the circular contours. Finally, we present: (a) ic-MRCI(Q) calculated minimum energy path of incoming oxygen to the diatom (O$_2$) is devoid of any ``reef'' feature; (b) NACTs and diabatic PES matrix elements as function of hyperangles ($\theta$, $\phi$) at a fixed hyperradius $\rho = 4$ Bohr for a four state sub Hilbert space.

	\section*{I. INTRODUCTION}
	
	\vspace{-0.2cm}
	
	\noindent
	Ozone (O$_3$) layer plays a vital role in stratospheric photochemistry shielding Earth from UV radiation~\cite{johnston1992atmospheric,matsumi2003photolysis} and dissociates to molecular oxygen and atomic oxygen. Despite the apparent simplicity of the triatomic O$_3$ molecule, its spectroscopy and dynamics have remained challenging for decades. The persistent discrepancies\cite{schinke2006dynamical} between predicted and observed low-temperature \(\text{O} + \text{O}_2\) isotopic exchange rates are shown up from the negative temperature dependence in experiment. Moreover, the selective enrichment of heavy ozone isotopomers both in the atmosphere and laboratory settings deviates from the conventional mass-dependent isotope effect. This anomaly is known as the mass-independent fractionation (MIF) effect\cite{MIF}, which also remains unresolved. A major challenge in bridging theory and experiment lies in the absence of highly accurate first principle based global adiabatic/diabatic potential energy surfaces (PESs). In case of ozone, the complexity of \textit{ab initio} calculations primarily arises from the slow convergence of correlation energies and nuclear gradients with respect to both the basis set size and the active space, owing to the strong coupling between adjacent electronic states. Over the past four decades, there have been continuous progresses primarily on the development of ground adiabatic PESs for ozone employing both semi-empirical and \textit{ab initio} methods. Earlier works are summarized in Section S1 of the Supplementary Material (SM), whereas the following paragraphs focus on advancements in global adiabatic and diabatic PESs.

	\vspace{0.5cm}
	\noindent
	One of the earliest attempts to compute an \textit{ab initio} global adiabatic ground state ($\tilde{X}^1A'$) PES for ozone was made by Siebert \textit{et al.}~\cite{siebert2002vibrational} (SSB PES) using complete active space self consistent field (CASSCF) followed by multi-reference configuration interaction (MRCI) method with a relatively small active space of (12e,9o) and correlation-consistent polarized valence quadruple-zeta (cc-pVQZ) basis set. The SSB PES was generated by spline interpolation of approximately 5000 \textit{ab initio} energies and successfully captured both the \textit{C}$_{2v}$ and \textit{D}$_{3h}$ minima, as well as the entrance/exit channel, \(\text{O}_2(^3\Sigma_g^-)\) + \(\text{O}(^3P)\). On the other hand, the presence of a controversial artificial barrier (or reef) of 45 cm$^{-1}$ at the entrance/exit channel of minimum energy path (MEP) and an underestimated energy gap (\textit{D}$_e$) of approximately 900 cm$^{-1}$ between the \textit{C}$_{2v}$ minima and the dissociation limit of ozone (\(\text{O}_2(^3\Sigma_g^-)\) + \(\text{O}(^3P)\)) was empirically modified (along its MEP) by Babikov \textit{et al.}~\cite{babikov2003metastable} following the calculations of Hern\'andez-Lamoneda \textit{et al.}~\cite{hernandez2002does} and Fleurat-Lessard \textit{et al.}~\cite{Fleurat2003} The modified SSB PES was subsequently employed to investigate the metastable~\cite{babikov2003metastable} and van der Waals states~\cite{grebenshchikov2003vdw,ivanov2011collisional} of O$_3$.  Baloïtcha et al.~\cite{baloitcha2005theory} developed diabatic PESs and transition dipole moment surfaces for the lowest five $^1A'$ electronic states of ozone using contracted MRCI calculations with a slightly reduced aug-cc-pVTZ basis set. Those surfaces were used to locate several avoided crossings near the ground-state minimum and enabled quantum dynamics simulations that reproduced photodissociation cross sections and product state distributions. Schinke \textit{et al.} studied ozone photodissociation in the wavelength range 200–1100 nm, covering the Hartley~\cite{qu2005photodissociation}, Huggins~\cite{qu2004huggins}, Chappuis~\cite{grebenshchikov2006absorption} and Wulf absorption bands~\cite{grebenshchikov2007new} using diabatic PESs constructed from ic-MRCI/aug-cc-pVTZ level of adiabatic PESs.	Holka et al.~\cite{holka2010} validated earlier calculations of Schinke \textit{et. al.}~\cite{siebert2002vibrational,grebenshchikov2003vdw,qu2004huggins,qu2005photodissociation,grebenshchikov2006absorption,grebenshchikov2007new} through a systematic investigation of electronic ground state PES of ozone using ic-MRCI singles and doubles (ic-MRCISD) as well as multireference averaged quadratic coupled cluster (ic-MR-AQCC) with a full valence CAS(18e,12o) and augmented correlation consistent basis sets extrapolated to the complete basis set (CBS) limit.
	
	\vspace{0.5cm}
	
	\noindent
	Later, Dawes \textit{et al.} extensively studied the dissociation of O$_3$ using both basis set extrapolation~\cite{dawes2011communication} and explicitly correlated methods~\cite{dawes2013communication}, employing a full-valence complete active space, CAS(18e,12o) as reference. This calculation achieved an agreement within $\sim$30 cm$^{-1}$ for \textit{D}\textsubscript{e} by including spin-orbit coupling (SOC) and scalar relativistic effects. Moreover, it was showed that the controversial reef feature could be replaced by a smooth shoulder when the ground state ($\tilde{X}^1A'$) was treated simultaneously with twelve (12) excited singlet states, six (6) of which correlate with the high-lying singlet dissociation limit. The resulting reef-free PES (DLLJG PES) qualitatively reproduced the negative temperature dependence of thermal rate constant for isotopic exchange reactions  in terms of experimental trend.~\cite{Li2014,Sun2015} Lepers \textit{et al.}~\cite{lepers2012long} developed a theoretical framework for long-range interactions of ozone and examined its spectroscopic and dynamical properties through electrostatic and van der Waals contributions. Tyuterev \textit{et al.}~\cite{tyuterev2013new} developed an analytical ground-state PES (TKTHS PES) for ozone by correcting the non-vanishing ``reef'' feature of DLLJG surface at some geometry near the transition state along the MEP and computed thermal rate constants for the $^{18}$O + $^{32}$O$_2$ $\rightarrow$ $^{34}$O$_2$ + $^{16}$O exchange reaction~\cite{Tyuterev2018}. Mankodi \textit{et al.}~\cite{mankodi2017dissociation} constructed ab initio PESs using restricted active space self consistent field (RASSCF) followed by Complete Active Space Second-Order Perturbation Theory (CASPT2)  with an active space of (12e,9o) and aug-cc-pVTZ basis set to compute collision-induced dissociation (CID) cross sections for the O$_2$ + O reaction by Quasi-Classical Trajectory (QCT) calculation. Powell \textit{et al.}~\cite{powell2017investigation} analyzed the O$_2$ + O $\rightleftharpoons$ O$_3$ reaction to examine the impact of various electronic structure methods on eliminating spurious reef features. Another PES is constructed by Varga et al.~\cite{varga2017} (PIP PES) for nine low-lying electronic states of ozone including singlet, triplet, and quintet using dynamical scaled external correlation (DSEC) corrected extended multi-state CASPT2 calculations to study the O($^3$P) + O$_2$($^3\Sigma_g^-$) collisions across a wide energy range. Alijah \textit{et al.}~\cite{Alijah2018} investigate the presence of topological phase in ozone, which affects highly excited rovibrational states. More recently, Egorov \textit{et al}.~\cite{egorov2023long} developed a long-range \textit{ab initio} PES for ground state of ozone using extended CAS(24e,15o) at the ic-MRCI(Q) level combined with extrapolation to the complete basis set [CBS(56Z)] limit, where SOC among the nine (9) multiplet electronic states near dissociation channel (\(\text{O}_2(^3\Sigma_g^-)\) + \(\text{O}(^3P)\)) was incorporated to improve \textit{D}\textsubscript{e}.	Wang \textit{et al}.~\cite{Wang2025} employed MRCI-based adiabatic PES (RKHS PES) to investigate the high-temperature rate of atom-exchange and dissociation dynamics of O$_3$.
	
	\vspace{0.5cm}
	\noindent
	For comprehensive understanding of ozone’s electronic structure, advanced computational methods are required to capture the strong electron-nuclear coupling. The identification and location of CIs provide essential insights into molecular processes. Over the past decades, Adhikari \textit{et al.} formulated Beyond Born-Oppenheimer (BBO) equations for multi-state multi-mode systems and paved a practical way to implement such a complete theory in the field of spectroscopy and reactive scattering, where nonadiabatic coupling terms (NACTs) are incorporated from the first principles, enabling a unified and more accurate description of molecular~\cite{mukherjee2019conical,ghosh2017beyond,paul2011ab,Hazra2022_CPC_23,5_state} as well as chemical processes~\cite{mukherjee2017beyond,mukherjee2024beyond,HeH2_Surf,sarkar2006extended}. While implementing such BBO based treatment to formulate diabatic Hamiltonian, it is necessary to demonstrate the path independence~\cite{pathdependence2023} of observables (cross section and rate constant) calculated from path dependent diabatic Hamiltonians and the existence of sub Hilbert space (SHS)~\cite{SHS2023} for any specific molecular or chemical processes. So far, we have constructed BBO based diabatic Hamiltonian for three (H$_3^+$\cite{H3+_Surf}, FH$_2$\cite{FH2_Surf} and H$_3$\cite{ghosh2025H3}) and four (HeH$_2^+$\cite{HeH2_Surf} and at present, O$_3$) state SHS, where scattering calculations have been carried out\cite{H3+_dynamics_2021,FH2_dynamics,ghosh2025H3,HeH2+_dynamics_2023} or will be performed on O$_3$ to explore the quantum effect on reaction attributes. While comparing our calculated cross section and rate constant for the above reactive processes with other theoretical results and experimental data, it would be quite interesting to examine the agreement/disagreement of the profiles (cross section/rate constant) obtained from the recently developed diabatic surfaces using CHIPR\cite{Guan_Chen_Varandas_2024} and coupled adiabatic PESs employing Frobenius companion matrices using neural network based algorithm\cite{Shu_Truhlar_2024}.
	
	\vspace{0.5cm}
	\noindent
	In this study, we investigate the intricacies of electronic structure calculations for ozone by exploring (a) various active spaces: CAS(12e,9o), (18e,12o) and (24e,15o) in combination with basis sets ranging from AVDZ $\rightarrow$ AV6Z; (b) the impact of different numbers of states in State Averaged Multi-Configurational Self-Consistent Field (SA-MCSCF) calculation to obtain global adiabatic PESs and in Coupled-Perturbed Multi-Configurational Self-Consistent Field (CP-MCSCF) method to evaluate nonadiabatic coupling terms (NACTs) of ozone and (c) internally contracted Multi-Reference Configuration Interaction method incorporating Davidson corrections [ic-MRCI(Q)] on the SA-MCSCF reference orbitals using converged CAS(18e,12o) and basis set (AVQZ) for the calculation of highly accurate global adiabatic PESs. The paper is structured as follows: Section II provides a brief overview on BBO theory and its equations for the construction of diabatic Hamiltonian from adiabatic quantities (PESs and NACTs). In Section III, we discuss on \textit{ab initio} calculations for convergence of molecular parameters using different CASs, basis sets and various number of electronic states in SA-MCSCF followed by ic-MRCI(Q) calculation within Jacobi framework. Section IV explores the location of CIs by defining appropriate contours both at \(\textit{C}_{2v}\) and \(\textit{D}_{3h}\) geometries. Finally, Section V presents Diabatic PESs and couplings at a fixed hyperradius, $\rho$ = 4 Bohr. Section VI presents the conclusions and future perspective of this work.
	
	\section*{II. THEORETICAL BACKGROUND}
	\subsection*{II.A. Adiabatic Representation of Schr\"{o}dinger Equation}
	In the Born-Oppenheimer-Huang treatment,~\cite{BO,BH} the total molecular wavefunction, \{$\Psi(r,R)$\}, for any $M$-dimensional electronic Hilbert Space can be expanded in terms of the electronic eigenfunctions (\{$\xi_i$\}s) with nuclear coordinate dependent coefficients known as adiabatic nuclear wavefunctions (\{$\psi_i^{\rm ad}$\}s) as:
	\begin{eqnarray}
		\Psi(\textbf{r},\textbf{R}) = \sum_{i=1}^M \psi_i^{\rm{ad}}(\textbf{R}) \xi_i(\textbf{r};\textbf{R}),
		\label{eq:BHO-expansion} 
	\end{eqnarray}
	where electronic and nuclear coordinates are denoted by $\textbf{r}$ and $\textbf{R}$, respectively.
	
	\noindent
	The total electron-nuclear Hamiltonian ($ \hat{H}(\textbf{r},\textbf{R}) $) is the sum of the nuclear kinetic energy operator ($ \hat{T}_{\rm nuc}(\textbf{R})$) and the electronic Hamiltonian ($ \hat{H}_{\rm{el}}(\textbf{r};\textbf{R})$).
	\begin{eqnarray}
		\hat{H}(\textbf{r},\textbf{R}) = \hat{T}_{\rm nuc}(\textbf{R}) + \hat{H}_{\rm{el}}(\textbf{r};\textbf{R}),
		\label{Tot_H}
	\end{eqnarray}
	
	\noindent
	The electronic eigenfunctions (\{$\xi_i$\}s) satisfy the electronic SE:
	\begin{eqnarray}
		\hat{H}_{\rm{el}}(\textbf{r};\textbf{R})\xi_i(\textbf{r};\textbf{R}) = u_i(\textbf{R})
		\xi_i(\textbf{r};\textbf{R}), \qquad  
		\langle \xi_i | \xi_j \rangle_{\textbf{r}} = \delta_{ij}
		\label{eq:Elec_SE}
	\end{eqnarray}
	where $u_i(\textbf{R})$ represents the adiabatic PESs.
	
	\noindent
	Using the form of molecular wave function [Eq.~(\ref{eq:BHO-expansion})], the total electron-nuclear Hamiltonian [Eq.~(\ref{Tot_H})] and the electronic eigenvalue equation [Eq.~(\ref{eq:Elec_SE})], the molecular SE \{$\hat{H}(\textbf{r},\textbf{R})	\Psi(\textbf{r},\textbf{R}) = E\Psi(\textbf{r},\textbf{R}$\} on projection with the electronic eigenfunction (\{$\xi_i$\}) provides the kinetically coupled matrix form of adiabatic nuclear SE~\cite{Baer2006_bbo_book_sect211}:
	\begin{eqnarray}
		-\frac{\hbar^2}{2m}\left(\boldsymbol{\nabla}_R+\boldsymbol{\tau}\right)^2\psi^{\rm{ad}} + \left(U-E\right)\psi^{\rm{ad}} 
		= 0,
		\label{eq:Adia_SE}
	\end{eqnarray}
	where $E$ signifies total energy, $U$ is the adiabatic potential energy matrix given by $ U_{ij} =u_i\delta_{ij} $ and $\boldsymbol{\tau}$ defines the nonadiabatic coupling matrix (NACM) as given by:
	\begin{eqnarray}
		\boldsymbol{\tau}_{ij} = \langle\xi_i\vert\boldsymbol{\nabla}_R\xi_j\rangle_\textbf{r} = -\langle\xi_j\vert\boldsymbol{\nabla}_R\xi_i\rangle_\textbf{r} = -\boldsymbol{\tau}_{ji},
		\label{eq:Nacm}
	\end{eqnarray}
	and thereby, $\tau$ is a skew-symmetric matrix. On the other hand, while taking derivative on the electronic SE [Eq.~(\ref{eq:Elec_SE})] with respect to nuclear coordinates ($\boldsymbol{\nabla}_R$) and projecting by electronic eigenfunctions ($\langle\xi_i\vert$), it appears that NACTs would be singular at point(s) or along seam(s) of degeneracy between electronic states as given below:		
	\begin{eqnarray}
		\boldsymbol{\tau}_{ij} = \frac{\langle\xi_i\vert\boldsymbol{\nabla}_R \hat{H}_{\rm{el}}\vert\xi_j\rangle_\textbf{r}}{u_j-u_i},
		\label{eq:HF}
	\end{eqnarray}
	which is a reminiscent of Hellman-Feynman theorem~\cite{clusius1941einfuhrung,feynman1939forces,epstein1954note}. Such singular terms not only introduce numerical inaccuracies but also could lead to non-conservation of energy~\cite{mukherjee2024quasi} in the adiabatic representation of SE. Therefore, it is necessary to transform the adiabatic SE [Eq.~(\ref{eq:Adia_SE})] to another representation known as diabatic so that one can get rid of such singular terms.
	
	\subsection*{II.B. Diabatic Schr\"{o}dinger Equation}
	The adiabatic nuclear SE [Eq. (\ref{eq:Adia_SE})] can be transformed into the diabatic framework by carrying out an orthogonal transformation, $\psi^{\rm{ad}}=A\psi^{\rm{dia}}$, 
	\begin{eqnarray}
		\left(-\dfrac{\hbar^2}{2m}\nabla_R^2 + W-E\right)\psi^{\rm{dia}} = 0,
		\label{eq:Dia_SE}
	\end{eqnarray}
	where $\psi^{\rm{dia}}$ denotes the diabatic nuclear wavefunction and $ W $ is the diabatic potential energy matrix,
	\begin{eqnarray}
		W = A^{\dagger}U A.
		\label{eq:Dia_pot}
	\end{eqnarray}
	In order to obtain such adiabatic-to-diabatic representation of SE [Eq.~(\ref{eq:Dia_SE})], the transformation matrix $ A $ (also known as ADT matrix) must satisfy the following ADT condition,~\cite{Baer1975_CPL_35,top1977incorporation,sarkar2006extended}
	\begin{eqnarray}
		\boldsymbol{\nabla}_RA + \boldsymbol{\tau}A = 0.
		\label{eq:ADT_eq}
	\end{eqnarray}
	
	\noindent
	Since $\boldsymbol{\tau}$ is a skew-symmetric matrix, it can be depicted by manipulating [Eq.~(\ref{eq:ADT_eq})] that the transformation matrix, $A$ has to be orthogonal by construction ($A^{\dagger}A = I$).
	
	\subsection*{II.C. ADT Equations}
	Without losing generality, the ADT matrix ($ A $) can be considered as a product of elementary rotation matrices constituted with mixing angles between the possible electronic states.~\cite{top1977incorporation,Alijah2000_JPCA_104,Sarkar2008_JPCA_112_curl,Mukherjee2019_IRPC_BBO,Hazra2022_CPC_23} Therefore, for $M$ electronic states, the ADT matrix ($A$) can be considered as a product of $\Lambda =\ ^MC_2 = \frac{M(M-1)}{2}$ elementary rotation matrices in $\Lambda !$ number of possible ways.
	
	\vspace{0.2cm}
	
	\begin{eqnarray}
		A = P_n\{A^{12}(\Theta_{12}).A^{13}(\Theta_{13})
		........A^{M-1,M}(\Theta_{M-1,M})\}, \qquad n=1,...,\Lambda !,
		\label{eq:Amat}
	\end{eqnarray}
	
	\vspace{0.2cm}
	
	\noindent
	where $P_n$ is the $n$th permutation between two rotation matrices and \{$\Theta_{ij}$\}s represent the ADT angles dependent on the nuclear coordinates. It may be noted that numerical solutions of any set of ADT equations originating from different permutations of elementary rotation matrices appear to be the same solution, \{$\Theta_{ij}$\}s.~\cite{ADT_JCTC_16}
	
	\vspace{0.2cm}
	
	\noindent
	The matrix elements of any of the $A^{mn}(\Theta_{mn})$ matrices can be
	defined as:
	
	\vspace{0.2cm}
	
	\begin{eqnarray}
		&\,& \left[A^{mn}(\Theta_{mn})\right]_{mm} = \cos\Theta_{mn} = \left[A^{mn}(\Theta_{mn})\right]_{nn};  
		\quad m\neq n\nonumber \\
		&\,& \left[A^{mn}(\Theta_{mn})\right]_{mn} = \sin\Theta_{mn} = -\left[A^{mn}
		(\Theta_{mn})\right]_{nm}; \quad m\neq n\nonumber \\
		&\,& \text{and} \quad \left[A^{mn}(\Theta_{mn})\right]_{ij} = \delta_{ij}; \quad \{i,j\}
		\neq \{m,n\}.
		\label{eq:Amat_ele}
	\end{eqnarray}
	
	\vspace{0.2cm}
	
	\noindent
	If the electronic sub-space ($N << M$) exits for a specific molecular or chemical processes, two representative elementary rotation matrices for a  4-state SHS ($N=4$) are given by,
	
	\vspace{0.2cm}
	
	\begin{eqnarray}
		A^{12}(\Theta_{12}) =
		\begin{pmatrix}
			\cos\Theta_{12} & \sin\Theta_{12} & 0 & 0\\
			-\sin\Theta_{12} & \cos\Theta_{12} & 0 & 0\\
			0 & 0 & 1 & 0 \\
			0 & 0 & 0 & 1
		\end{pmatrix},\hspace{0.3cm}
		A^{34}(\Theta_{34}) =
		\begin{pmatrix}
			1 & 0 & 0 & 0 \\
			0 & 1 & 0 & 0 \\
			0 & 0 & \cos\Theta_{34} & \sin\Theta_{34} \\
			0 & 0 & -\sin\Theta_{34} & \cos\Theta_{34} 
		\end{pmatrix}	
		\label{eq:ADT_4s}
	\end{eqnarray}
	
	\vspace{0.2cm}
	
	\noindent
	With the model form of $A$ matrix [Eq.~(\ref{eq:Amat})-(\ref{eq:ADT_4s})] and the skew-symmetric form of NACM [Eq.~(\ref{eq:Nacm})], the ADT condition [Eq.~(\ref{eq:ADT_eq})] turns into the following six (6) first order coupled differential equations,
	
	\vspace{0.2cm}
	
	\begin{eqnarray}
		\boldsymbol{\nabla}\Theta_{ij} = \sum_{m=1}^{m=4}c^{(m)}\tau_{(m) }, \qquad i = 1, 2, 3;\hspace{0.1cm} j > i,
		\label{eq:ADT_eq_3s}
	\end{eqnarray}
	
	\vspace{0.2cm}
	
	\noindent
	where the coefficients \{$c^{(m)}$\} are trigonometric functions of ADT angles \{$ \Theta_{ij} $\}. The explicit form of the ADT equations and the diabatic PESs are presented in the section S3 and S4 of SM. The NACM elements ($\boldsymbol{\tau}_{ij}$) are calculated at every point of nuclear CS using the \textit{ab initio} based CP-MCSCF method as implemented in the MOLPRO quantum chemistry package~\cite{MOLPRO_2020} and then, employed to calculate the ADT angles ($\{\Theta_{ij}\}$) and the elements of the diabatic potential energy matrix ($\{W_{ij}\}$). Adhikari and co-workers developed a generalized algorithm, ``ADT''~\cite{ADT_JCTC_16} by formulating the ADT quantities, namely the explicit form of ADT equations, NACTs and diabatic matrix elements through symbolic manipulation for $N$ dimensional SHS with $K$- degrees of freedom (DOF). This package can finally provide the numerical solution of ADT angles using the inputs of \textit{ab initio} calculated NACTs to obtain diabatic Hamiltonian. A more comprehensive discussion of the theoretical framework of BBO can be found elsewhere.\cite{Baer1975_CPL_35,top1977incorporation,Adhikari2000_PRA_62_BOII,sarkar2006extended,Sarkar2008_JPCA_112_curl,Baer2002_PhysRep_358,Mukherjee2019_IRPC_38,Dutta2020_PCCP_perspective}
	
	\subsection*{II.D. NACTs: Curl condition and Quantization}
	
	While exploring the existence of $N$-state SHS, we first numerically examined the validity of two conditions at each grid point or along a contour over the nuclear space as described below:
	
	\vspace{0.2cm}
	\noindent
	\textbf{(a) Curl condition:} When the cross derivatives on the scaler form of ADT equations [Eq.~(\ref{eq:ADT_eq})] with respect to the pair of coordinates, $\theta$ and $\phi$ are carried out and subtracted from each other, the analytic continuation of ADT matrix elements \( \Big(\frac{\partial^2 A_{ij}}{\partial\theta \partial\phi} = \frac{\partial^2 A_{ij}}{\partial\phi \partial\theta}\Big)\) provide the curl conditions (also referred to as the molecular field equation):~\cite{Baer2006_bbo_book_sect131,Baer1975_CPL_35,Sarkar2008_JPCA_112_curl}
	
	\vspace{0.2cm}
		
	\begin{eqnarray}
		F_{\theta \phi}^{ij} = \Big[\frac{\partial}{\partial \theta} \tau^{ij}_\phi 
		- \frac{\partial}{\partial \phi} \tau^{ij}_\theta\Big] -
		\Big[(\tau_\phi \tau_\theta)_{ij} - (\tau_\theta \tau_\phi)_{ij}\Big] = 0,
		\label{eq:cc}
	\end{eqnarray}
	
	\vspace{0.2cm}
	
	\noindent
	where $ i,j $ are the two electronic states for a $N$-state SHS [$\{i,j\} \epsilon N$], $\tau_\theta^{ij}$ and $\tau_\phi^{ij}$ represent the scalar components of NACTs along $\theta$ and $\phi$ [$\theta$, $\phi$ represent hyperangles for a fixed hyperradius, $\rho$]. Such condition ensures that mathematical curl ($Z_{\theta \phi}^{ij}$) and the ADT curl ($C_{\theta\phi}^{ij}$) are equal: 
	
	\vspace{0.2cm}
	
	\begin{eqnarray}
		Z_{\theta\phi}^{ij} = C_{\theta\phi}^{ij}, 
		\label{eq:curl_condition} 
	\end{eqnarray}
	
	\vspace{0.2cm}
	
	\noindent
	where
	\begin{align}		
		Z_{\theta \phi}^{ij} & = \frac{\partial}{\partial \theta} \tau^{ij}_\phi - \frac{\partial}{\partial \phi} \tau^{ij}_\theta, \nonumber\\
		C_{\theta \phi}^{ij}  & = (\tau_\phi \tau_\theta)_{ij} - (\tau_\theta \tau_\phi)_{ij}. \nonumber
	\end{align}
	
	\vspace{0.2cm}

	\noindent
	The existence of a chosen SHS at a specific point in the nuclear CS is verified using [Eq.~(\ref{eq:cc}) and Eq.~(\ref{eq:curl_condition})], confirming that the NACTs are free from any non-removable components. Consequently, the calculation of ADT matrix, which incorporates the effect of entire magnitude of NACTs for solving the ADT equations and for the construction of diabatic PES matrix, will be both theoretically ``exact'' and numerically ``accurate''.	
	
	\vspace{0.2cm}
	\noindent
	\textbf{(b) Quantization of NACTs:}
	As per the Cauchy's residue theorem, a line integral along a circular contour encapsulating a singular function accumulates a value, an integer multiple of $\pi$. Since the NACTs could often be singular in the SHS, the following integral~\cite{yarkony1996} has an important role in the process of diabatization:
	\begin{eqnarray}		
		\int_{0}^{2\pi} \tau_\phi^{ij}(\theta)d\phi = 0\hspace{0.2cm} or\hspace{0.2cm} n\pi\hspace{0.2cm} or\hspace{0.2cm} n.(2\pi),
		\label{eq:quantization}
	\end{eqnarray}	
	where $n$ is the number of CIs enclosed by the contour defined by $\phi$ at fixed $\theta$. If the conditions given by Eq. (\ref{eq:curl_condition}) and Eq. (\ref{eq:quantization}) are satisfied at each specific grid point and over the circular nuclear coordinate, respectively, it can be assumed that the involved electronic states form a SHS. Otherwise, one need to expand the size of the SHS to verify those conditions until satisfied.

	\section*{III. \textit{AB INITIO} CALCULATIONS}	
\begin{figure}[!htp]
		\centering
		\includegraphics[width=0.45\linewidth]{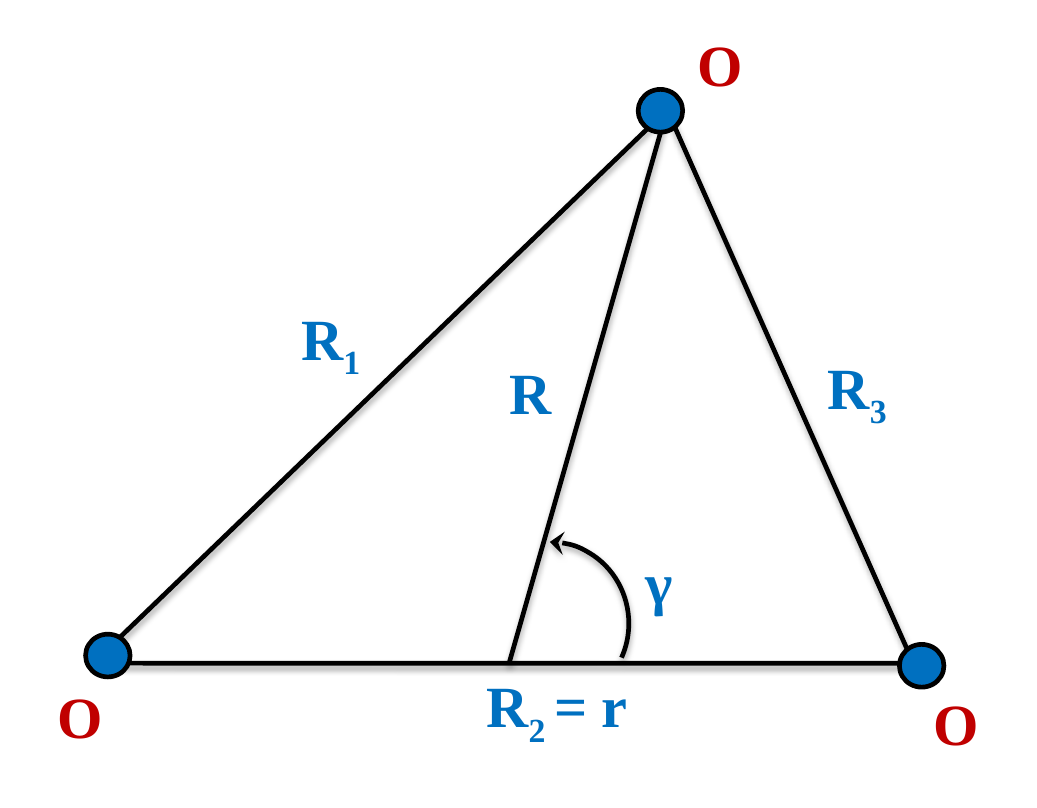}
		\caption{Schematic diagram of the Jacobi coordinates for the O$_3$ system.}
		\label{fig:jacobi}
	\end{figure}

	Since the Jacobi coordinates are physically realizable and offer distinct advantages, we have generated the adiabatic PESs for capturing the features at various geometries and the NACTs for locating the CIs of the triatomic species, O$_3$  as defined by the variables $r$, $R$, and $\gamma$ as depicted in FIG. \ref{fig:jacobi}. The separation between the peripheral atoms of the O$_2$ molecule is denoted as $r$. The third O atom is presented at a distance $R$ from the center of mass of the diatom (O$_2$), where the angle between the vectors ($\textbf{r},\textbf{R}$) is denoted by $\gamma$. The side lengths from the third oxygen atom to the peripheral oxygen atoms are $R_1$ and $R_3$, where the separation between the two peripheral oxygen atoms is termed as $R_2$ (= $r$). All calculations have been performed using the MOLPRO quantum chemistry package~\cite{MOLPRO_2020}
	
	\subsection*{III.A.	Convergence of Active Space and Basis Set for \textit{ab initio} calculation of O$_3$}
	
	\noindent
	The molecular parameters and energetics at selected geometries on the adiabatic ground-state PES of O$_3$ are computed systematically using different active spaces, basis sets and numbers of states in the SA-MCSCF and thereafter, ic-MRCI(Q) calculations, as summarized in Table~\ref{tab:O3_mcscf} and \ref{tab:O3_mrci}, respectively. The ``convergence'' as well as ``accuracy'' of such parameters and the overall smoothness of the adiabatic PESs are examined by: (a) exploring three variants of CASs, namely, (i) outer valence CAS(12e,9o) using only the 2p orbitals; (ii) full valence CAS(18e,12o) including the 2s and 2p orbitals; (iii) extended CAS(24e,15o) including core and valence orbitals (1s, 2s and 2p); (b) expanding the basis set (aug-cc-pVXZ, X = 2, 3, .., 6) from AVDZ to AV6Z and eventually to the complete basis set (CBS) limit; (c) extending the number of singlet states in the SA-MCSCF calculation (3S $\rightarrow$ 13S) and also with three triplet (3T) and three quintet (3Q) states along with three singlet states (3S); and (d) finally, incorporating multireference effects along with Davidson correction. These systematic refinements improve substantially the accuracy of the molecular parameters as well as smoothness of adiabatic PESs within computational feasibility.
	
	\vspace{0.25cm}
	
	\noindent
	\textbf{III.A.1. SA-MCSCF}
	
	\vspace{0.2cm}
	
	\noindent
	 For the first two variants of CAS [(12e,9o) and (18e,12o)] with AVDZ basis set, 3S-SA-MCSCF to 7S-SA-MCSCF calculations depict that the higher CAS(18e,12o) could able to provide not only improved molecular parameters at different geometries but also quite smooth adiabatic PESs both at the interaction and asymptotic regions. The additional improvement obtained by using a larger active space CAS(24e,15o) with the same basis set (AVDZ) and upto 7S-SA-MCSCF calculation is negligibly ``small'' while being computationally more demanding. On the process of averaging over increasing number of states, it appears that both 7S-SA-MCSCF and 13S-SA-MCSCF calculations with CAS(18e,12o) using AVDZ basis set provide smooth adiabatic PESs, but the energetics at specific geometries are not improved significantly. Subsequently, calculations using higher basis sets (AVDZ $\rightarrow$ AV6Z) have been carried out at the 7S-SA-MCSCF level with CAS(18e,12o) for betterment of energetics. In contrast, for the larger active space CAS(24e,15o), the 4S-SA-MCSCF calculations can produce sufficiently smooth adiabatic PESs, where such calculation (4S-SA-MCSCF) for higher basis set (AVDZ $\rightarrow$ AV6Z) can improve energetics incrementally. Although the 4S-SA-MCSCF calculations with CAS(24e,15o) yield converged geometrical parameters and energetics at the AVQZ basis set, the 7S-SA-MCSCF calculations with CAS(18e,12o) achieve convergence in geometrical parameters and energetics only around the interaction region (near \textit{C}$_{2v}$ and \textit{D}$_{3h}$ minima) at the AVQZ level. For the latter active space [CAS(18e,12o)], the energetics of the B1 (O$_3$ $\rightarrow$ O$_2$ + O) and B2 (O$_2$ $\rightarrow$ O + O) dissociation processes increase incrementally with increasing basis set size (AVDZ $\rightarrow$ AV6Z). Nevertheless, nearly converged or converged results for B1 and B2 processes obtained with CAS(18e,12o)/7S-SA-MCSCF and CAS(24e,15o)/4S-SA-MCSCF calculations using the AV6Z basis set, are still far away from the experimentally measured values~\cite{Arnold1994,Herzberg1966}.
	 
	 \vspace{0.2cm}
	 
	 \noindent	 
	 The convergence of molecular parameters and energetics was further examined by including three triplet (3T) and three quintet (3Q) states in addition to three singlet (3S) states in the SA-MCSCF calculations for both CAS(18e,12o) and CAS(24e,15o). Although the resulting PESs are sufficiently smooth in both cases, the predicted geometries and energetics do not show any improvement compared to singlet only approaches.

	\vspace{0.2cm}
	
	\noindent
	The progressive convergence of molecular parameters at different geometries as well as energetics for various processes are depicted in Table~\ref{tab:O3_mcscf} with the change of active space, basis set, and number of states involved in SA-MCSCF approaches with same as well as different spin states.
	 
	 \begin{table}[h!]
	 	\centering
	 	\caption{SA-MCSCF computed energy (eV) differences relative to O + O$_2$ and other configurations of O\textsubscript{3} on the ground ($\tilde{X}^1A'$) adiabatic PES for various active spaces and basis sets.}
	 	\label{tab:O3_mcscf}
	 	\small
	 	\renewcommand{\arraystretch}{0.9} 
	 	\setlength{\tabcolsep}{7pt}
	 	\begin{tabular}{ccccccccc}
	 		\toprule
	 		\textbf{Active Space} & \textbf{Basis Set} & \textbf{States} & $\Delta E_1^a$ & $\Delta E_2^b$ & $\Delta E_3^c$ & $\textit{C}_{2v}$ ($R_1, \alpha$)$^d$ & $\textit{D}_{3h}$ ($R_1$)$^d$ & O$_2$ ($R_2$)$^d$ \\
	 		\midrule
	 		(12e, 9o) & AVDZ & 3S & 1.52 & 0.12 & 3.74 & 1.28, 115.2 & 1.47 & 1.22 \\
	 		&       & 4S  & 0.32 & 0.03 & 2.63 & 1.28, 115.2 & 1.47 & 1.22 \\
	 		&       & 7S  & 0.29 & -1.29 & 3.87 & 1.28, 115.2 & 1.47 & 1.22 \\
	 		
	 		\hline
	 		& AVDZ  & 3S  & 1.43 & 0.25 & 3.85 & 1.28, 115.2 & 1.47 & 1.22 \\
	 		&       & 4S  & 1.41 & 0.17 & 3.88 & 1.28, 115.2 & 1.47 & 1.22 \\
	 		&       & 7S  & 1.45 & 0.12 & 3.92 & 1.29, 116.6 & 1.48 & 1.22 \\
	 		&       & 13S & 1.47 & 0.10 & 3.94 & 1.29, 116.6 & 1.48 & 1.22 \\
	 		(18e, 12o) &  & 3S+3T+3Q & 1.41 & 0.04 & 3.84 & 1.30, 116.5 & 1.47 & 1.22 \\
	 		& AVTZ  & 7S  & 1.34 & 0.24 & 4.05 & 1.29, 116.6 & 1.47 & 1.22 \\
	 		& AVQZ  & 7S  & 1.37 & 0.26 & 4.04 & 1.29, 116.8 & 1.47 & 1.22 \\
	 		& AV5Z  & 7S  & 1.36 & 0.29 & 4.08 & 1.29, 116.8 & 1.47 & 1.22 \\
	 		& AV6Z  & 7S  & 1.36 & 0.40 & 4.09 & 1.29, 116.8 & 1.47 & 1.22 \\
	 		
	 		\hline
	 		& AVDZ  & 3S  & 1.43 & 0.25 & 3.88 & 1.28, 115.2 & 1.47 & 1.22 \\
	 		&       & 4S & 1.43 & 0.17 & 3.91 & 1.29, 115.8 & 1.46 & 1.22 \\
	 		&       & 7S & 1.46 & 0.12 & 3.91 & 1.30, 115.9 & 1.46 & 1.22 \\
	 		&       & 3S+3T+3Q & 1.41 & 0.04 & 3.85 & 1.31, 116.5 & 1.47 & 1.22 \\
	 		(24e, 15o) & AVTZ & 4S & 1.29 & 0.27 & 4.03 & 1.29, 116.7 & 1.47 & 1.22 \\
	 		& AVQZ & 4S & 1.33 & 0.27 & 4.06 & 1.29, 116.8 & 1.47 & 1.22 \\
	 		& AV5Z & 4S & 1.33 & 0.27 & 4.08 & 1.29, 116.8 & 1.47 & 1.22 \\
	 		& AV6Z & 4S & 1.33 & 0.28 & 4.07 & 1.29, 116.8 & 1.47 & 1.22 \\
	 		\hline
	 		\textbf{Expt.} & -- & -- & -- & 1.14$^e$ & 5.12~\cite{Wang2024} & 1.27, 116.7~\cite{Arnold1994} & -- & 1.21~\cite{Herzberg1966} \\
	 		\bottomrule
	 	\end{tabular}
	 	
	 	\vspace{0.5cm}
	 	\raggedright
	 	{\small $^a$ $E(\textit{D}_{3h}) - E(\textit{C}_{2v})$ \\
	 		$^b$ $E(O_2 + O) - E(\textit{C}_{2v})$ \\
	 		$^c$ $E(O + O + O) - E(O_2 + O))$ \\
	 		$^d$ $R_1, R_2, R_3$: $O_1-O_2$, $O_2-O_3$  and $O_3-O_1$ distances (\AA), respectively; $\alpha$: $\angle O_2-\hat{O_1}-O_3$(deg). \\
	 		$^e$ experimental dissociation energy (D$_0$) = (1.0617 $\pm$ 0.0004 eV)~\cite{Ruscic2010} with the zero point energies (ZPEs) of O$_3$ (0.1771 $\pm$ 0.0002 eV)~\cite{egorov2023long} and O$_2$ (0.0965 $\pm$ 10$^{-5}$ eV)~\cite{egorov2023long}.}
	 \end{table}
	 
	 \vspace{0.25cm}
	 
	 \noindent
	 \textbf{III.A.2. ic-MRCI(Q)}
	 
	 \vspace{0.2cm}
	 
	 \noindent
	 Since the molecular parameters for defining global adiabatic PESs at the interaction as well as asymptotic region are crucially important, it appears various SA-MCSCF calculated results even with larger active spaces [CAS(18e,12o) and CAS(24e,15o)] as well as with increasing basis sets (AVDZ $\rightarrow$ AV6Z) 
	 remain far away from the available experimental values, namely, dissociation energy of B1 (O$_3$ $\rightarrow$ O$_2$ + O) and B2 (O$_2$ $\rightarrow$ O + O)  processes.  On the other hand, as the precise description of bond dissociation processes demands the inclusion of dynamic correlation, we carry out internally contracted multi-reference configuration interaction calculations with fixed-reference Davidson correction [ic-MRCI(Q)]~\cite{Werner1988,Knowles1988} for four low-lying electronic states, where those four (low-lying) reference states are taken from (i) 7S-SA-MCSCF with CAS(18e,12o) and (ii) 4S-SA-MCSCF with CAS(24e,15o), using different basis sets. The resulting geometrical parameters as well as energetics show significant improvement and show in good agreement with previous theoretical~\cite{krishna1987theoretical,xantheas1991potential,atchity1997global,muller1998systematic,siebert2002vibrational,holka2010,lepers2012long,dawes2013communication,tyuterev2013new,mankodi2017dissociation,varga2017,egorov2023long} and experimental measurements~\cite{Arnold1994,Herzberg1966} (see Table~\ref{tab:O3_mrci}). 
	 
	 \vspace{0.2cm}
	 
	 \noindent
	 The ic-MRCI(Q) calculations show: (a) The geometrical parameters near the vicinity of \textit{C}$_{2v}$ and \textit{D}$_{3h}$ minima as well as the energy separation between these two minima remain nearly unchanged compared to the SA-MCSCF results, which is due to the similar types of correlation energies at those geometries (see Table~\ref{tab:CBS_corr}); (b) Using AVQZ basis set, the energy gap between the \textit{C}$_{2v}$ minimum and the asymptote of \(\text{O}_2(^3\Sigma_g^-)\) + \(\text{O}(^3P)\) channel improves significantly by reaching 1.101~eV with  CAS(18e,12o), compared to 0.253~eV obtained at the 7S-SA-MCSCF level. For the extended CAS(24e,15o), this gap increases to 1.110~eV relative to the 4S-SA-MCSCF value (0.270 eV), where the experimentally measured value~\cite{Arnold1994} is 1.144 eV. Moreover, the dissociation energy of molecular O$_2$ is found to be 5.106~eV with CAS(18e,12o), compared to the corresponding 7S-SA-MCSCF calculated value, 4.04~eV. Again, for the larger CAS(24e,15o), such dissociation energy is 5.145~eV, compared to 4.06~eV obtained at the 4S-SA-MCSCF level. The dissociation energy for B2 process calculated with CAS(18e,12o) and AVQZ basis (5.106 eV) shows slightly closer agreement with experiment (5.117~eV)~\cite{Wang2024} compare to the corresponding case with CAS(24e,15o) and AVQZ basis (5.145 eV); (c) Numerical instabilities of electronic structure calculations using 7S-SA-MCSCF followed by ic-MRCI(Q) approaches become progressively more pronounced at large $R (\geq 7$–8 bohr) for the CAS(18e,12o), which makes difficult to obtain reliable dissociation energies in this region for both the B1 and B2 processes. To overcome this issue, a long-range correction is introduced using the methodology proposed by Tyuterev \textit{et al.}~\cite{tyuterev2013new}, where at present, the 7-SA-MCSCF calculations are first performed using the CAS(12e,9o) and therein, both the 1s and 2s orbitals are optimized with active 2p orbitals. The resulting orbitals are re-optimized at the 7S-SA-MCSCF level using CAS(18e,12o) within the frozen-core approximation (i.e., 1s orbitals kept frozen). These reference orbitals are employed in ic-MRCI(Q) calculations using CAS(18e,12o) with frozen-core approximation. On the contrary, it appears that such treatment is not required for the CAS(24e,15o) calculations; (d) Upon further increase of basis set size [AVQZ $\rightarrow$ AV5Z $\rightarrow$ AV6Z], the dissociation energy for the B$_1$ process continues to improve [1.101 $\rightarrow$ 1.119 $\rightarrow$ 1.130 eV for CAS(18e,12o) and 1.110 $\rightarrow$ 1.130 $\rightarrow$ 1.136 eV for CAS(24e,15o)] and approaches the experimental value (1.144 eV). However, the higher basis sets [AVQZ $\rightarrow$ AV5Z $\rightarrow$ AV6Z] tend to ``slightly'' overestimate the experimental value of O$_2$ dissociation energy, 5.117 eV in the B$_2$ process, yielding values of 5.106 $\rightarrow$ 5.139 $\rightarrow$ 5.147 eV for CAS(18e,12o) and 5.145 $\rightarrow$ 5.183 $\rightarrow$ 5.192~eV for CAS(24e,15o). For the CBS extrapolation based on both the CAS(18e,12o) and CAS(24e,15o), the AVQZ $\rightarrow$ AV5Z $\rightarrow$ AV6Z $\rightarrow$ CBS(56Z) scheme is adopted, where long-range correction is incorporated for the former CAS(18e,12o) at large $R$ as described above~\cite{tyuterev2013new}. The extrapolation to the CBS limit is made using the E\textsubscript{cbs} + A/X$^3$ functional, where E\textsubscript{cbs} corresponds to the total energy (reference + correlation). This strategy reproduces the experimental dissociation energy for the B1 process (O$_3$ $\rightarrow$ O$_2$ + O) reasonably well, but ``slightly'' overestimates the experimental value of B2 process (O$_2$ $\rightarrow$ 2O). (e) From the analysis on computational time and memory requirement (see Table~S1 of SM), it is observed that the CAS(18e,12o) and AVQZ basis set is the optimal one for generating multi-state global adiabatic PESs, which not only provide closely ``accurate'' description of both the interaction region as well as the asymptotic dissociation limits, but also computationally feasible. Although the CAS(24e,15o) increases the ``accuracy'' of the PES, but it is highly computationally demanding.; (f) To further assess the accuracy of the PESs, vibrational calculations are performed using the CAS(18e,12o) with different basis sets, as summarized in Table~\ref{tab:O3_vib}. These calculations yield converged equilibrium geometries ($r_e$, $\alpha_e$) and harmonic (vibrational) frequencies ($\omega_i$s) of ozone for the CAS(18e,12o) with AVQZ basis set and provided results that are in closest agreement with the available experimental~\cite{lee1990vibrational} and other theoretical data~\cite{holka2010,tyuterev2004variational,tyuterev2013new}.
	 
	 \begin{table}[h!]
	 	\centering
	 	\caption{Comparison of molecular parameters and energetics at specific geometries on the ground-state adiabatic ($\tilde{X}^1A'$) PES of O\textsubscript{3}, computed at the ic-MRCI(Q) level using different active spaces and basis sets, together with other theoretical and experimental data.}
	 	\label{tab:O3_mrci}
	 	\renewcommand{\arraystretch}{0.67} 
	 	\setlength{\tabcolsep}{10pt}
	 	\begin{tabular}{lcccccc}
	 		\toprule
	 		\textbf{References} & $\Delta E_1^a$ & $\Delta E_2^b$ & $\Delta E_3^c$ & $\textit{C}_{2v}$ ($R_1, \alpha$)$^d$ & $\textit{D}_{3h}$ ($R_1$)$^d$ & O$_2$ ($R_2$)$^d$ \\
	 		\hline
	 		\addlinespace[0.2cm]
	 		M\"uller \textit{et al.}~\cite{muller1998systematic} & 1.279 & 1.062 & -- & 1.267, 117.2 & 1.438 & -- \\
	 		Siebert \textit{et al.}~\cite{siebert2002vibrational} & 1.340 & 1.027 & -- & 1.275, 116.8 & 1.441 & -- \\
	 		Holka \textit{et al.}~\cite{holka2010} & -- & 1.110 & -- & 1.270, 116.8 & -- & 1.203 \\
	 		Dawes \textit{et al.}~\cite{dawes2013communication} & 1.330 & 1.160 & -- & 1.270, 116.8 & 1.440 & -- \\
	 		Tyuterev \textit{et al.}~\cite{tyuterev2013new} & -- & 1.119 & -- & 1.275, 116.8 & -- & 1.206 \\
	 		Varga \textit{et al.}~\cite{varga2017} & 1.190 & 1.160 & -- & 1.270, 119.4 & 1.430 & 1.210 \\
	 		Egorov \textit{et al.}~\cite{egorov2023long} & -- & 1.143 & -- & 1.269, 116.9 & -- & 1.205 \\
	 		\addlinespace[0.2cm]
	 		\cline{1-7}
	 		
	 		\addlinespace[0.2cm]
	 		\textbf{Present Work}$^e$ &  &  &  &  &  &  \\
	 		
	 		\addlinespace[0.2cm]
	 		\textit{CAS (18e,12o)} & & & & & & \\
	 		\addlinespace[0.2cm]
	 		
	 		AVDZ & 1.467 & 0.722 & 4.635 & 1.293, 116.6 & 1.483 & 1.217 \\
	 		AVTZ & 1.342 & 1.019 & 5.013 & 1.288, 116.8 & 1.470 & 1.217 \\
	 		\rowcolor{gray!15}
	 		AVQZ & 1.365 & 1.101 & 5.106 & 1.288, 116.8 & 1.470 & 1.217 \\
	 		AV5Z & 1.369 & 1.119 & 5.139 & 1.288, 116.8 & 1.467 & 1.217 \\
	 		AV6Z & 1.369 & 1.130 & 5.147 & 1.288, 116.8 & 1.466 & 1.217 \\
	 		CBS(56Z) & 1.370 & 1.143 & 5.155 & 1.288, 116.8 & 1.466 & 1.217 \\
	 		
	 		\addlinespace
	 		\textit{CAS (24e,15o)} & & & & & & \\
	 		\addlinespace[0.2cm]
	 		
	 		AVTZ & 1.372 & 1.088 & 5.106 & 1.290, 116.7 & 1.470 & 1.217 \\
	 		AVQZ & 1.379 & 1.110 & 5.145 & 1.284, 115.3 & 1.470 & 1.217 \\
	 		AV5Z & 1.380 & 1.130 & 5.183 & 1.284, 115.3 & 1.470 & 1.217 \\
	 		AV6Z & 1.380 & 1.136 & 5.192 & 1.284, 115.3 & 1.470 & 1.217 \\
	 		CBS(56Z) & 1.380 & 1.142 & 5.201 & 1.288, 116.8 & 1.466 & 1.217 \\
	 		\addlinespace[0.2cm]
	 		\cline{1-7}
	 		\textbf{Expt.} & -- & 1.144$^f$ & 5.117~\cite{Wang2024} & 1.273, 116.8~\cite{Arnold1994} & -- & 1.207~\cite{Herzberg1966} \\
	 		\bottomrule
	 	\end{tabular}
	 	
	 	\vspace{0.5cm}
	 	\raggedright
	 	{\small $^a$ $E(\textit{D}_{3h}) - E(\textit{C}_{2v})$ \\
	 		$^b$ $E(O_2 + O) - E(\textit{C}_{2v})$ \\
	 		$^c$ $E(O + O + O) - E(O_2 + O))$ \\
	 		$^d$ $R_1, R_2, R_3$: $O_1-O_2$, $O_2-O_3$  and $O_3-O_1$ distances (\AA), respectively; $\alpha$: $\angle O_2-\hat{O_1}-O_3$(deg). \\
	 		$^e$ Computed using 7-SA-MCSCF followed by ic-MRCI(Q) \\
	 		$^f$ experimental dissociation energy (D$_0$) = (1.0617 $\pm$ 0.0004 eV)~\cite{Ruscic2010} with the zero point energies (ZPEs) of O$_3$ (0.1771 $\pm$ 0.0002 eV)~\cite{egorov2023long} and O$_2$ (0.0965 $\pm$ 10$^{-5}$ eV)~\cite{egorov2023long}.}
	 \end{table}
	 
	 \begin{table}[h!]
		\centering
		\caption{Variation of correlation energies (in eV) for different geometries of O\textsubscript{3} at the ic-MRCI(Q) level on the ground adiabatic ($\tilde{X}^1A'$) PES using CAS(18e,12o) with increasing basis set size (AVTZ $\rightarrow$ AV6Z).}
		\label{tab:CBS_corr}
		\renewcommand{\arraystretch}{0.7}
		\setlength{\tabcolsep}{12pt}
		\begin{tabular}{ccccc}
			\toprule
			\textbf{Geometry} & AVTZ & AVQZ & AV5Z & AV6Z \\
			\midrule
			\textit{C}$_{2v}$ Minima  & -15.74 & -16.83 & -17.22 & -17.38 \\
			\textit{D}$_{3h}$ Minima  & -15.73 & -16.83 & -17.22 & -17.38 \\
			O$_2$ + O Channel & -14.93 & -15.99 & -16.36 & -16.50 \\
			O + O + O Asymptote & -14.00 & -14.96 & -15.26 & -15.41 \\
			\bottomrule
		\end{tabular}
	 \end{table}
	 
	 \begin{table}[h!]
	 	\centering
	 	\caption{Comparison of optimized equilibrium geometries ($r_e$ and $\alpha_e$) and harmonic frequencies of ozone for different basis sets using CAS(18e,12o) at the ic-MRCI(Q) level, along with previous theoretical and experimental data. Bond lengths are in \AA, angles in degrees ($^\circ$), and harmonic frequencies ($\omega$) in cm$^{-1}$.}
	 	\label{tab:O3_vib}
	 	\renewcommand{\arraystretch}{0.8}
	 	\setlength{\tabcolsep}{8pt}
	 	
	 	\begin{tabular}{lccccc}
	 		\toprule
	 		\textbf{Basis / Source} 
	 		& $r_e^a$ (\AA) 
	 		& $\alpha_e^b$ ($^\circ$) 
	 		& $\omega_1$ ($a_1$) 
	 		& $\omega_2$ ($a_1$) 
	 		& $\omega_3$ ($b_2$) \\
	 		\midrule
	 		AVDZ  & 1.272 & 116.8 & 1088.9 & 696.7 & 1021.2 \\
	 		AVTZ  & 1.272 & 116.8 & 1120.8 & 708.1 & 1074.2 \\
	 		AVQZ  & 1.272 & 116.8 & 1141.2 & 718.6 & 1103.1 \\
	 		AV5Z  & 1.272 & 116.8 & 1145.8 & 720.5   & 1109.4 \\
	 		AV6Z  & 1.272 & 116.8 & 1147.0 & 721.6 & 1112.0 \\
	 		\midrule
	 		Holka \textit{et al.}~\cite{holka2010} & 1.275 & 116.8 & 1133.2 & 712.5 & 1089.4 \\
	 		Tyuterev \textit{et al.}~\cite{tyuterev2013new} & --    & --    & 1135.8  & 714.2  & 1093.7  \\
	 		Empirical~\cite{tyuterev2004variational} & -- & -- & 1132.1 & 714.4 & 1087.1 \\
	 		\midrule
	 		Expt.~\cite{lee1990vibrational} 
	 		& 1.272 & 116.8 & 1134.9 & 716.0 & 1089.2 \\
	 		\bottomrule
	 	\end{tabular}
	 	
	 	\vspace{0.3cm}
	 	\raggedright
	 	{\small
	 		$^a$ Optimized O--O bond length \\
	 		$^b$ Optimized O--O--O bond angle
	 	}
	 \end{table}
	
	\vspace{0.2cm}
	
	\noindent
	The optimal choice of the active space as well as the basis set is the crucial aspect for the \textit{ab initio} calculation of O$_3$ to generate not only multi-state global adiabatic PESs but also NACTs between those states for the construction of diabatic Hamiltonian, and thereafter, to calculate cross sections and rate constants through scattering calculation on such coupled PESs. In order to describe the overall strategy for the construction of multi-surface diabatic Hamiltonian, we have presented the details in Section S2 of SM.
	
	\newpage
	\subsection*{III.B. Adiabatic PESs}
	\textbf{III.B.1. 7S-SA-MCSCF PESs} \\
	We employ the Jacobi coordinate system to generate multi-state adiabatic PESs over \(r\)-\(R\) space by varying each of the coordinates from 1 a.u. to 12 a.u. (0.53\AA\ – 6.35\AA) with a grid of \textit{ab initio} points 111 $\times$ 111 for \(\gamma = 0^\circ\)(collinear), \(45^\circ\)(intermediate bending), and \(90^\circ\) (\(\text{C}_{2v}\)) geometries of ozone using 7S-SA-MCSCF method with CAS(18e,12o) and AVQZ basis set. Since the lowest four singlet states ($\tilde{X}^1A'$, $1^1A''$, $1^1A'$ and $2^1A''$) are interacting with each other within 4 eV at various values of $\gamma$ through symmetry driven and accidental CIs and those states form a SHS (see subsection IV.A.), we intend to construct four (4) state diabatic Hamiltonian (4$\times$4). The low-lying four singlet adiabatic PESs for $\gamma$ = 0$^\circ$, 45$^\circ$ and 90$^\circ$ are displayed in FIGs.~S2, S3 and S4 in section S6 of the SM as functions of Jacobi coordinates ($r$, $R$) with 1 $\le$ $r$ $\le$ 12 a.u. and 1 $\le$ $R$ $\le$ 12 a.u.

	\vspace{0.2cm}
		
	\noindent	
	At $\gamma = 90^\circ$, the ground adiabatic PES ($\tilde{X}^1A'$) of ozone exhibits two distinct features: (i) a global minimum with \textit{C}$_{2v}$ structure at \{$r$ = 1.45 Bohr ,\ $R$ = 1.27 Bohr\} and (ii) a \textit{D}$_{3h}$ ring with relatively less stable minimum at \{$r$ = 2.78 Bohr,\ $R$ = 2.40 Bohr\}, where the minimum is 1.37 eV higher than \textit{C}$_{2v}$ one. Consequently, the ground ($\tilde{X}^1A'$) adiabatic PES depicts four (4) equilibrium structures: a higher-energy \textit{D}$_{3h}$ ring minimum and three lower-energy \textit{C}$_{2v}$ open minima arising due to nuclear permutation, where the other two \textit{C}$_{2v}$ minima can be found at $\gamma$ = 45$^\circ$ and 135$^\circ$. The transition from \textit{C}$_{2v}$ to \textit{D}$_{3h}$ structure passes through a trasition state (TS1) at \{$r$ = 3.0 Bohr,\ $R$ = 2.4 Bohr\}, while the strained \textit{D}$_{3h}$ ring dissociates into \(\text{O}_2(^3\Sigma_g^-)\) + \(\text{O}(^3P)\) via another trasition state (TS2) at \{$r$ = 4.57 \AA,\ $R$ = 2.28 \AA\}. The \(\text{O}_2(^3\Sigma_g^-)\) + \(\text{O}(^3P)\) channel is represented by a well, which lies 0.26 eV above the \textit{C}$_{2v}$ minimum. The dissociation of molecular O$_2$ [\(\text{O}_2(^3\Sigma_g^-)\) $\rightarrow$ \(2\text{O}(^3P)\)] occurs at 4.12 eV with respect to entrance channel. On the other hand, the three excited states ($1^1A''$, $1^1A'$, and $2^1A''$) lack the \textit{D}$_{3h}$ ring minimum, but the minima at \textit{C}$_{2v}$ configurations are above the \(\text{O}_2(^3\Sigma_g^-)\) + \(\text{O}(^3P)\) channel. Note that in Table~\ref{tab:O3_params}, the dissociation channel of ozone (\(\text{O}_2(^3\Sigma_g^-)\) + \(\text{O}(^3P)\)) is endothermic w.r.t. the \textit{C}$_{2v}$ minima of the ground adiabatic state, but such channel is exothermic from the corresponding minima of the respective excited adiabatic states ($1^1A''$, $1^1A'$ and $2^1A''$). On the contrary, the dissociation energy of O$_2$ from the first three  adiabatic states are essentially same, but it differs in the third excited ($2^1A''$) state.
	
	\begin{table}[h!]
		\centering
		\caption{Dissociation energy of B1 (O$_3$ $\rightarrow$ O$_2$ + O) and B2 (O$_2$ $\rightarrow$ 2O) processes calculated using CAS(18e,12o) and AVQZ basis set.}
		\label{tab:O3_params}
		\renewcommand{\arraystretch}{0.8}
		\setlength{\tabcolsep}{12pt}
		\begin{tabular}{llcc}
			\toprule
			\textbf{PES} & \textbf{Method} & $\Delta E_2^a$ & $\Delta E_3^b$ \\
			\midrule
			$\tilde{X}^1A'$ & 7S-SA-MCSCF & 0.260 & 4.040 \\
			& ic-MRCI(Q)  & 1.101 & 5.098 \\
			\midrule
			$1^1A''$ & 7S-SA-MCSCF & -1.176 & 4.032 \\
			& ic-MRCI(Q)  & -0.378 & 5.121 \\
			\midrule
			$1^1A'$ & 7S-SA-MCSCF & -1.619 & 4.083 \\
			& ic-MRCI(Q)  & -0.851 & 5.115 \\
			\midrule
			$2^1A''$ & 7S-SA-MCSCF & 0.702 & 1.001 \\
			& ic-MRCI(Q)  & 1.237 & 2.152 \\		
			\bottomrule
		\end{tabular}
		
		\vspace{0.3cm}
		\raggedright
		{\small
			$^a$ $E(O_2 + O) - E(\textit{C}_{2v})$ \\
			$^b$ $E(O + O + O) - E(O_2 + O))$ \\
		}
	\end{table}

	\vspace{0.5cm}
	
\begin{figure}[!htp]
		\centering
		\centering
		\hspace*{-3cm}
		\begin{subfigure}{0.4\linewidth}
			\centering
			\begin{overpic}[width=1.4\linewidth]{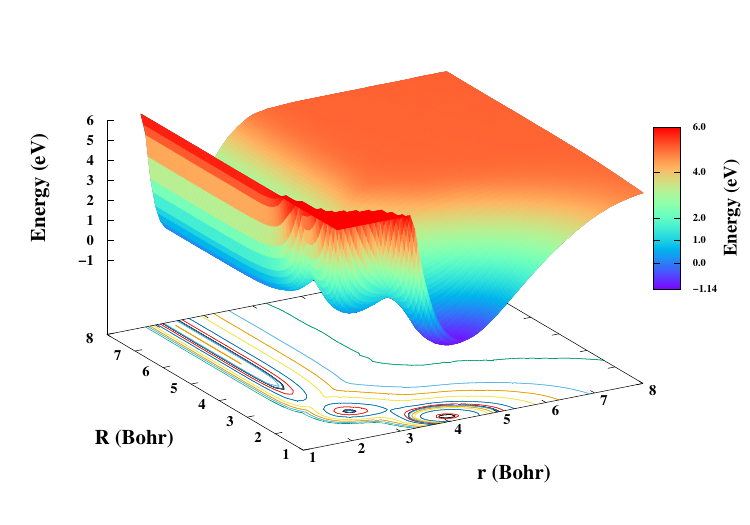}
				\put(1.5,70){\textbf{(a)} \hspace{4cm} u\textsubscript{1}} 
			\end{overpic}
			\phantomcaption
			\label{fig:u1_jacobi_mrci}
		\end{subfigure}
		\hspace{2.5cm}
		\begin{subfigure}{0.4\linewidth}
			\centering
			\begin{overpic}[width=1.4\linewidth]{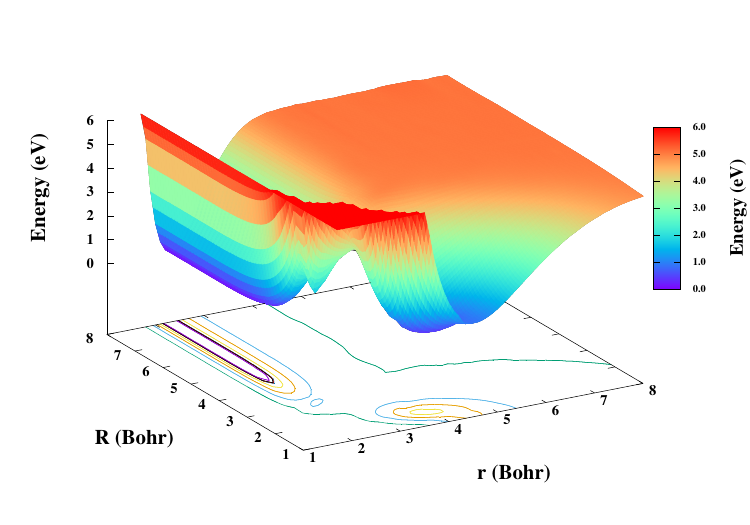}
				\put(1.5,70){\textbf{(b)} \hspace{4cm} u\textsubscript{2}}
			\end{overpic}
			\phantomcaption
			\label{fig:u2_jacobi_mrci}
		\end{subfigure}
		\hspace*{-3cm}
		\begin{subfigure}{0.4\linewidth}
			\centering
			\begin{overpic}[width=1.4\linewidth]{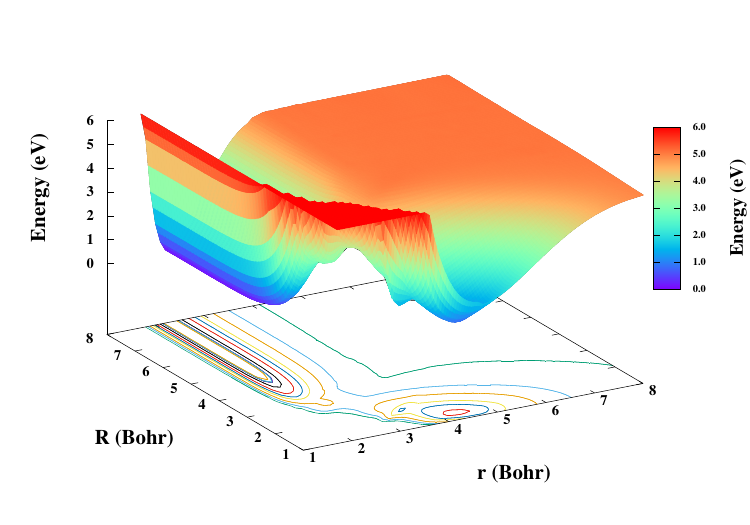}
				\put(1.5,70){\textbf{(c)} \hspace{4cm} u\textsubscript{3}}
			\end{overpic}
			\phantomcaption
			\label{fig:u3_jacobi_mrci}
		\end{subfigure}
		\hspace{2.5cm}
		\begin{subfigure}{0.4\linewidth}
			\centering
			\begin{overpic}[width=1.4\linewidth]{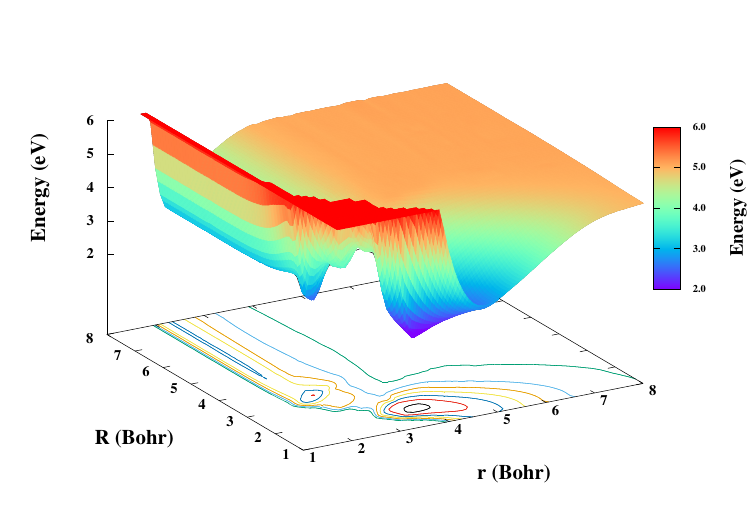}
				\put(1.5,70){\textbf{(d)} \hspace{4cm} u\textsubscript{4}}
			\end{overpic}
			\phantomcaption
			\label{fig:u4_jacobi_mrci}
		\end{subfigure}			
		\caption{Four low lying adiabatic PESs of O$_3$ [Figures (a) to (d)] at $\gamma = 90^\circ$ computed using 7S-SA-MCSCF followed by ic-MRCI(Q) with a CAS(18e,12o) and AVQZ basis set.  In reference to the reactant channel, O + O$_2$ setting at 0, \textit{C}$_{2v}$ and \textit{D}$_{3h}$ minima are at -1.101 eV and -0.264 eV respectively, respectively and the dissociation energy of O$_2$ is indicated as 5.10 eV.}
		\label{fig:adia_mrci_jacobi_90}
	\end{figure}

\begin{figure}[!htp]
		\centering
		\hspace*{-3cm}
		\begin{subfigure}{0.4\linewidth}
			\centering
			\begin{overpic}[width=1.5\linewidth]{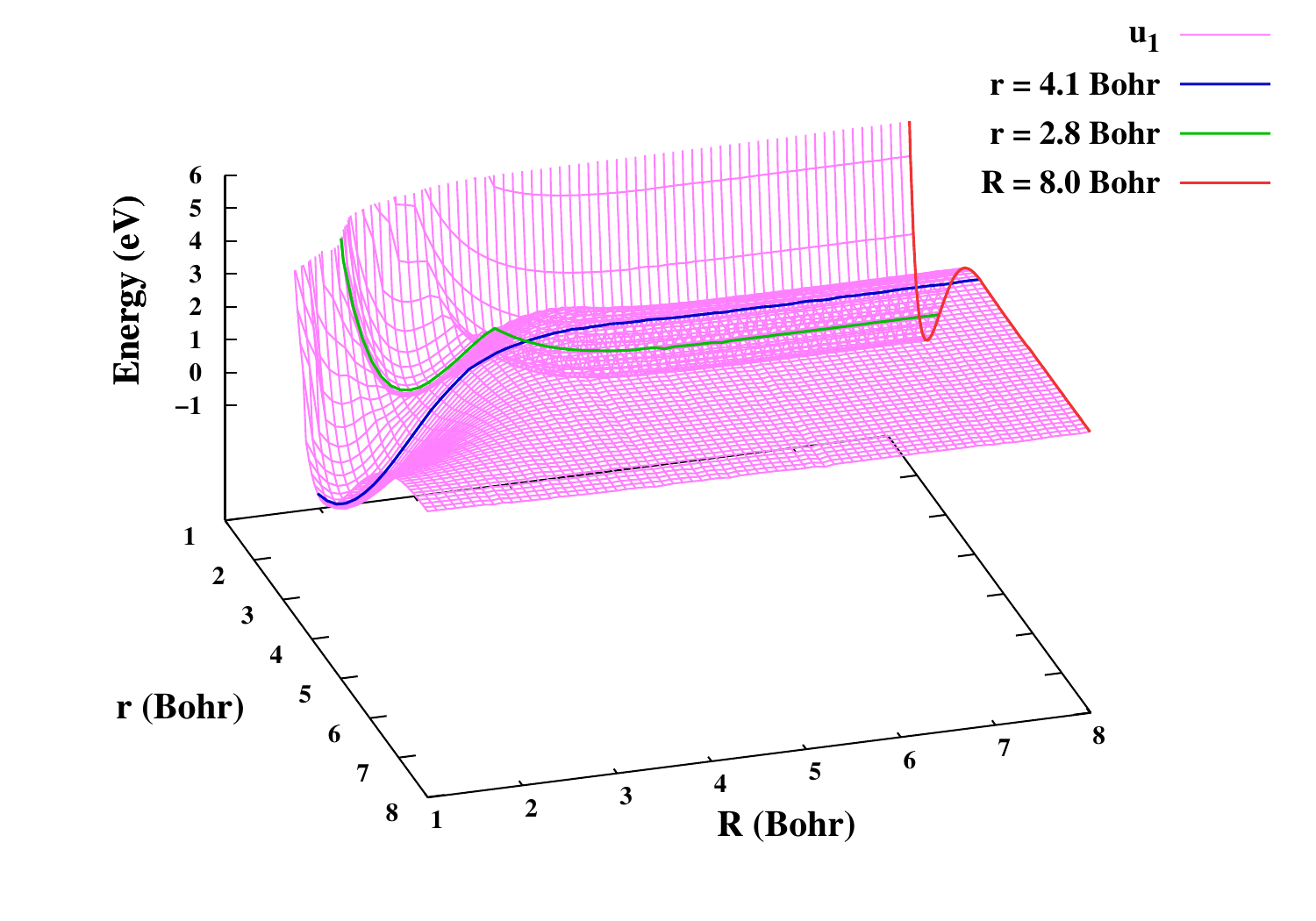}
				\put(15,75){\textbf{(a)}} 
			\end{overpic}
			\label{fig:C2v_min}
		\end{subfigure}
		\hspace{3.5cm}
		\begin{subfigure}{0.4\linewidth}
			\centering
			\begin{overpic}[width=1.4\linewidth]{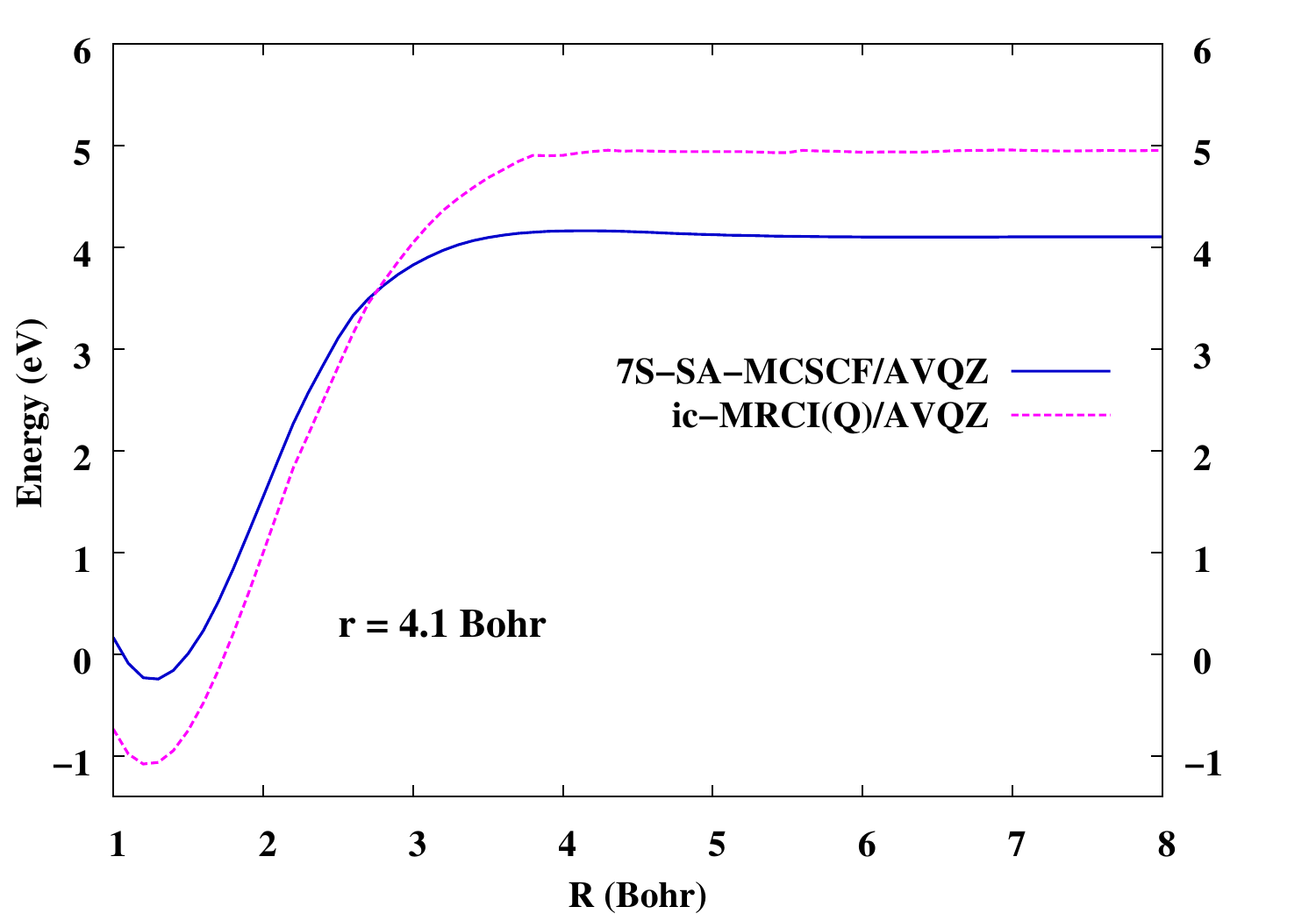}
				\put(-2,75){\textbf{(b)}}
			\end{overpic}
			\label{fig:C2v_min}
		\end{subfigure}		
		\vspace{0.5cm} 
		\hspace*{-2cm}		
		\begin{subfigure}{0.4\linewidth}
			\centering
			\begin{overpic}[width=1.4\linewidth]{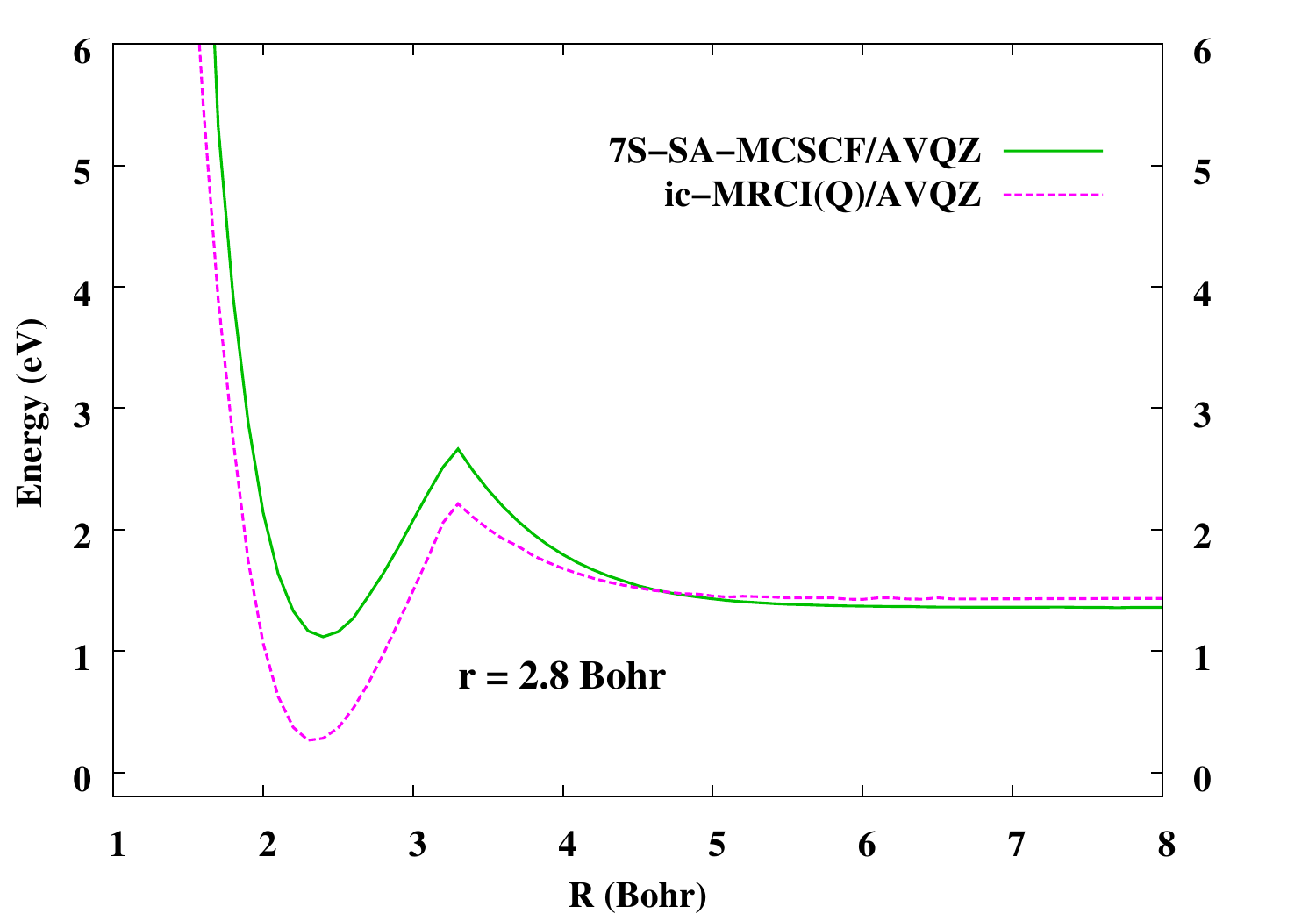}
				\put(5,75){\textbf{(c)}}
			\end{overpic}
			\label{fig:d3h_min}
		\end{subfigure}
		\hspace{2.5cm}
		\begin{subfigure}{0.4\linewidth}
			\centering
			\begin{overpic}[width=1.4\linewidth]{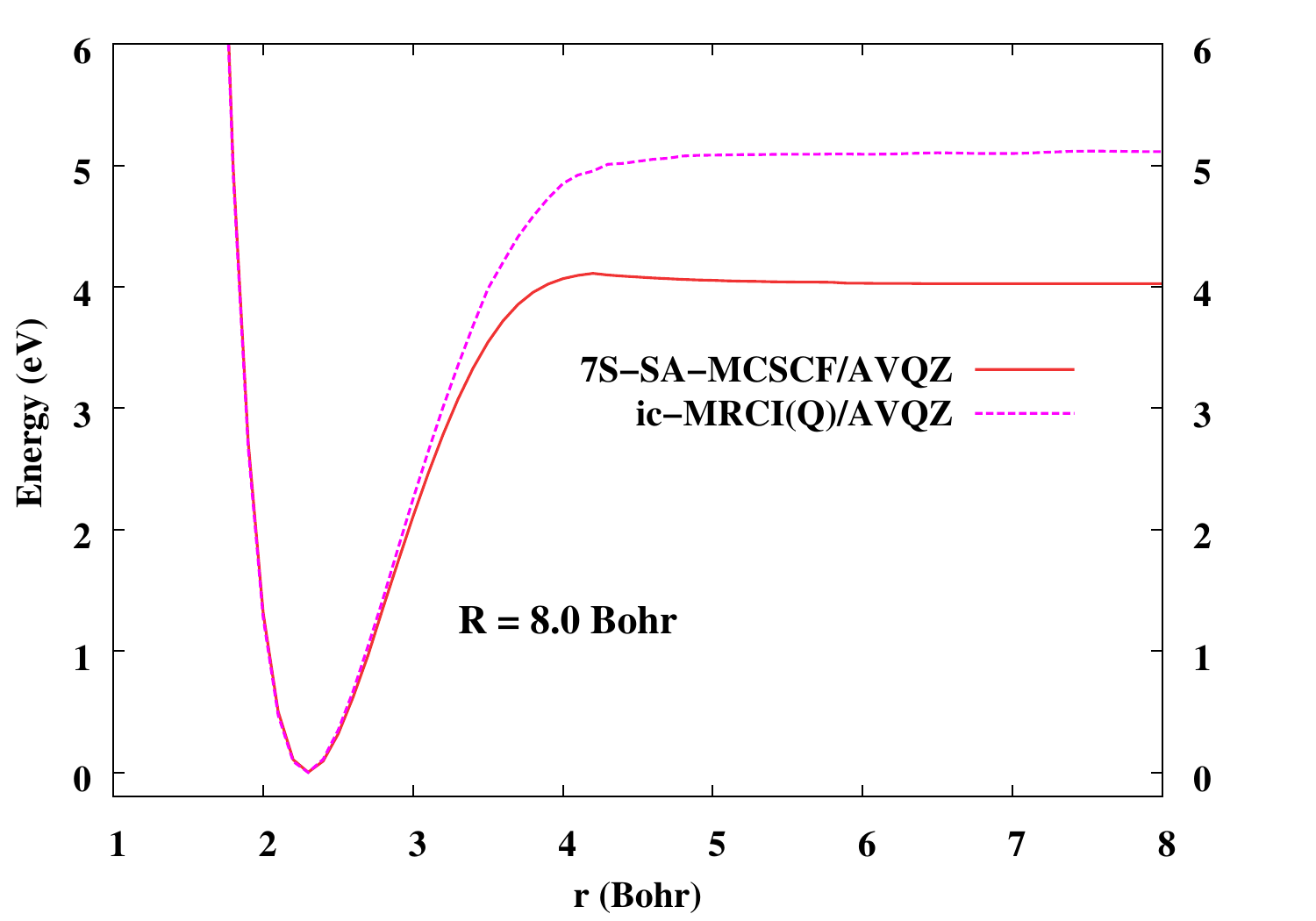}
				\put(-2,75){\textbf{(d)}}
			\end{overpic}
			\label{fig:O2_diss}
		\end{subfigure}		
		\caption{Figure (a) represents the ground adiabatic $\tilde{X}^1A'$ (u$_1$) PES computed using 7-SA-MCSCF method with CAS(18e,12o) and the AVQZ basis set. Three marked PECs on the PES correspond to \( r = 4.1 \) Bohr (blue), \( r = 2.8 \) Bohr (green), and \( R = 8.0 \) Bohr (red), representing the \textit{C}$_{2v}$, \textit{D}$_{3h}$ and O$_2$ dissociation regions via the B2 process, respectively. Figures (b), (c), and (d) presents the comparison between the 7-SA-MCSCF and ic-MRCI(Q) data over the marked PECs using same CAS and basis. All energies are referenced to the asymptotic \(\text{O}_2(^3\Sigma_g^-)\) + \(\text{O}(^3P)\) channel.}
		\label{fig:mrci_PECs}
	\end{figure}

	\newpage
	\noindent
	\textbf{III.B.2. ic-MRCI(Q) Calculation} \\	
	\noindent
	Using the low-lying four reference states ($\tilde{X}^1A'$, $1^1A''$, $1^1A'$ and $2^1A''$) obtained from 7S-SA-MCSCF, we perform ic-MRCI(Q) calculation employing the same active space, CAS(18e,12o) and the basis set, AVQZ  at $\gamma = 90^\circ$ as a function of $r$ and $R$. Those four low-lying PESs calculated at the ic-MRCI(Q) level are depicted in four separate panels as shown in FIG~\ref{fig:adia_mrci_jacobi_90}, whereas the improvements over the specific regions of the ground adiabatic state using ic-MRCI(Q) method are illustrated through 1D potential energy curves (PECs) in FIG.~\ref{fig:mrci_PECs}. All molecular parameters and energetics of the ground adiabatic PES computed at the ic-MRCI(Q) level are summarized in Table~\ref{tab:O3_mrci}, while Table~\ref{tab:TSs} compares the locations of the TSs with earlier theoretical results~\cite{kalemos2008,siebert2002vibrational,Chen2011,varga2017,mankodi2017dissociation,Wang2025}. The locations of the \textit{D}$_{3h}$ \{$r$ = 2.78 Bohr, $R$ = 2.40 Bohr\} and \textit{C}$_{2v}$ \{$r$ = 1.45 Bohr, $R$ = 1.27 Bohr\} minima, as well as the energy separation between them (1.365 eV), remain nearly unchanged relative to the 7S-SA-MCSCF level (1.370). However, at the ic-MRCI(Q) level, the energy gap between the \textit{C}$_{2v}$ minima and the dissociation limit of O$_3$ [\(\text{O}_2(^3\Sigma_g^-)\) + \(\text{O}(^3P)\)] increases to 1.101 eV compared to the 7S-SA-MCSCF value, 0.26 eV. The dissociation of O$_2$ occurs at 5.098 eV relative to the entrance channel, significantly higher than the 7S-SA-MCSCF value (4.04 eV). For the excited singlet states, the corresponding energy gaps are listed in Table~\ref{tab:O3_params}.
	
	\begin{table}[h!]
		\centering
		\caption{Convergence of the location of TSs on the ground ($\tilde{X}^1~A'$) adiabatic PES of O\textsubscript{3} using different basis sets.}
		\label{tab:TSs}
		\renewcommand{\arraystretch}{0.7} 
		\setlength{\tabcolsep}{10pt}      
		\begin{tabular}{ccccc}
			\toprule
			\textbf{TS} & R$_1$ (\AA) & R$_2$ (\AA) & $\alpha$ ($^\circ$) & $\Delta$E (eV)~$^a$ \\
			\midrule
			TS$_1~^b$ & & & & \\
			Kalemos \textit{et al.}~\cite{kalemos2008} & 1.46 & 1.46 & 87.0 & 1.38 \\
			Chen \textit{et al.}~\cite{Chen2011} & 1.41 & 1.41 & 84.0 & 1.28 \\
			SSB PES~\cite{siebert2002vibrational} & 1.42 & 1.42 & 83.1 & 1.34 \\
			PIP PES~\cite{varga2017} & 1.39 & 1.39 & 81.0 & 1.26 \\
			RKHS PES~\cite{Wang2025} & 1.41 & 1.41 & 80.5 & 0.94 \\
			Present work~$^d$ & 1.39 & 1.39 & 80.7 & 1.05 \\
			\midrule
			TS$_2~^c$ & & & & \\
			Mankodi \textit{et al.}~\cite{mankodi2017dissociation} & 1.80 & 1.80 & 44.0 & 1.70 \\
			Present work~$^d$ & 1.71 & 1.71 & 43.6 & 2.76 \\
			\bottomrule
		\end{tabular}
		
		\vspace{0.5cm}		
		\raggedright
		{\small $^a$ Energy differences are relative to the $\text{O}_2(^3\Sigma_g^-) + \text{O}(^3P)$ entrance channel. \\
			$^b$ TS \{$\textit{C}_{2v}$ $\rightarrow$ $\textit{D}_{3h}$\}. \\
			$^c$ TS \{$\textit{D}_{3h}$ $\rightarrow$ $\text{O}_2(^3\Sigma_g^-) + \text{O}(^3P)$\}. \\
			$^d$ Computed with 7S-SA-MCSCF followed by ic-MRCI(Q) using AVQZ basis set.}	
	\end{table}	

\begin{figure}[!htp]
		\centering
		\includegraphics[width=0.6\linewidth]{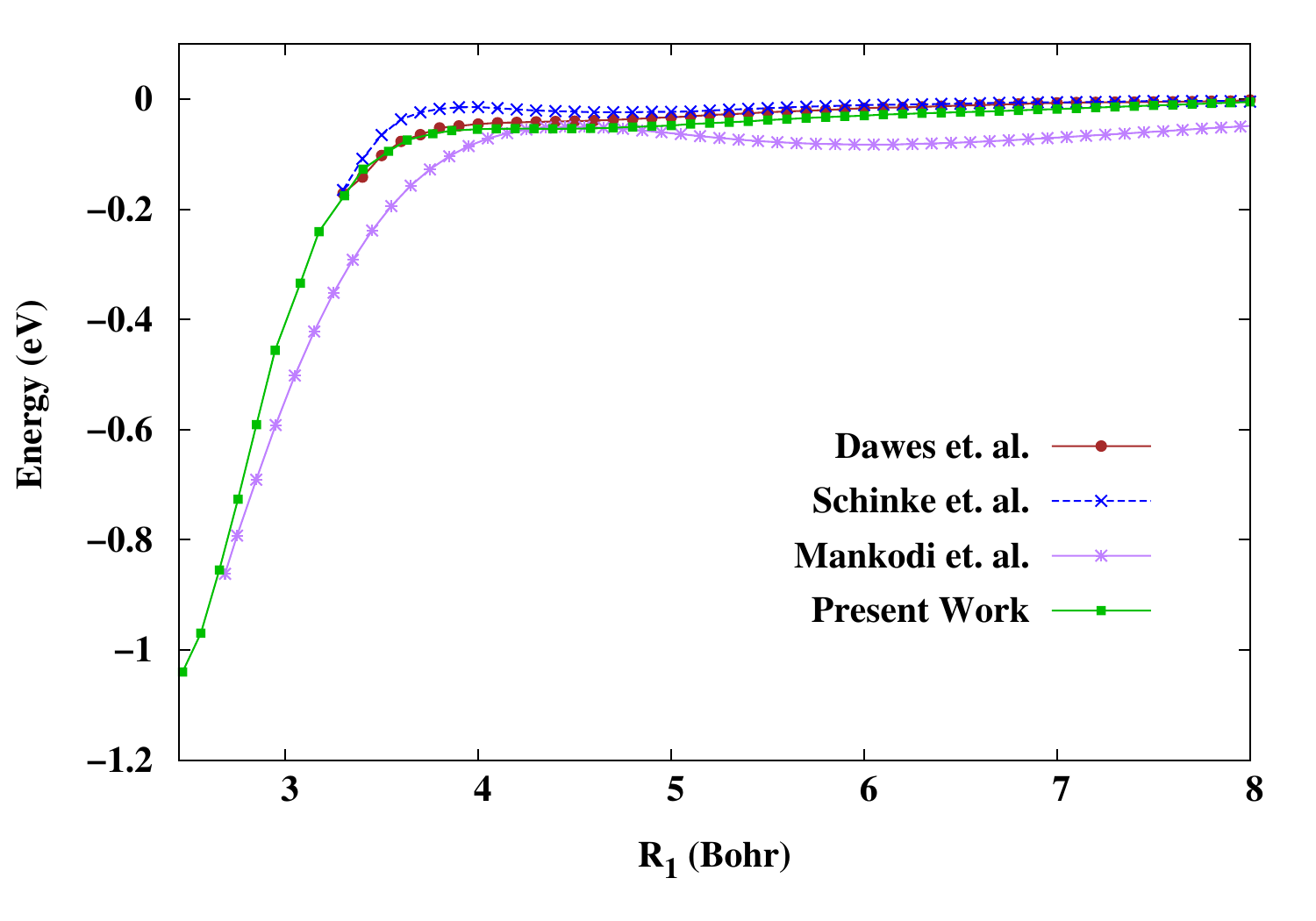}
		\caption{Comparison of the Dawes \textit{et. al.} (brown), Schinke \textit{et. al.} (blue), Mankodi \textit{et. al.} (purple) PECs with present PEC along the MEP as a function of one of the O-O bond length, R$_1$ (Bohr) while keeping other O-O bond (R$_2$) fixed at 2.3 Bohr. All energies are referenced to the asymptotic \(\text{O}_2(^3\Sigma_g^-)\) + \(\text{O}(^3P)\) channel.}
		\label{fig:MEP}
	\end{figure}

	\noindent
	Since the accuracy of global adiabatic and thereafter, diabatic PESs is essential for performing reliable dynamics studies of reactive processes, it is necessary to explore one of the most investigated features of the $\tilde{X}^1A'$ PES, namely, the so-called ``reef'' feature along the minimum energy path (MEP). Such profile appears at the asymptote as the atomic oxygen approaches the O\(_2\) molecule. Although various high-level earlier calculations~\cite{siebert2002vibrational,dawes2011communication,varga2017,Tyuterev2018} consistently predict the presence of this ``reef'', its empirical removal~\cite{babikov2003metastable} has been shown to yield improved agreement between computed and experimental thermal rate coefficients. Moreover, this ``reef'' feature is directly linked to the experimentally observed negative temperature dependence of the thermal rate constant, k(T). In this context, we have used the same methodology (7S-SA-MCSCF approach followed by ic-MRCI(Q)) with same active space [CAS(18e,12o)] and basis set (AVQZ), and calculate the ground adiabatic PES as functions of $\gamma$ and $R$ by keeping the O$_2$ bond length ($r$) fixed at 2.3 Bohr. Figure \ref{fig:MEP} presents a comparison between the presently computed MEP along with previously reported ones as a function of one of the O–O bond distance ($R_1$). Our calculated result shows that the ``reef'' is extremely small (of the order of 10$^{-4}$ eV), which is practically negligible.
	
	\subsection*{III.C. Location of CIs: Quantization of NACTs}
\begin{figure}[!htp]
		\centering
		\includegraphics[width=0.5\linewidth]{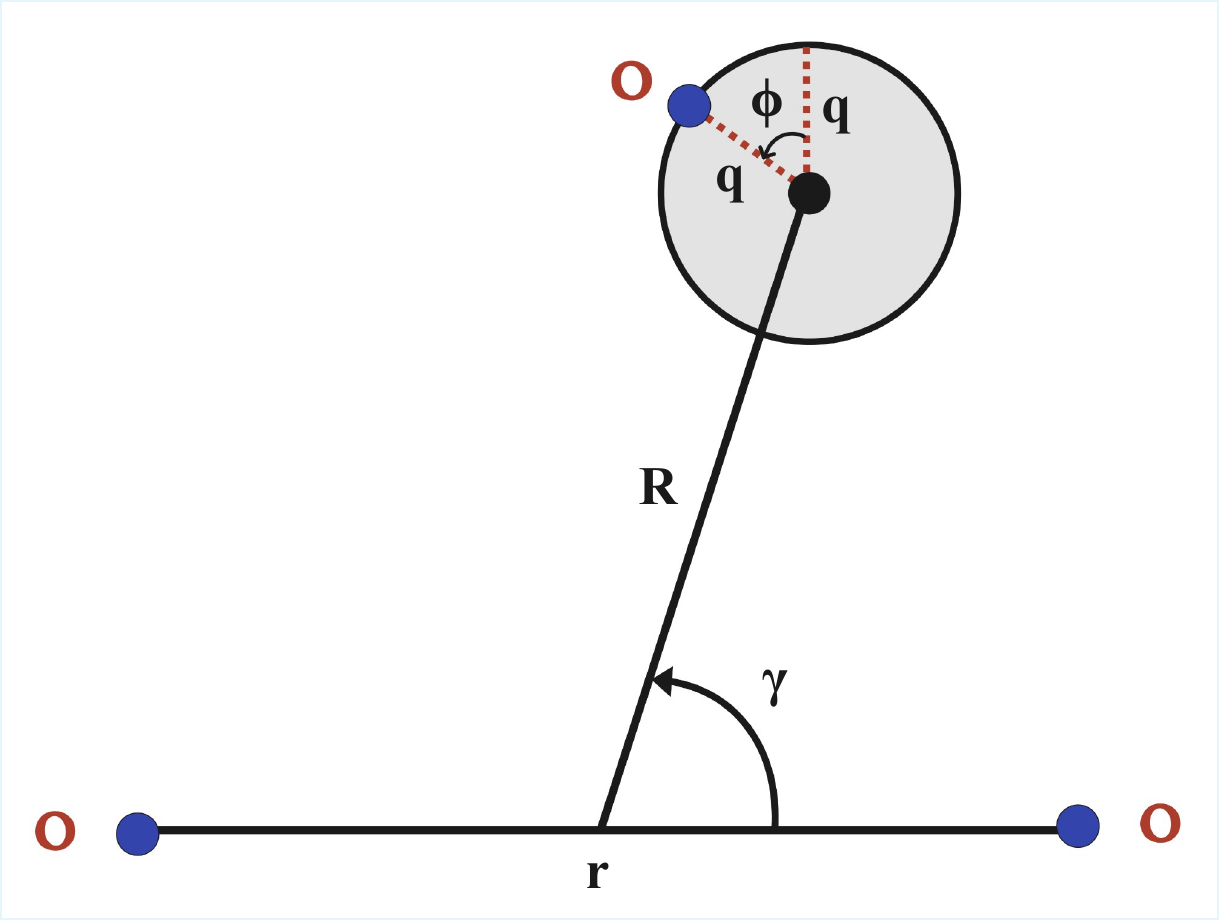}
		\caption{Schematic diagram of a contour (red circle) around a CI in Jacobi coordinates for the O$_3$ system.}
		\label{fig:loop}
	\end{figure}

	\noindent
	To locate both the symmetry driven and accidental CIs between the lowest four (4) singlet adiabatic states of ozone, we compute the corresponding NACTs along circular contours encapsulating the CIs. FIG.~\ref{fig:loop} illustrates such a contour, where two O atoms remain fixed at distance $r$, the third atom (O) moves along a circular path defined by \{$q$, $\phi$\} enclosing a CI and the center of the contour is located at a distance \( R \) from the midpoint of the O--O bond. Calculations are carried out through 7S-SA-MCSCF followed by CP-MCSCF level using the same CAS(18e,12o) with increasing basis set size (AVDZ $\rightarrow$ AVQZ). The convergences of location and minimum energy separation of CIs obtained with different basis sets near the \textit{D}$_{3h}$ and \textit{C}$_{2v}$ minima ($\gamma = 90^\circ$) are summarized in Table~\ref{tab:CI_location}.
	
	\begin{table}[h!]
		\centering
		\renewcommand{\arraystretch}{0.8} 
		\setlength{\tabcolsep}{8pt}       
		\caption{Locations (Bohr) and minimum energy separations ($\times$$10^{-4}$ eV) between the 1–2, 2–3, and 3–4 CIs near the \textit{C}$_{2v}$ and \textit{D}$_{3h}$ minima with increasing basis set size at the 7S-SA-MCSCF/CAS(18e,12o) level.}
		\label{tab:CI_location}
		\renewcommand{\arraystretch}{0.8} 
		\setlength{\tabcolsep}{8pt}      
		\vspace{0.2cm}
		\begin{tabular}{cccccccc}
			\toprule
			\textbf{Geometry} & \textbf{Basis} & \multicolumn{2}{c}{\textbf{1–2 CI}} & \multicolumn{2}{c}{\textbf{2–3 CI}} & \multicolumn{2}{c}{\textbf{3–4 CI}} \\
			\cmidrule(lr){3-4} \cmidrule(lr){5-6} \cmidrule(lr){7-8}
			& & ($\mathbf{r}$, $\mathbf{R}$) & $\mathbf{\Delta E_{12}}^a$ & ($\mathbf{r}$, $\mathbf{R}$) & $\mathbf{\Delta E_{23}}^b$ & ($\mathbf{r}$, $\mathbf{R}$) & $\mathbf{\Delta E_{34}}^c$ \\
			\midrule
			\multirow{3}{*}{\textit{C}$_{2v}$}
			& AVDZ & (4.01, 2.10) & 0.68   & (2.38, 2.56)  & 6.23  &  (3.86, 1.98)  &  0.11  \\
			& AVTZ & (3.98, 2.06) & 7.89  &  (2.33, 2.54)  & 2.97  &  (3.93, 2.03)  &  5.14  \\
			& AVQZ & (3.97, 2.05) & 0.10 & (2.30, 2.52)   & 1.50   & (3.85, 1.96) & 0.40 \\
			\midrule
			\multirow{3}{*}{\textit{D}$_{3h}$}
			& AVDZ & --- & --- & (3.06, 2.65) & 0.01 & --- & --- \\
			& AVTZ & --- & --- & (3.06, 2.65) & 0.03 & --- & --- \\
			& AVQZ & --- & --- & (3.06, 2.65) & 0.03 & --- & --- \\
			\bottomrule
		\end{tabular}
	
		\vspace{0.5cm}
		\raggedright
		{\small $^a~\Delta E_{12} = E_2 - E_1$ \\
		$^b~\Delta E_{23} = E_3 - E_2$ \\
		$^c~\Delta E_{34} = E_4 - E_3$}
	\end{table}
	\pagebreak
	
	\vspace{0.2cm}
	
	\noindent
	While locating the conical intersections (CIs) at the \textit{D}$_{3h}$ and \textit{C}$_{2v}$ geometries ($\gamma = 90^\circ$), we analyzed the adiabatic potential energy curves (PECs) of O$_3$ for the lowest four electronic states as functions of $R$ at various fixed values of $r$, using 7S-SA-MCSCF method with CAS(18e,12o) and AVQZ basis set. Once the CI locations are identified, as illustrated in Figure~S7 of the SM for specific values of $r$ and $R$, both the adiabatic PECs and the NACTs are computed along a circular contour around those points to calculate ADT angles and to explore quantization of NACTs. The center of the contour is located at a specific distance of $R$ from the midpoint of O--O with a separation of $r$. The increased radius of the contour leads to the coupling with the states outside the low-lying four states at much higher energy, while decreasing the radius causes the contour to pass directly over the CIs instead of enclosing those. Therefore, the choice of the $q$ value is really crucial for locating the CIs. For a fixed contour radius $q$ around a point defined by specific values of \{$r$,$R$\} at $\gamma = 90^\circ$, the geometries at $\phi = 0$ and $\pi$ correspond to \textit{C}$_{2v}$ symmetry, whereas all other values of $\phi$ correspond to \textit{C}$_s$ symmetry.
	
	\vspace{0.2cm}
	
	\noindent
	In such analysis, we have used the calculated NACTs between the first four (4) states and thereafter, four (4) state ADT calculations are performed (see the relevant equations in S2 of SM) to depict the quantization of NACTs, and the nature of ADT matrix for different sets of \{$r$, $R$\} values as discussed below.
	
	\vspace{0.5cm}
		
	\noindent
	\textbf{(a) \textit{D}$_{3h}$ Geometry:} $r$ = 3.06 Bohr, $R$ = 2.65 Bohr, $q$ = 0.25 Bohr and $\gamma$ = 90$^\circ$
	
	\vspace{0.2cm}
	
\begin{figure}[!htp]
		\centering
		\hspace*{-3cm}
		\begin{subfigure}{0.4\linewidth}
			\centering
			\begin{overpic}[width=1.4\linewidth]{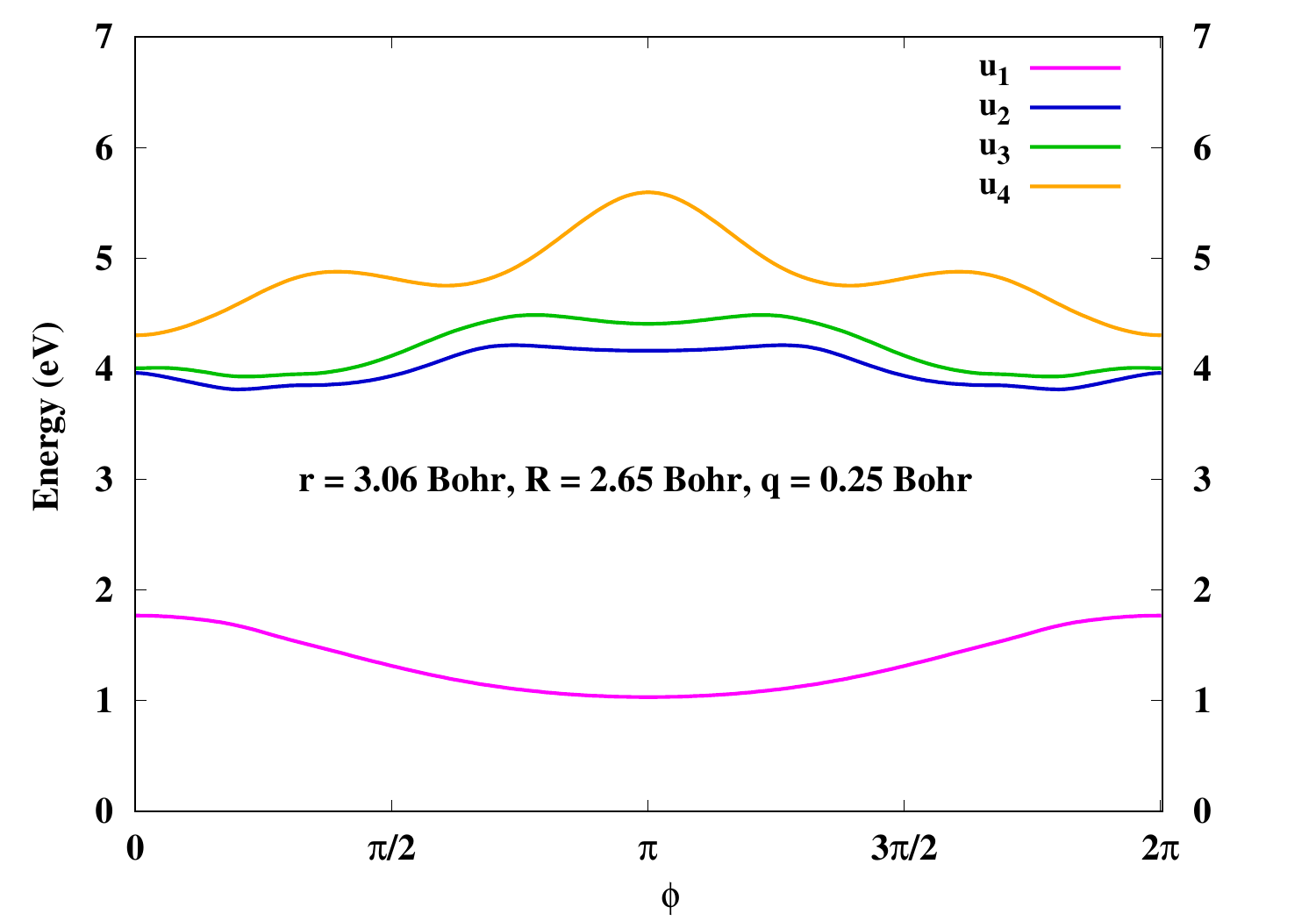}
				\put(1.5,70){\textbf{(a)}} 
			\end{overpic}
			\phantomcaption
			\label{fig:D3h_adiapes}
		\end{subfigure}
		\hspace{2.5cm}
		\begin{subfigure}{0.4\linewidth}
			\centering
			\begin{overpic}[width=1.4\linewidth]{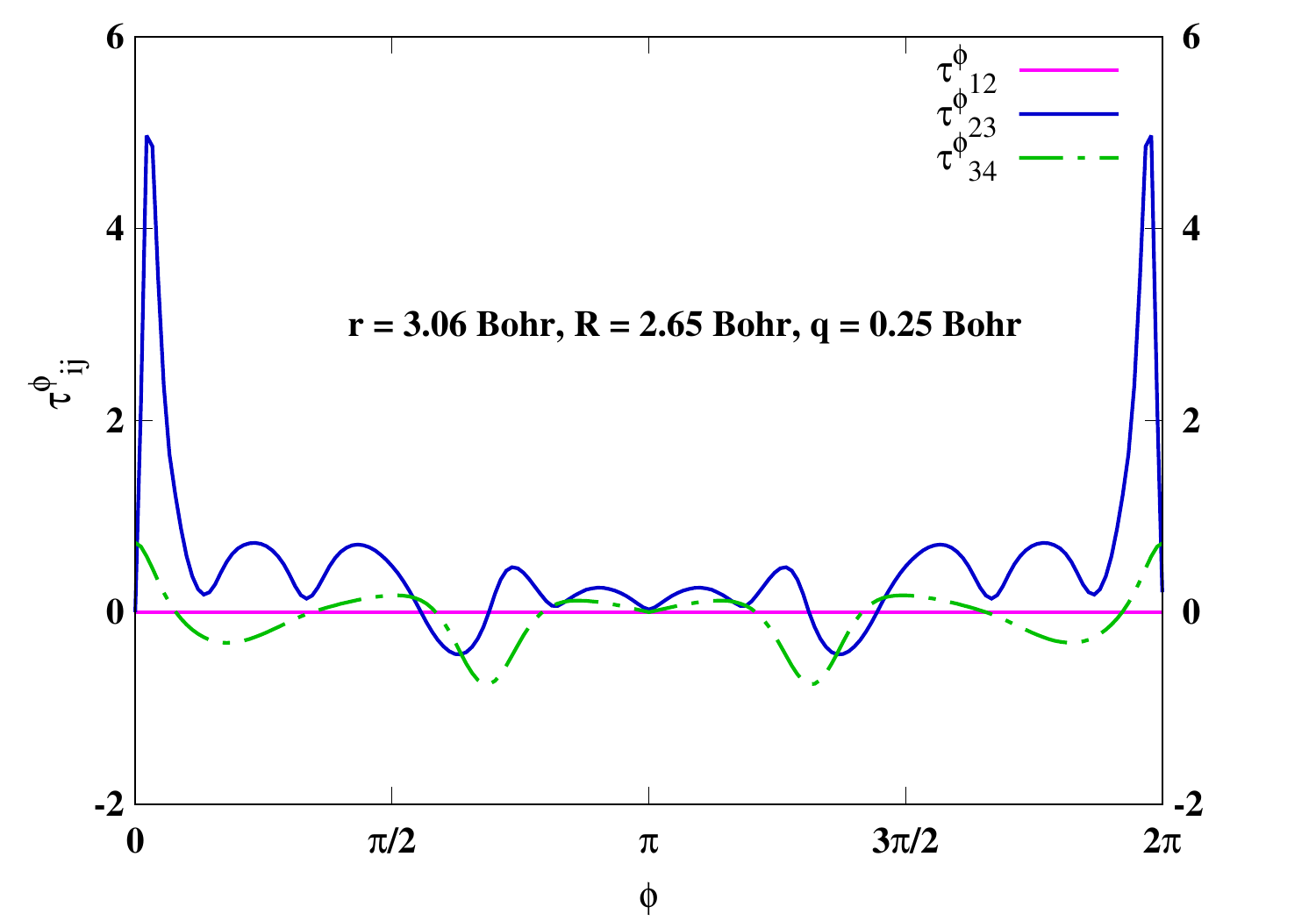}
				\put(1.5,70){\textbf{(b)}}
			\end{overpic}
			\phantomcaption
			\label{fig:D3h_tau}
		\end{subfigure}
		\hspace*{-3cm}
		\begin{subfigure}{0.4\linewidth}
			\centering
			\begin{overpic}[width=1.4\linewidth]{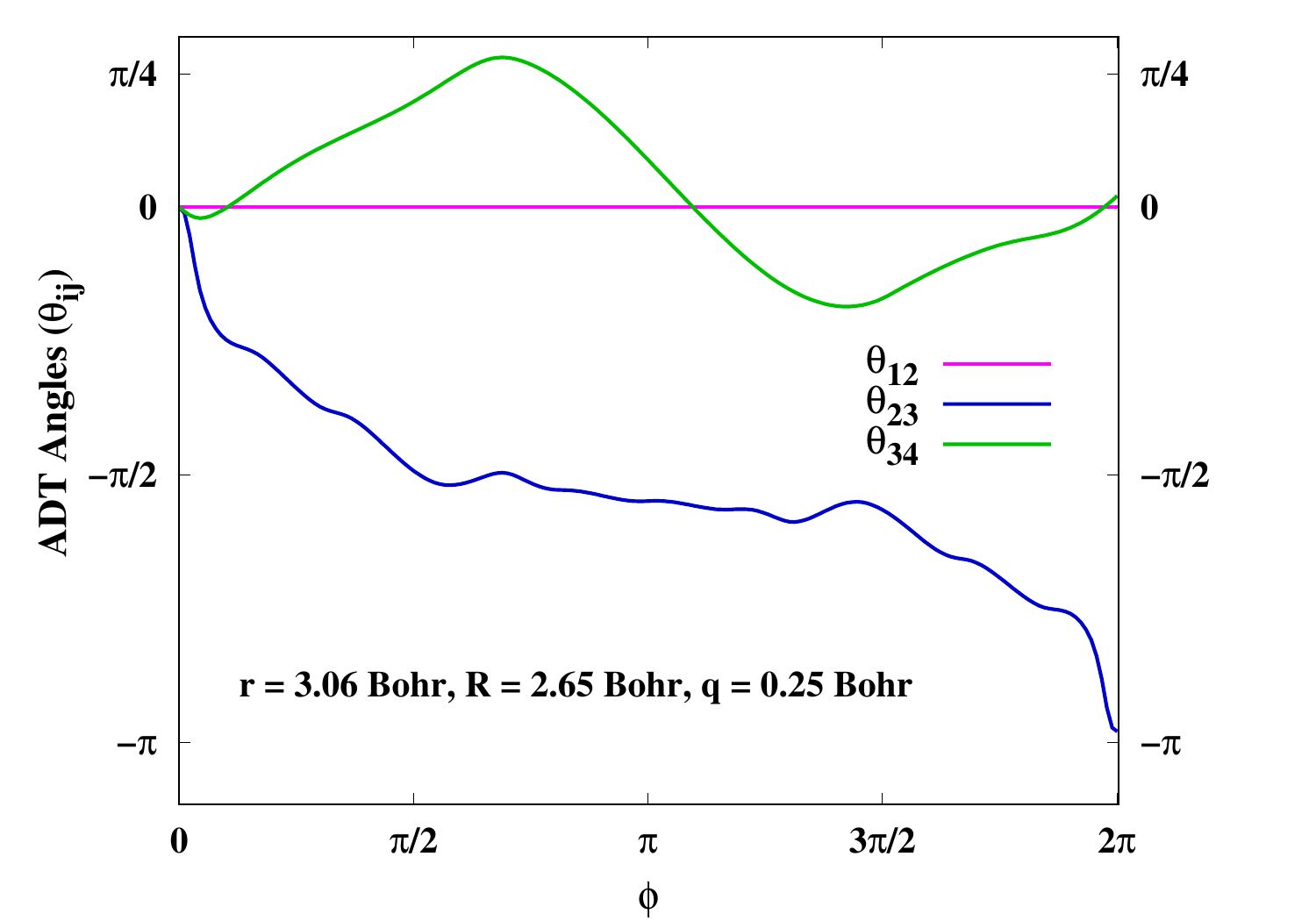}
				\put(1.5,70){\textbf{(c)}}
			\end{overpic}
			\phantomcaption
			\label{fig:D3h_angles}
		\end{subfigure}
		\hspace{2.5cm}
		\begin{subfigure}{0.4\linewidth}
			\centering
			\begin{overpic}[width=1.4\linewidth]{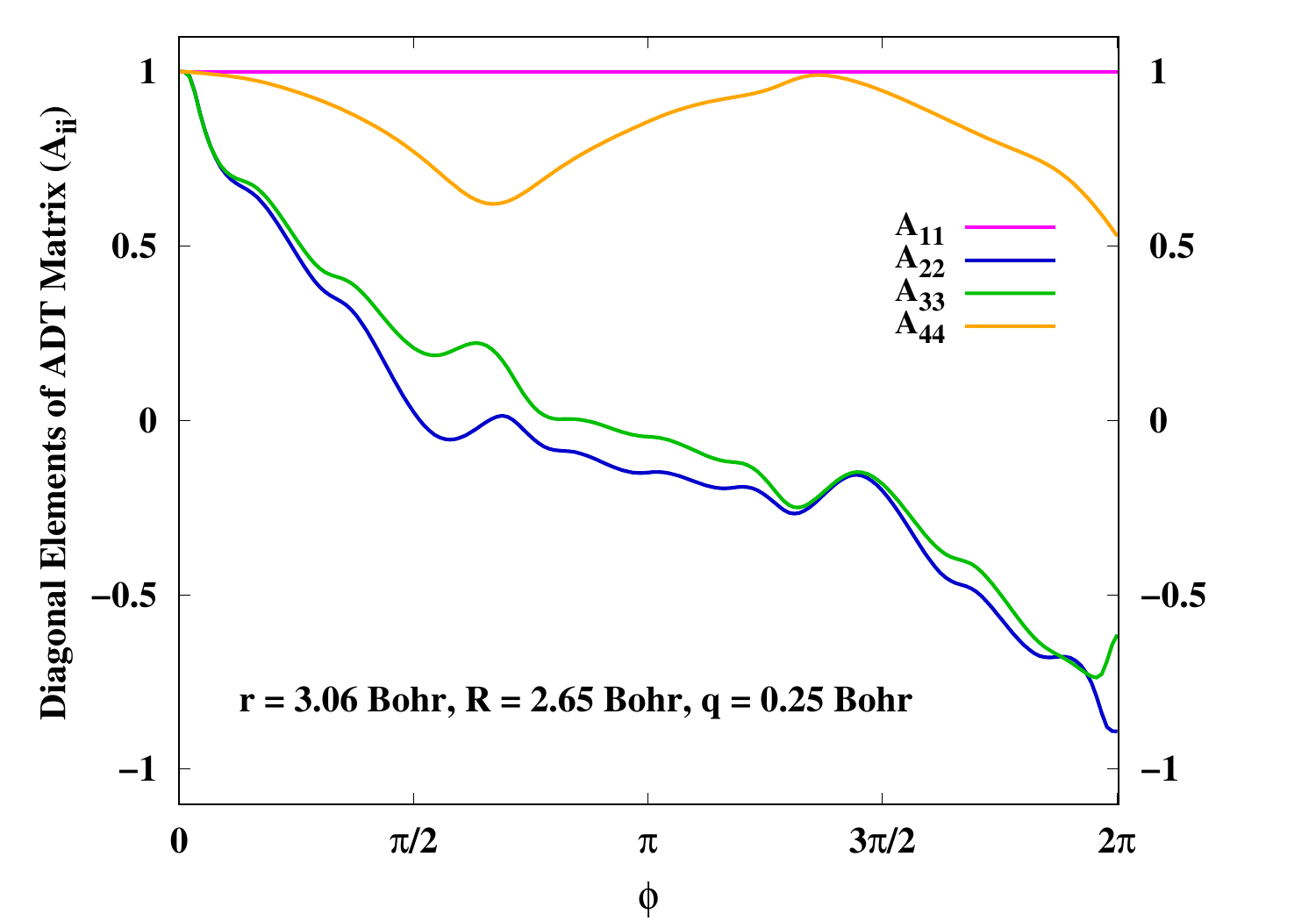}
				\put(1.5,70){\textbf{(d)}}
			\end{overpic}
			\phantomcaption
			\label{fig:D3h_amat}
		\end{subfigure}		
		\label{fig:D3h_23CI}
		\caption{Figure (a) depicts the PECs around a \textit{D}$_{3h}$ point (\(r = 3.06\) and \(R = 2.65\) Bohr). Figure (b) illustrates the $\phi$ component of NACTs ($\tau^{\phi}_{ij}$) computed along the defined contour. Figures (c) and (d) represent the ADT angle ($\theta_{ij}$) and diagonal elements of the ADT matrix (\(A_{ii}\)), respectively. These quantities are computed using CAS(18e,12o) with AVQZ basis set along a circular contour of radius \(q = 0.25\) Bohr.}
	\end{figure}

	\noindent
	Figure~\ref{fig:D3h_adiapes} presents the adiabatic PECs along a circular contour of radius $q = 0.25$ Bohr around the \textit{D}$_{3h}$ point ($r = 3.06$ Bohr, $R = 2.65$ Bohr), whereas in Figure~\ref{fig:D3h_tau}, the NACT profiles show two pronounced peaks for $\tau_{23}^\phi$ near $\phi = 0$ and $2\pi$ and a non-negligible value for $\tau_{34}^\phi$. Since the ADT angle $\theta_{23}$ undergoes a phase change of $\pi$ along the closed contour (Figure~\ref{fig:D3h_angles}) accompanied by a sign inversion in the profiles of $A_{22}$ and $A_{33}$ (Figure~\ref{fig:D3h_amat}), this confirms the presence of a symmetry-driven CI inside the contour between two E$^\prime$ states, $u_2$ and $u_3$. On the other hand, the wide energy separation between $u_3$ and $u_4$ indicates a pseudo Jahn Teller (PJT) type of interation between these two states.
	
	\vspace{0.2cm}
	
	\noindent
	\textbf{(b) \textit{C}$_{2v}$ Geometry:}
	
	\vspace{0.2cm}
	
	\noindent
	(i) $r$ = 4.10, $R$ = 3.40 and q = 0.1 Bohr
	
	\vspace{0.2cm}
	
\begin{figure}[!htp]
		\centering
		\hspace*{-3cm}
		\begin{subfigure}{0.4\linewidth}
			\centering
			\begin{overpic}[width=1.4\linewidth]{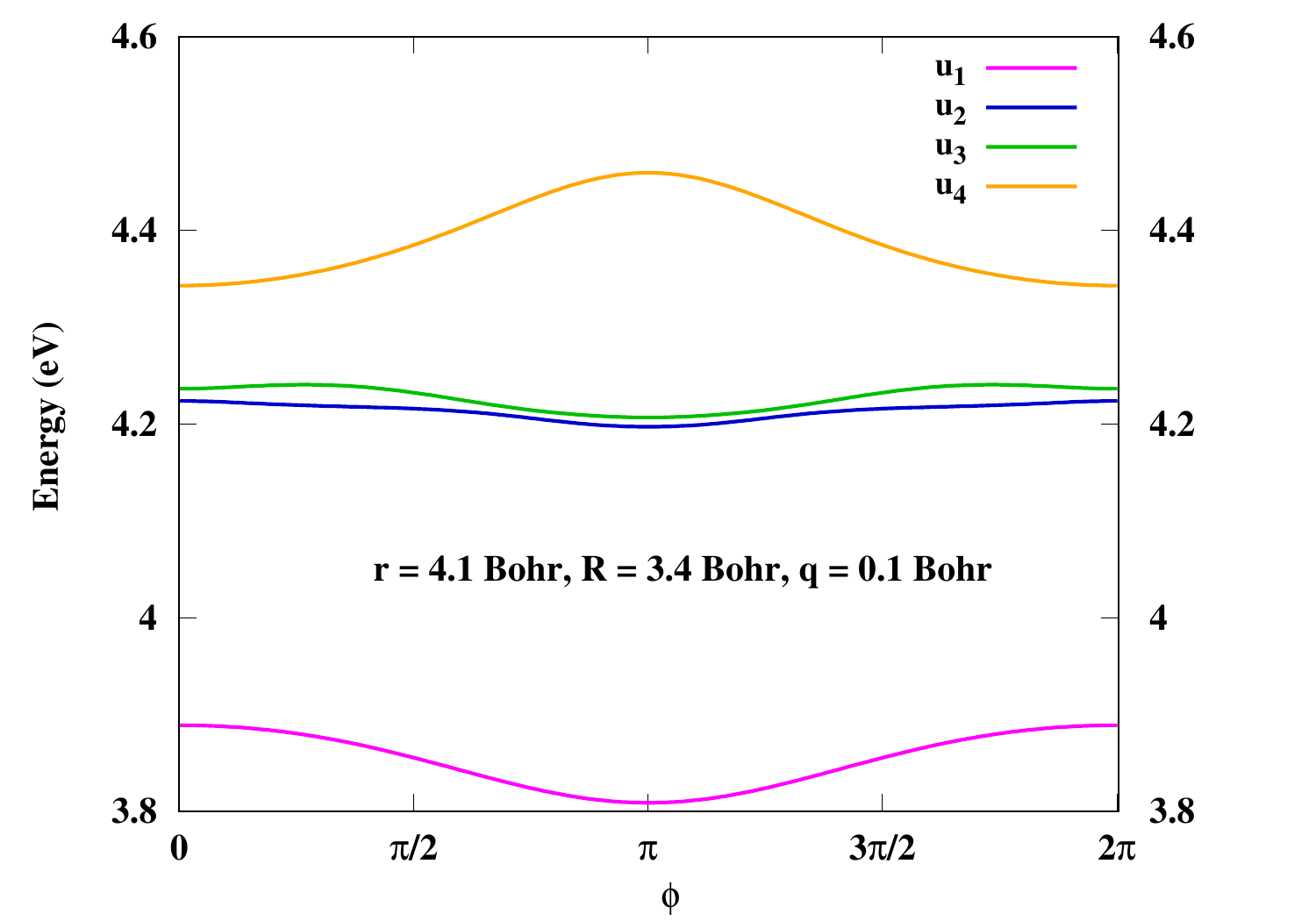}
				\put(1.5,70){\textbf{(a)}} 
			\end{overpic}
			\phantomcaption
			\label{fig:C2v_23_adiapes}
		\end{subfigure}
		\hspace{2.5cm}
		\begin{subfigure}{0.4\linewidth}
			\centering
			\begin{overpic}[width=1.4\linewidth]{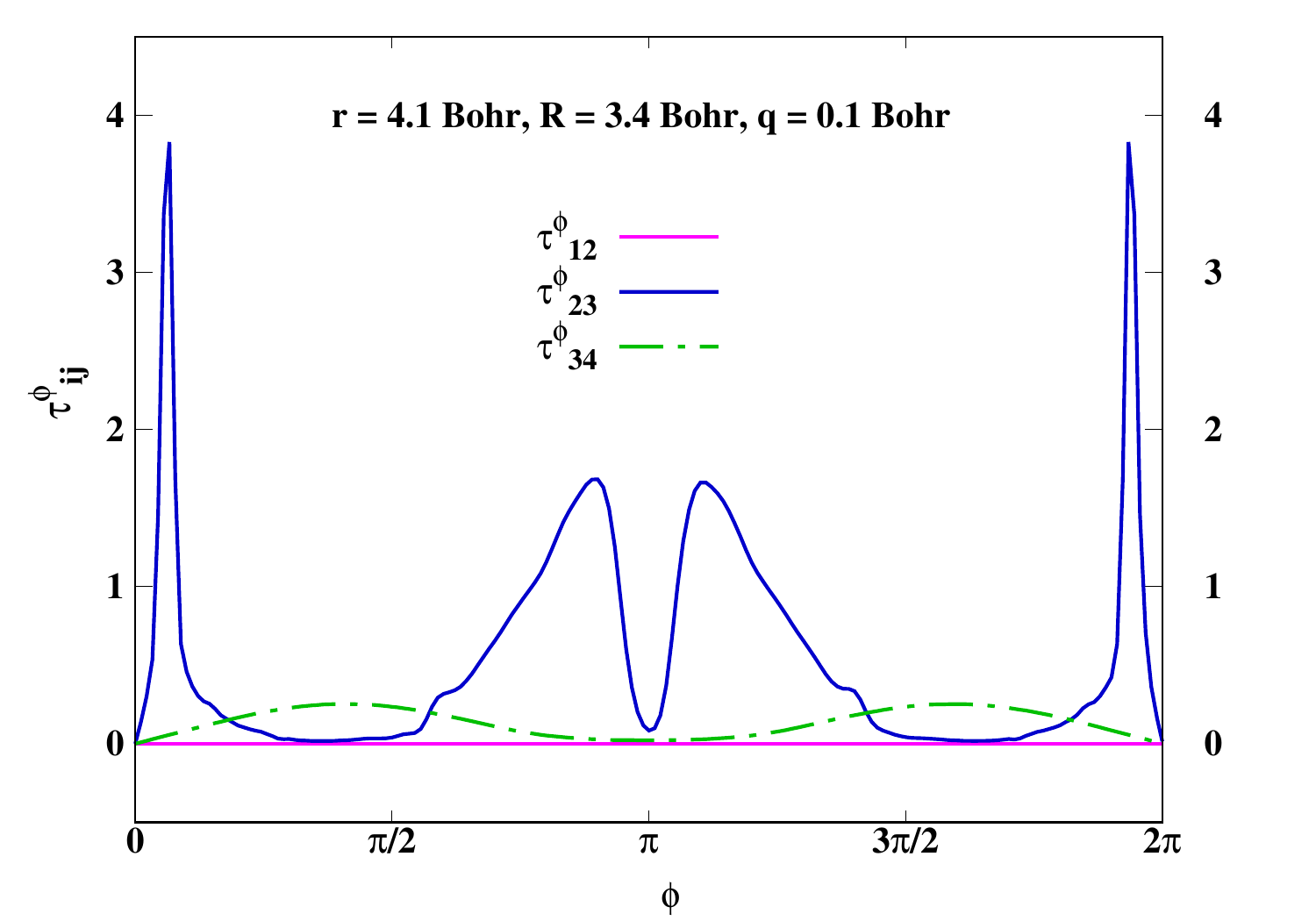}
				\put(1.5,70){\textbf{(b)}}
			\end{overpic}
			\phantomcaption
			\label{fig:C2v_23_tau}
		\end{subfigure}
		\hspace*{-3cm}
		\begin{subfigure}{0.4\linewidth}
			\centering
			\begin{overpic}[width=1.4\linewidth]{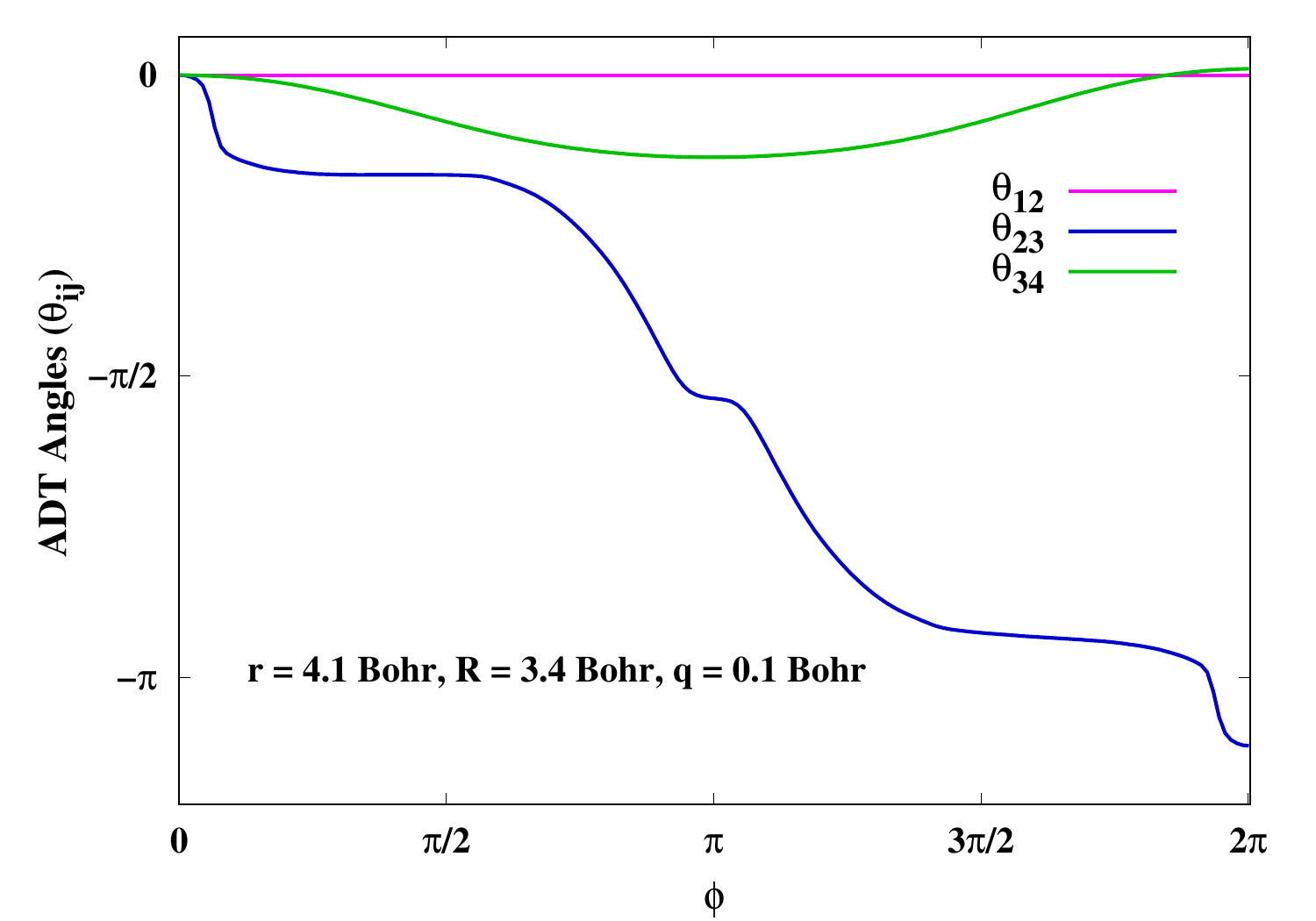}
				\put(1.5,70){\textbf{(c)}}
			\end{overpic}
			\phantomcaption
			\label{fig:C2v_23_angles}
		\end{subfigure}
		\hspace{2.5cm}
		\begin{subfigure}{0.4\linewidth}
			\centering
			\begin{overpic}[width=1.4\linewidth]{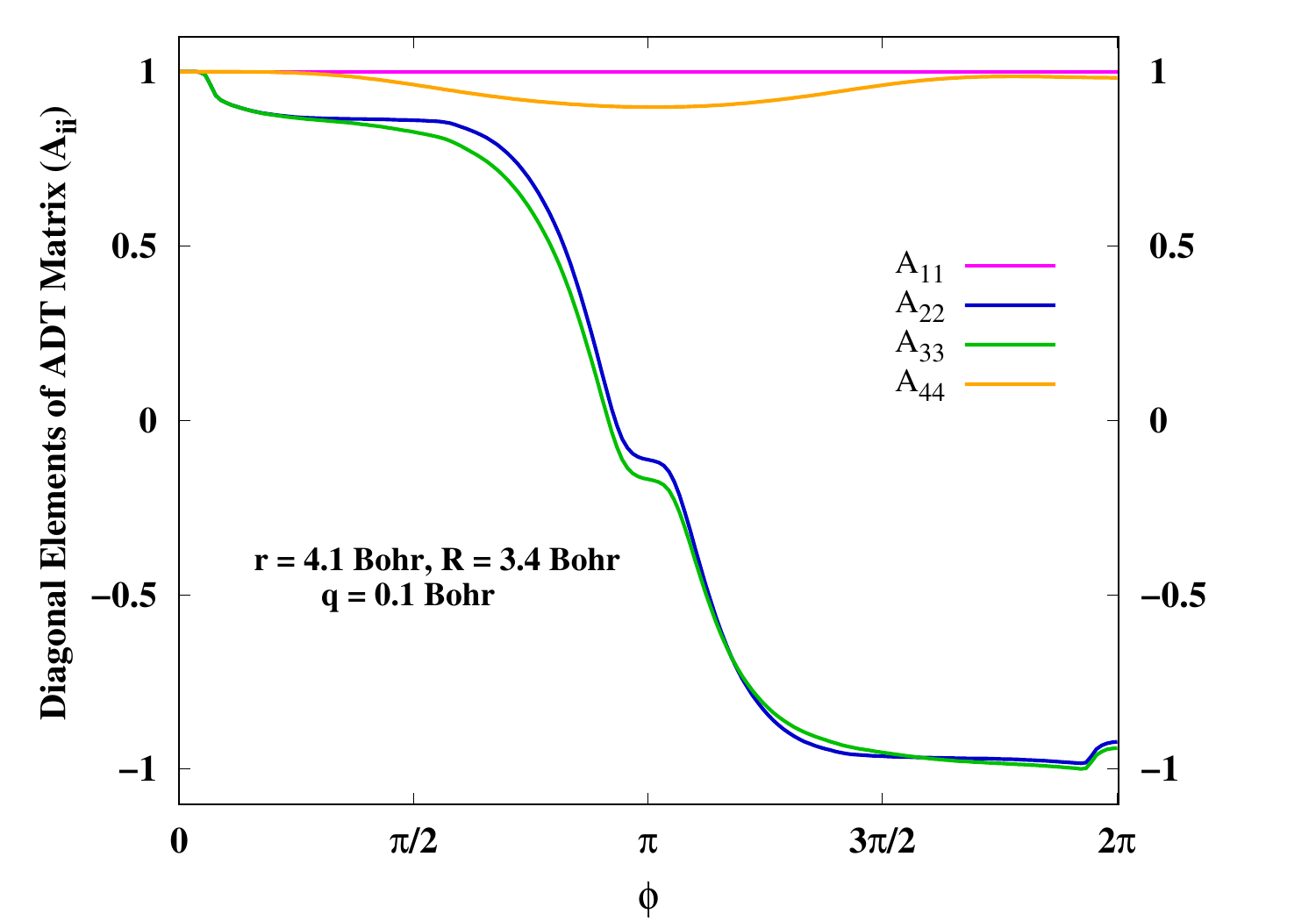}
				\put(1.5,70){\textbf{(d)}}
			\end{overpic}
			\phantomcaption
			\label{fig:C2v_23_amat}
		\end{subfigure}			
		\label{fig:C2v_23CI}
		\caption{Figure (a) depicts the PECs around a \textit{C}$_{2v}$ point (\(r = 4.1\) and \(R = 3.4\) Bohr). Figure (b) illustrates the $\phi$ component of NACTs ($\tau^{\phi}_{ij}$) computed along the defined contour. Figures (c) and (d) represent the ADT angle ($\theta_{ij}$) and diagonal elements of the ADT matrix (\(A_{ii}\)), respectively. These quantities are computed using CAS(18e,12o) with AVQZ basis set along a circular contour of radius \(q = 0.1\) Bohr.}
	\end{figure}

	\noindent
	For this geometry, FIG.~\ref{fig:C2v_23_adiapes} and~\ref{fig:C2v_23_tau} show the adibatic PECs and NACTs along the circular contour, respectively. The coupling term $\tau_{23}^\phi$ displays two sharp peaks at $\phi = 0$ and $2\pi$ along with broad features around $\phi = \pi$, where in those regions, $u_2$ and $u_3$ approach closer to each other. In FIG.~\ref{fig:C2v_23_angles} and \ref{fig:C2v_23_amat}, the associated ADT angle $\theta_{23}$ reaches $\pi$ and the diagonal matrix elements $A_{22}$ and $A_{33}$ exhibit a sign change, respectively, which indicate the presence of an accidental CI between $u_2$ and $u_3$. On the other hand, $\tau_{34}^\phi$ shows a negligible value, where the corresponding ADT angle $\theta_{34}$ and matrix elements $A_{33}$ and $A_{44}$ approach to zero and depict no sign change, respectively. This suggest a PJT interaction between $u_3$ and $u_4$ with large energy separation (see FIG.~\ref{fig:C2v_23_adiapes}).
	
	\vspace{0.2cm}
	
	\vspace{0.2cm}
	
	\noindent
	(ii) $r$ = 3.85, $R$ = 1.96 and q = 0.1 Bohr
	
	\vspace{0.2cm}
	
\begin{figure}[!htp]
		\centering
		\hspace*{-3cm}
		\begin{subfigure}{0.4\linewidth}
			\centering
			\begin{overpic}[width=1.4\linewidth]{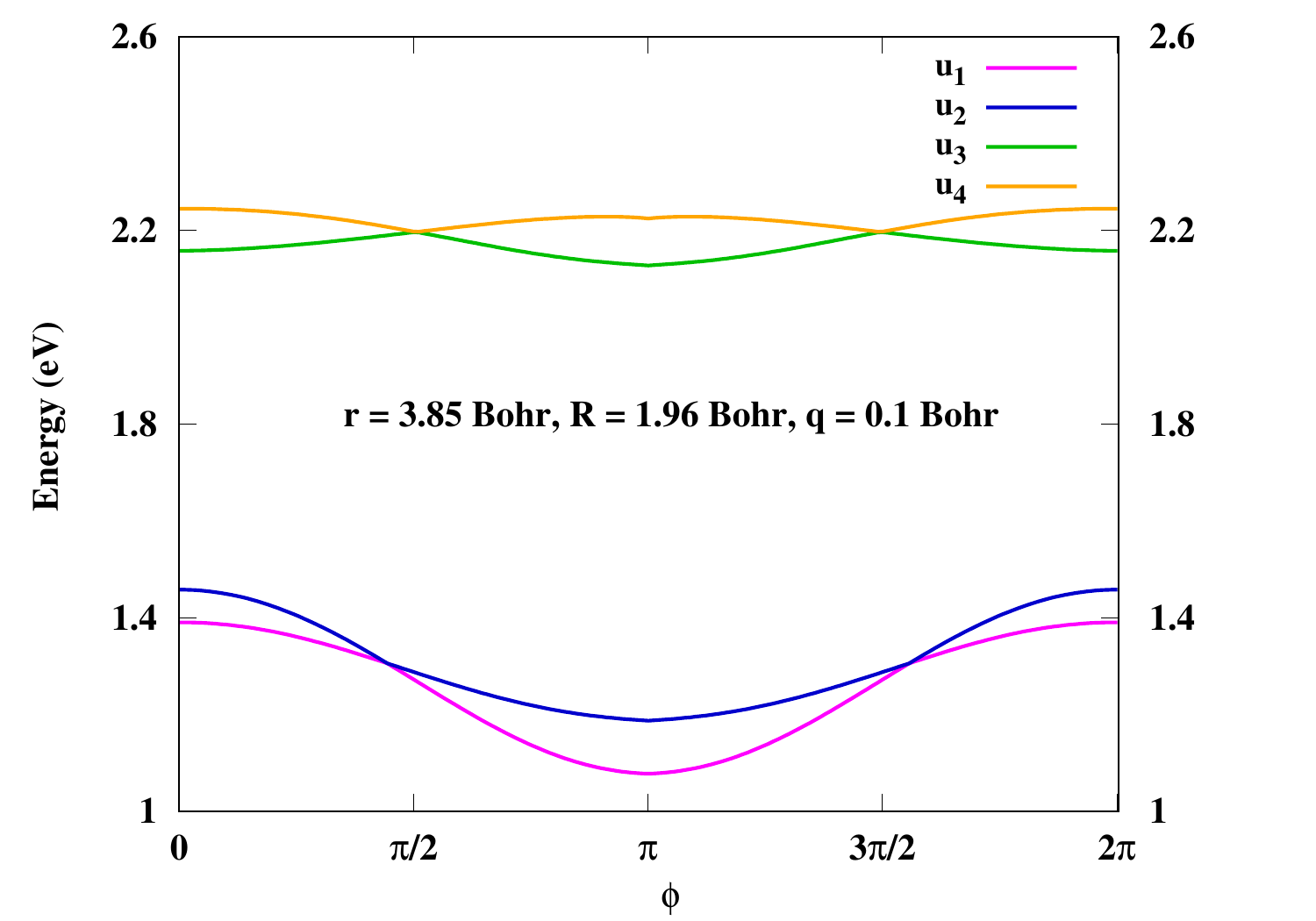}
				\put(5,75){\textbf{(a)}} 
			\end{overpic}
			\phantomcaption
			\label{fig:C2v_1234_adiapes}
		\end{subfigure}
		\hspace{2.5cm}
		\begin{subfigure}{0.4\linewidth}
			\centering
			\begin{overpic}[width=1.4\linewidth]{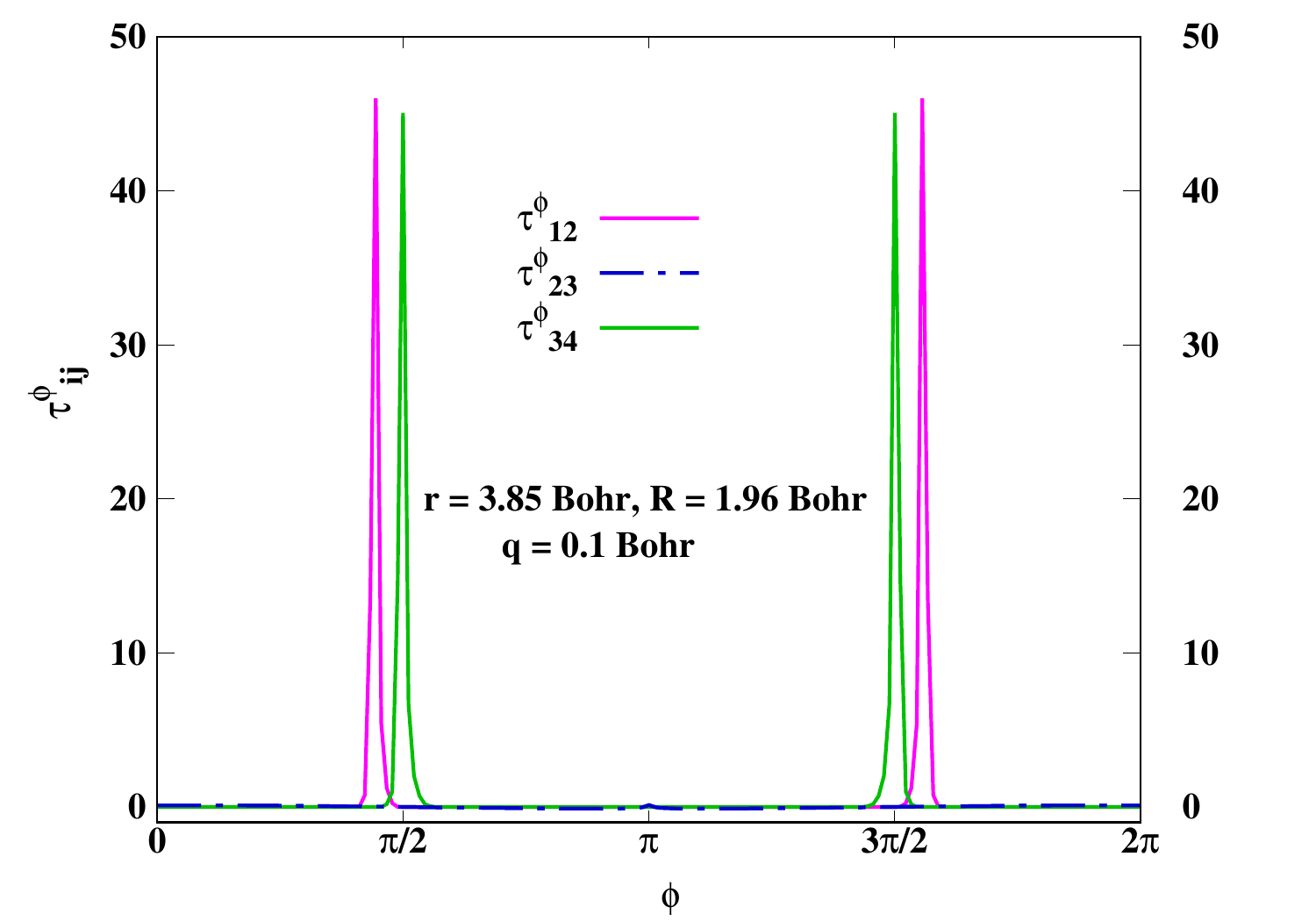}
				\put(0,70){\textbf{(b)}}
			\end{overpic}
			\phantomcaption
			\label{fig:C2v_1234_tau}
		\end{subfigure}
		\hspace*{-3cm}
		\begin{subfigure}{0.4\linewidth}
			\centering
			\begin{overpic}[width=1.4\linewidth]{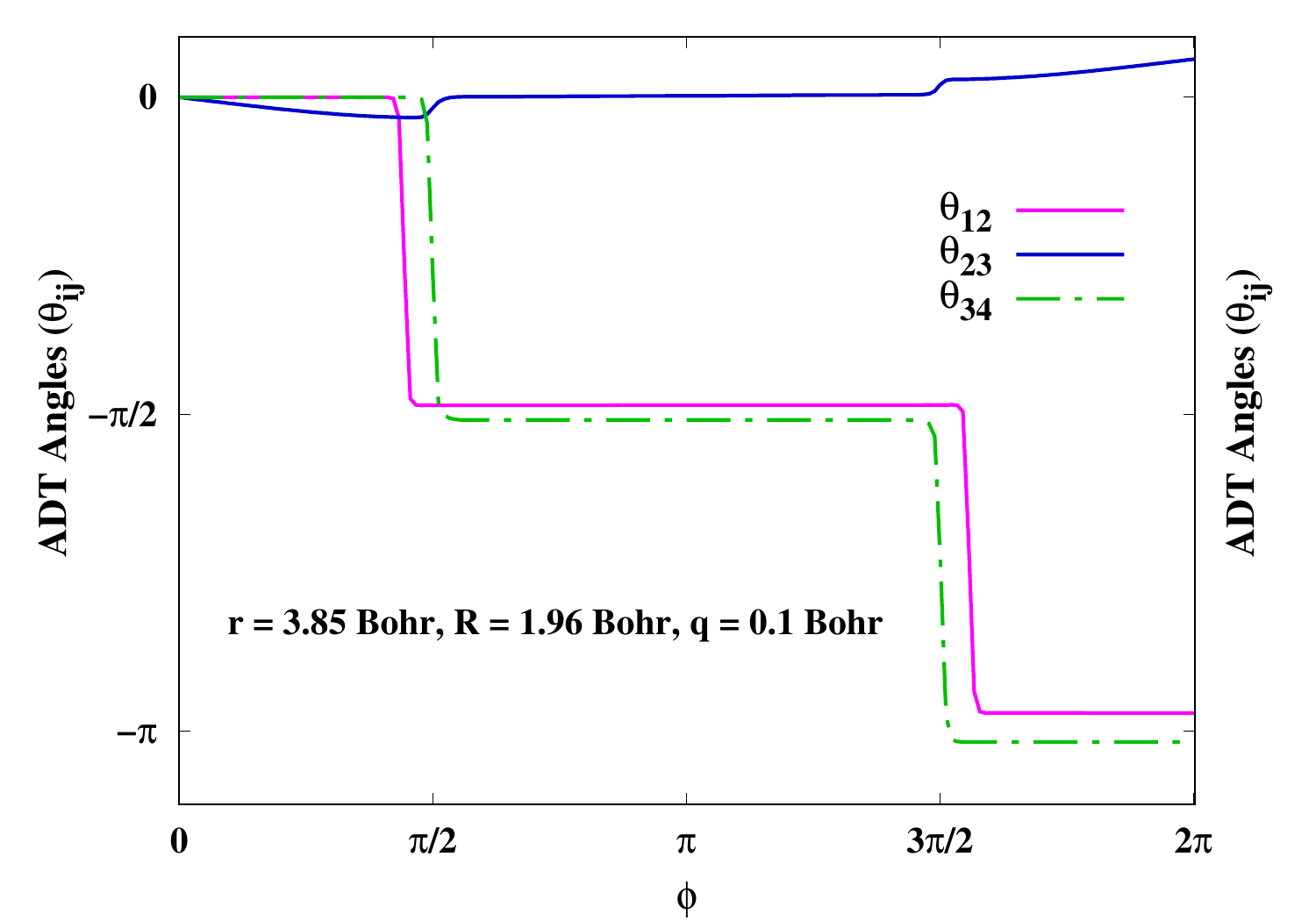}
				\put(1.5,70){\textbf{(c)}}
			\end{overpic}
			\phantomcaption
			\label{fig:C2v_1234_angle}
		\end{subfigure}
		\hspace{2.5cm}
		\begin{subfigure}{0.4\linewidth}
			\centering
			\begin{overpic}[width=1.4\linewidth]{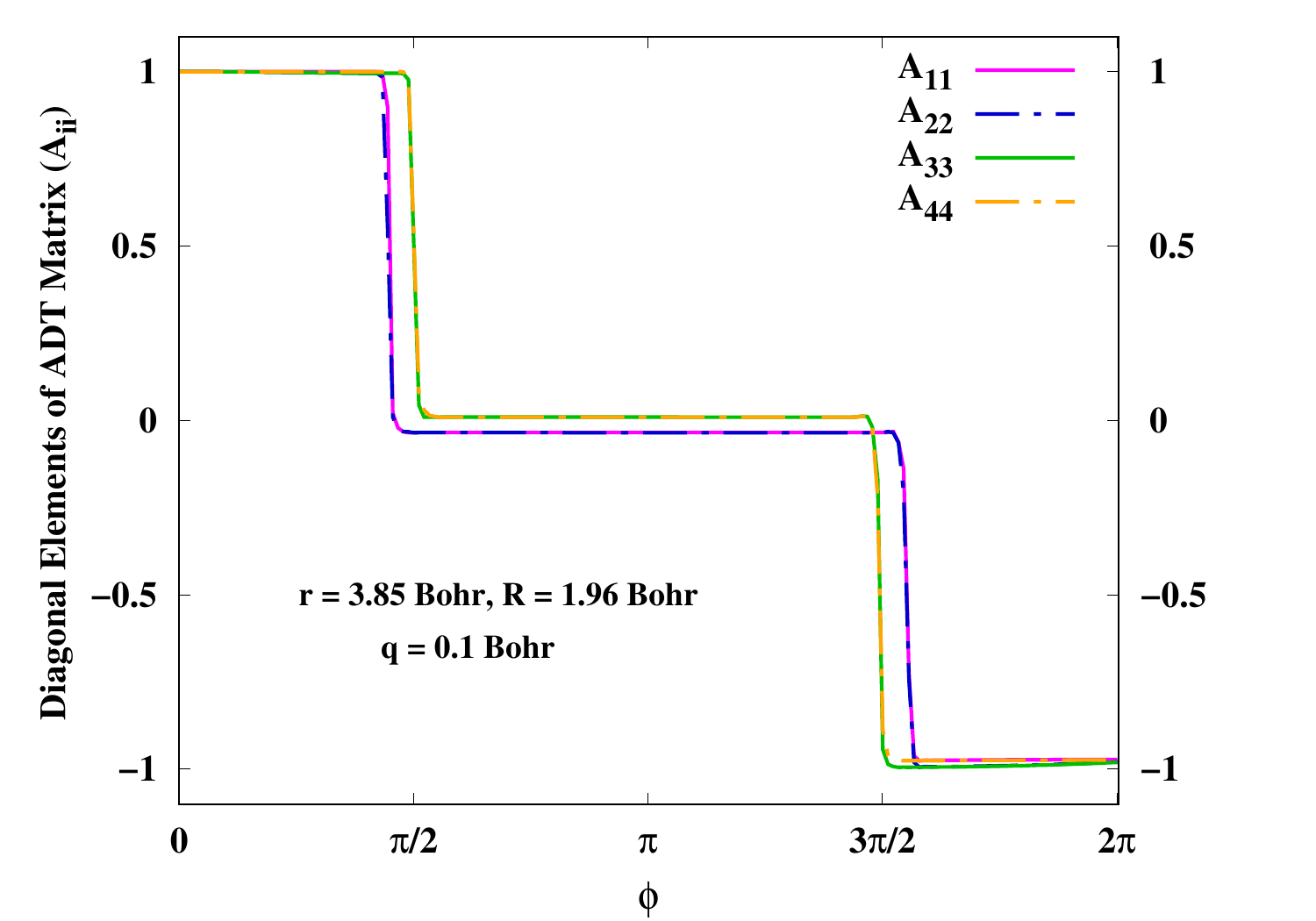}
				\put(1.5,70){\textbf{(d)}}
			\end{overpic}
			\phantomcaption
			\label{fig:C2v_1234_amat}
		\end{subfigure}			
		\label{fig:C2v_1234CI}
		\caption{Figure (a) depicts the PECs around a \textit{C}$_{2v}$ point (\(r = 3.85\) and \(R = 1.96\) Bohr). Figure (b) illustrates the $\phi$ component of NACTs ($\tau^{\phi}_{ij}$) computed along the defined contour. Figures (c) and (d) represent the ADT angle ($\theta_{ij}$) and diagonal elements of the ADT matrix (\(A_{ii}\)), respectively. These quantities are computed using CAS(18e,12o) with AVQZ basis set along a circular contour of radius \(q = 0.1\) Bohr.}
	\end{figure}

	\noindent
	FIG.~\ref{fig:C2v_1234_adiapes} presents the PECs of O$_3$ as a function of the circular angle $\phi$ around the \textit{C}$_{2v}$ point (\(r = 3.85\) and \(R = 1.96\) Bohr), highlighting the accidental CIs between 1–2 and 3–4 electronic states. The corresponding functional forms of NACTs ($\tau_{ij}^\phi$) are shown in FIG.~\ref{fig:C2v_1234_tau}, where both $\tau_{12}^\phi$ and $\tau_{34}^\phi$ display sharp peaks in regions where the respective states become degenerate, but $\tau_{23}^\phi$ shows a negligible value. The presence of accidental CIs enclosed by this circular contour is further evidenced by the ADT angles and matrix elements illustrated in FIG.~\ref{fig:C2v_1234_angle} and FIG.~\ref{fig:C2v_1234_amat}, respectively. Specifically, the ADT angles $\theta_{12}$ and $\theta_{34}$ reach $\pi$ and the diagonal ADT matrix elements ($A_{ii}$) undergo a sign change at the end of the closed contour and thereby, confirming the existence of accidental CIs between the 1–2 and 3–4 states. On the other hand, the negligible magnitude of $\tau_{23}^\phi$ and the zero value of ADT angle $\theta_{23}$ at the end of the close contour indicate a PJT interaction between $u_2$ and $u_3$ (indicated by their large energy separation).
	
	\vspace{0.6cm}
	
	\noindent
	\textbf{(c) \textit{C}$_{s}$ Geometry:}
	
	\vspace{0.2cm}
	
	\noindent
	(i) $r$ = 3.80, $R$ = 2.21, $\gamma$ = 82.69$^\circ$ and $q$ = 0.1 Bohr
	
	\vspace{0.2cm}
	
\begin{figure}[!htp]
		\centering
		\hspace*{-3cm}
		\begin{subfigure}{0.4\linewidth}
			\centering
			\begin{overpic}[width=1.4\linewidth]{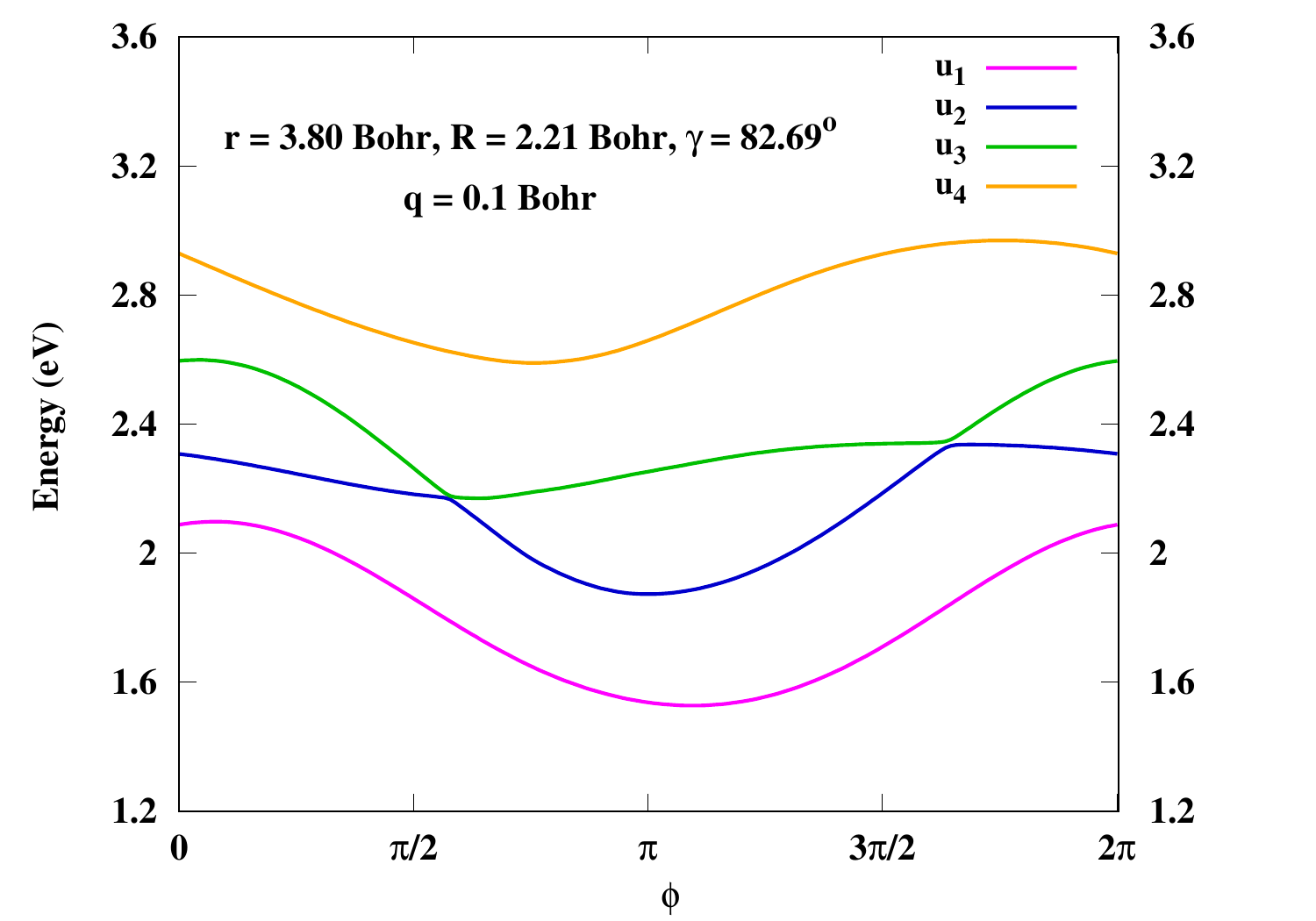}
				\put(5,75){\textbf{(a)}} 
			\end{overpic}
			\phantomcaption
			\label{fig:Cs_23_adiapes}
		\end{subfigure}
		\hspace{2.5cm}
		\begin{subfigure}{0.4\linewidth}
			\centering
			\begin{overpic}[width=1.4\linewidth]{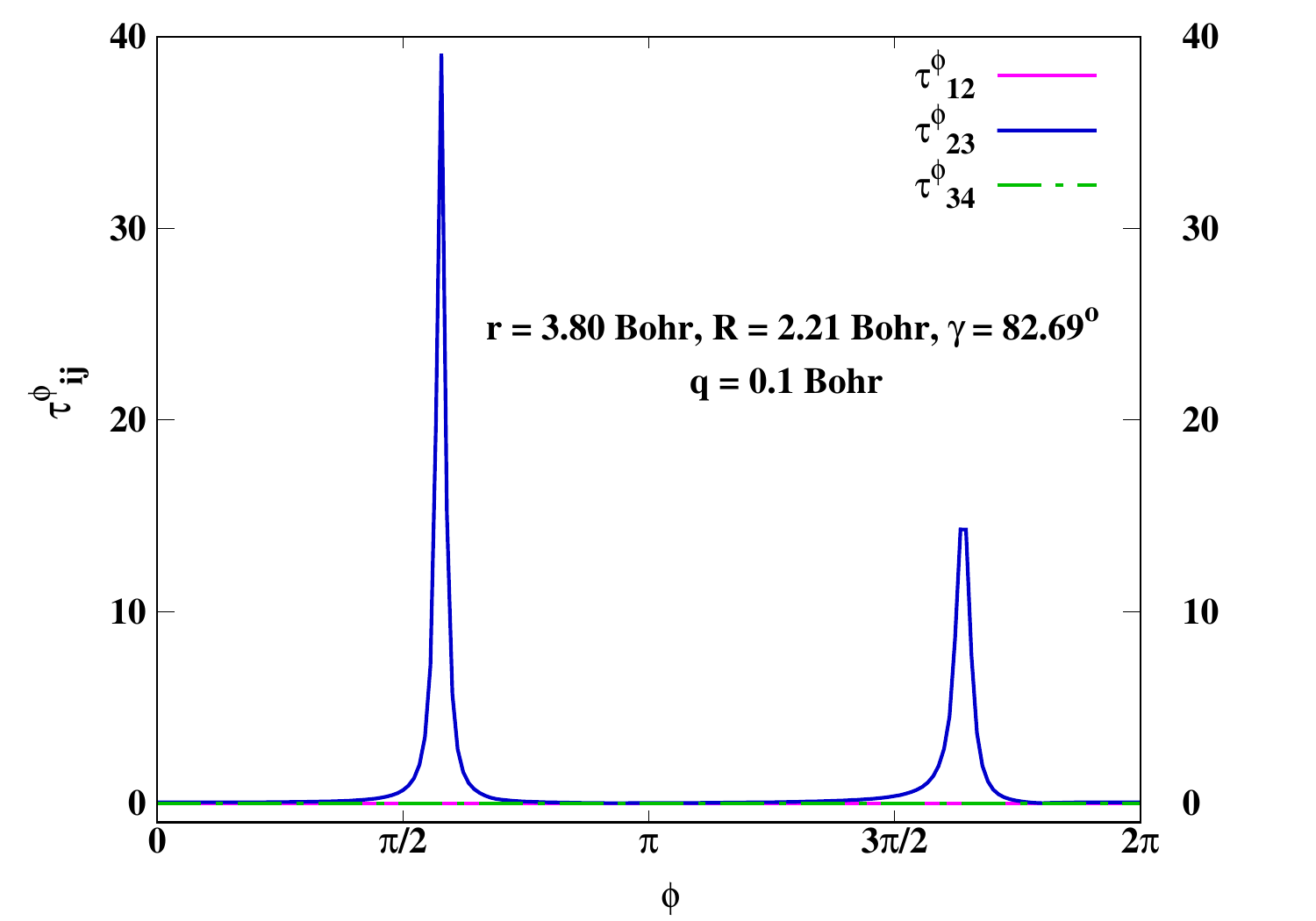}
				\put(0,70){\textbf{(b)}}
			\end{overpic}
			\phantomcaption
			\label{fig:Cs_23_tau}
		\end{subfigure}
		\hspace*{-3cm}
		\begin{subfigure}{0.4\linewidth}
			\centering
			\begin{overpic}[width=1.4\linewidth]{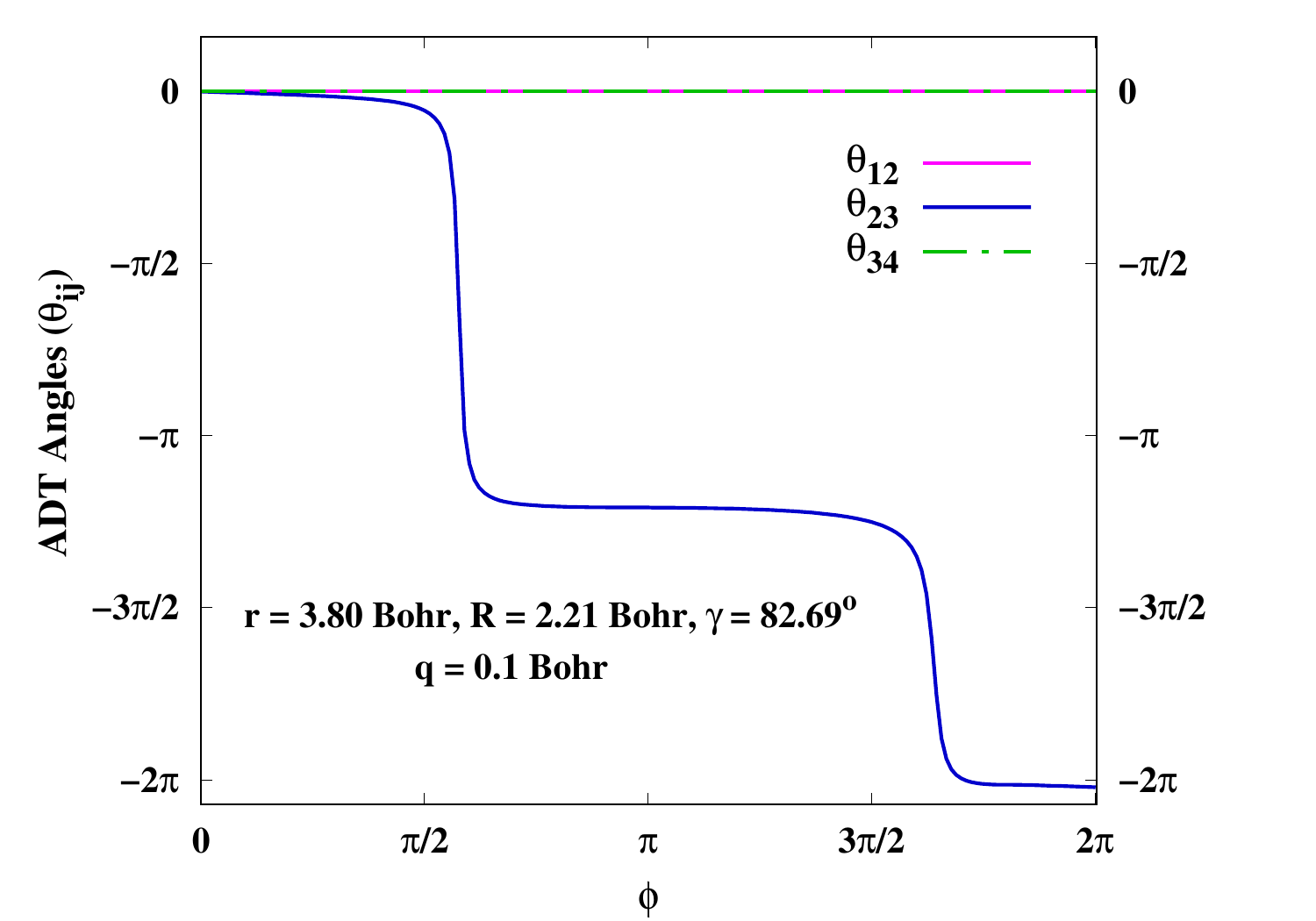}
				\put(1.5,70){\textbf{(c)}}
			\end{overpic}
			\phantomcaption
			\label{fig:Cs_23_angles}
		\end{subfigure}
		\hspace{2.5cm}
		\begin{subfigure}{0.4\linewidth}
			\centering
			\begin{overpic}[width=1.4\linewidth]{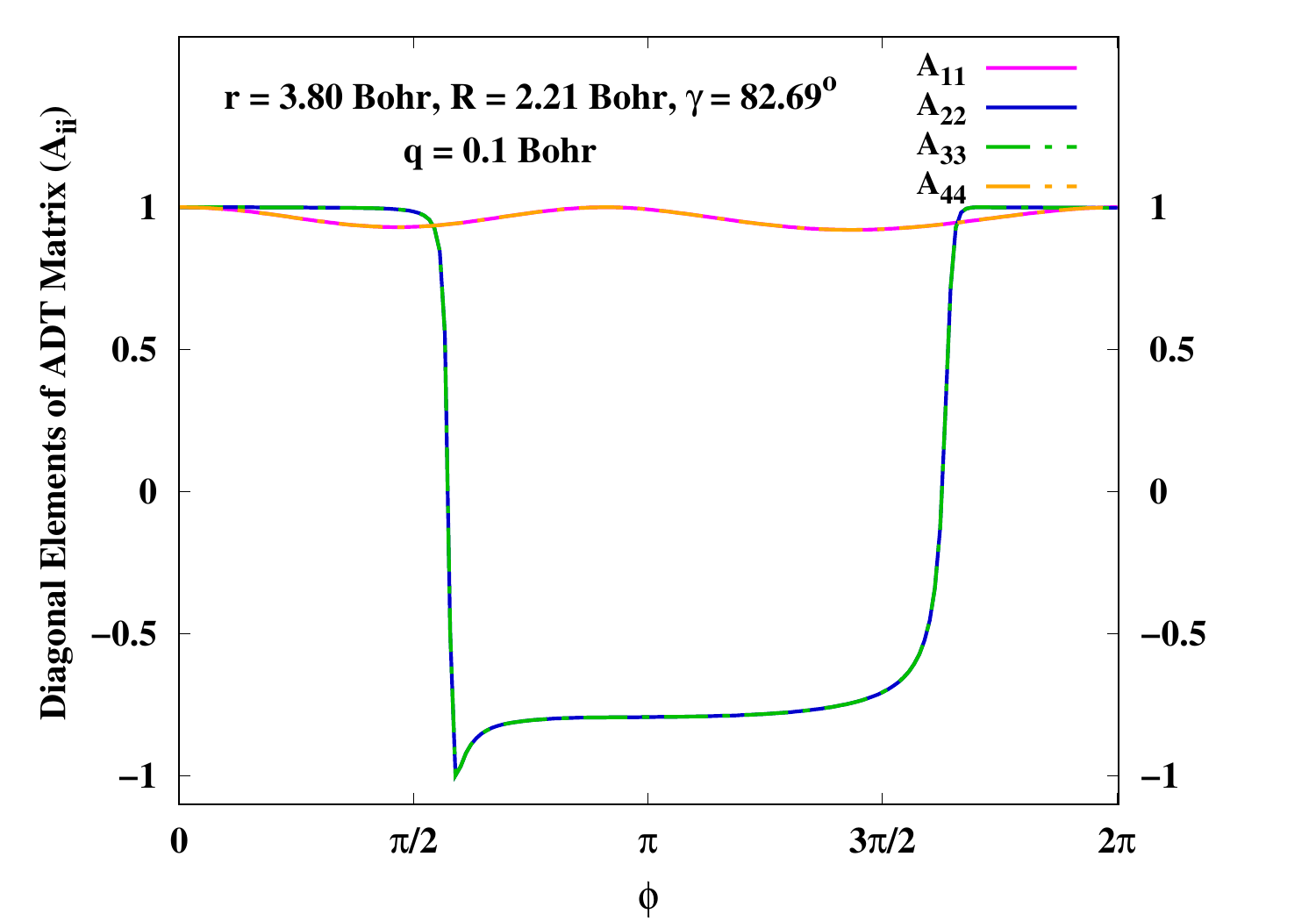}
				\put(1.5,70){\textbf{(d)}}
			\end{overpic}
			\phantomcaption
			\label{fig:Cs_23_amat}
		\end{subfigure}			
		\label{fig:Cs_23CI}
		\caption{Figure (a) depicts the PECs around a \textit{C}$_{s}$ point (\(r = 3.80\) and \(R = 2.21\) Bohr). Figure (b) illustrates the $\phi$ component of NACTs ($\tau^{\phi}_{ij}$) computed along the defined contour. Figures (c) and (d) represent the ADT angle ($\theta_{ij}$) and diagonal elements of the ADT matrix (\(A_{ii}\)), respectively. These quantities are computed using CAS(18e,12o) with AVQZ basis set along a circular contour of radius \(q = 0.1\) Bohr.}
	\end{figure}

	\noindent
	FIG.~\ref{fig:Cs_23_adiapes} and~\ref{fig:Cs_23_tau} show the adiabatic PECs and NACTs along the circular contour around the \textit{C}$_{s}$ geometry: $r$ = 3.80, $R$ = 2.21 and $\gamma$ = 82.69$^\circ$, respectively. The coupling term $\tau_{23}^\phi$ displays two sharp peaks where $u_2$ and $u_3$ approach closer to each other. In FIG.~\ref{fig:Cs_23_angles} and \ref{fig:Cs_23_amat}, the associated ADT angle $\theta_{23}$ reaches $2\pi$ and the diagonal matrix elements $A_{22}$ and $A_{33}$ exhibit two sign changes, respectively, which indicate the presence of two accidental \textit{C}$_{s}$ CIs between $u_2$ and $u_3$ inside the contour.
	
	\vspace{0.2cm}
	
	\noindent
	(ii) $r$ = 3.85, $R$ = 1.97, $\gamma$ = 88.57$^\circ$ and $q$ = 0.1 Bohr
	
	\vspace{0.2cm}
	
\begin{figure}[!htp]
		\centering
		\hspace*{-3cm}
		\begin{subfigure}{0.4\linewidth}
			\centering
			\begin{overpic}[width=1.4\linewidth]{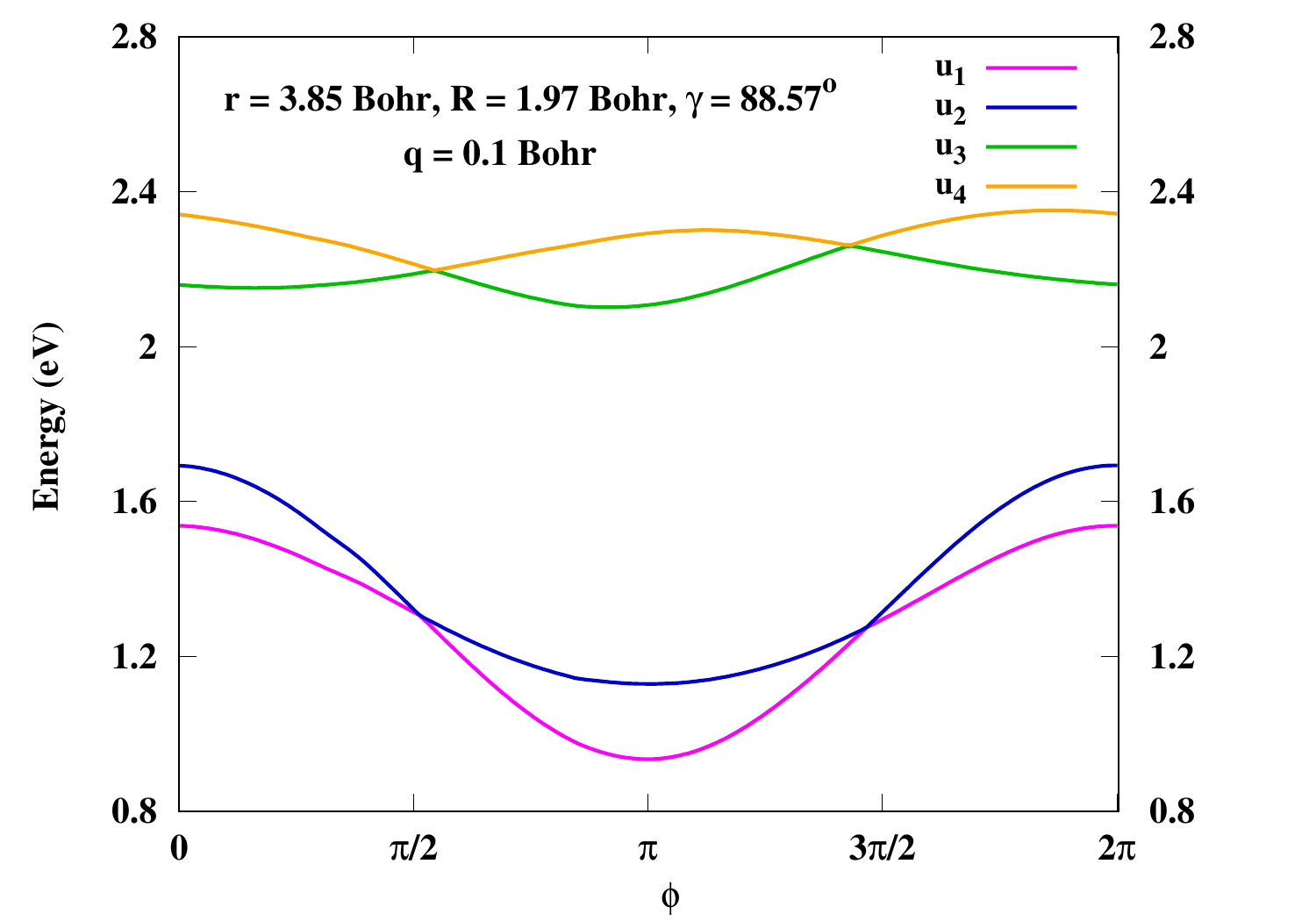}
				\put(5,75){\textbf{(a)}} 
			\end{overpic}
			\phantomcaption
			\label{fig:Cs_1234_adiapes}
		\end{subfigure}
		\hspace{2.5cm}
		\begin{subfigure}{0.4\linewidth}
			\centering
			\begin{overpic}[width=1.4\linewidth]{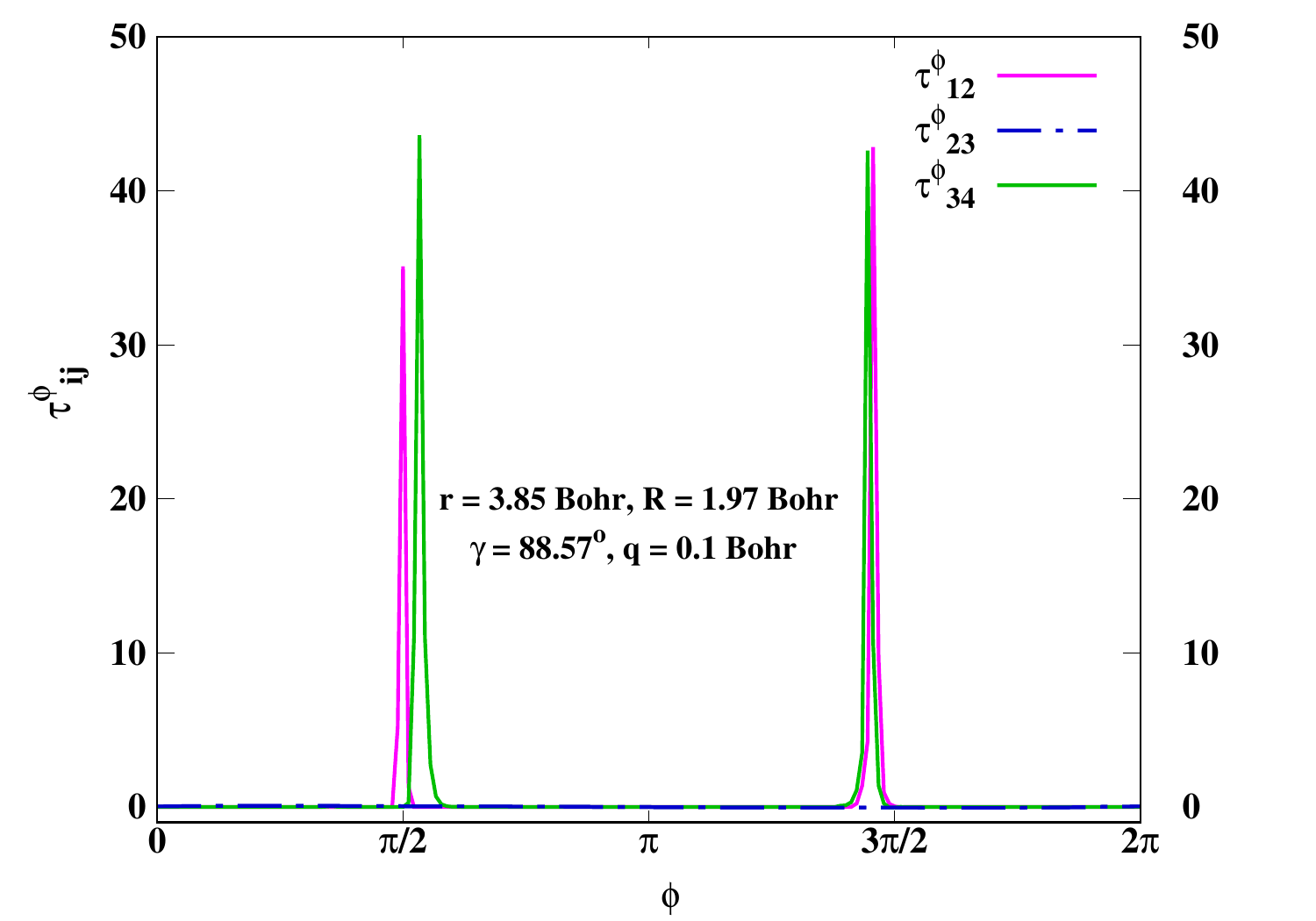}
				\put(0,70){\textbf{(b)}}
			\end{overpic}
			\phantomcaption
			\label{fig:Cs_1234_tau}
		\end{subfigure}
		\hspace*{-3cm}
		\begin{subfigure}{0.4\linewidth}
			\centering
			\begin{overpic}[width=1.4\linewidth]{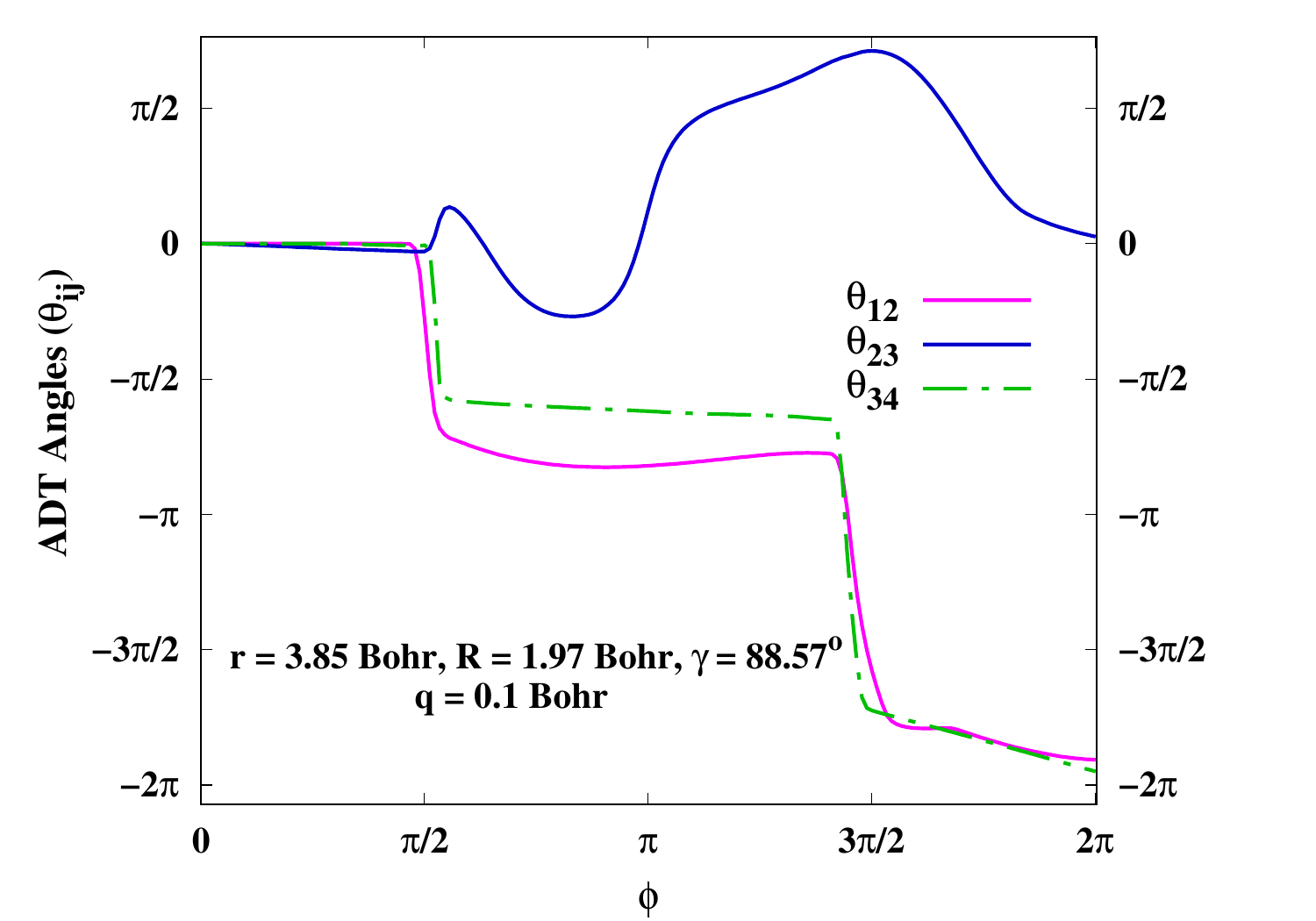}
				\put(1.5,70){\textbf{(c)}}
			\end{overpic}
			\phantomcaption
			\label{fig:Cs_1234_angle}
		\end{subfigure}
		\hspace{2.5cm}
		\begin{subfigure}{0.4\linewidth}
			\centering
			\begin{overpic}[width=1.4\linewidth]{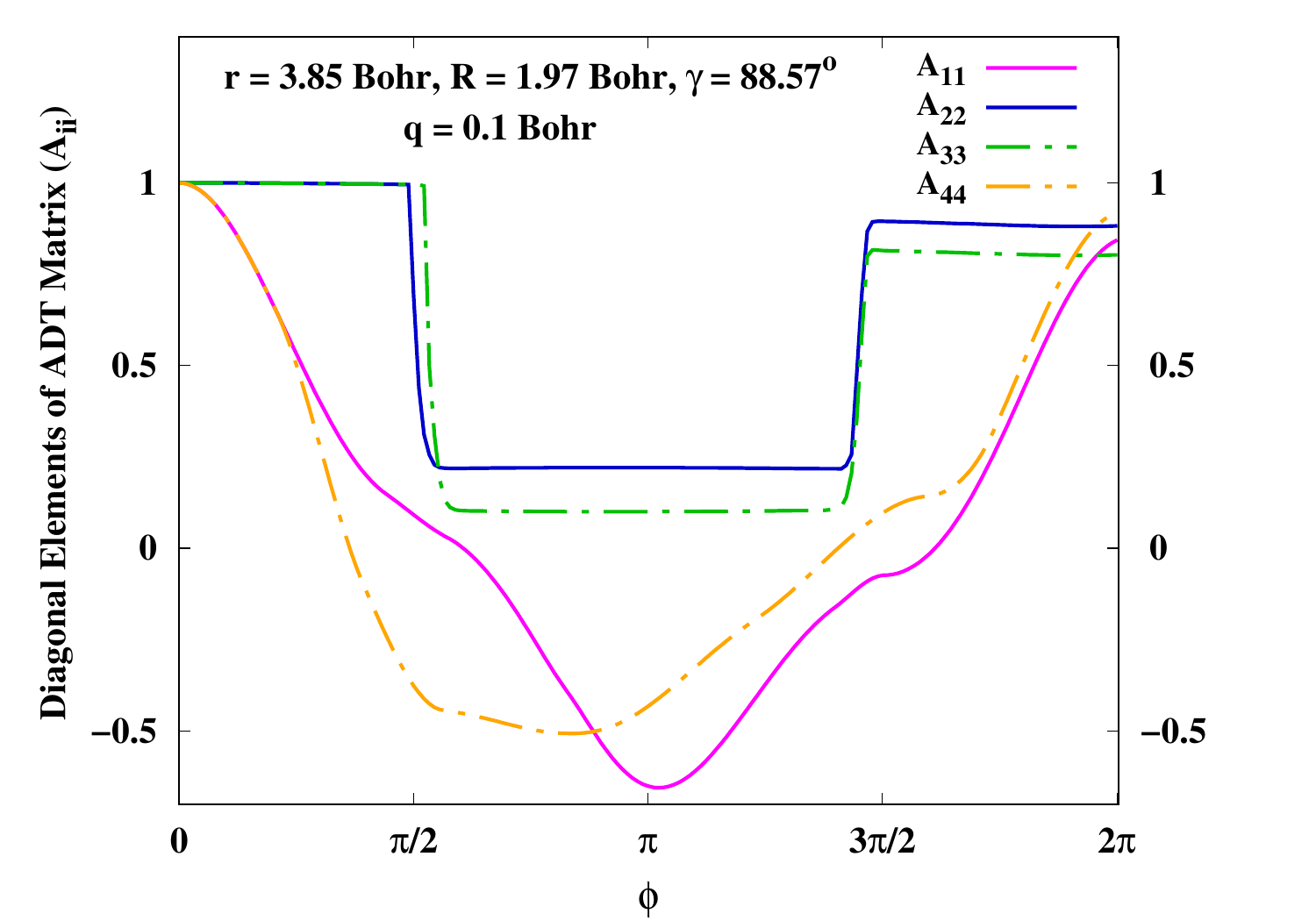}
				\put(1.5,70){\textbf{(d)}}
			\end{overpic}
			\phantomcaption
			\label{fig:Cs_1234_amat}
		\end{subfigure}			
		\label{fig:Cs_1234CI}
		\caption{Figure (a) depicts the PECs around a \textit{C}$_{s}$ point (\(r = 3.85\) and \(R = 1.97\) Bohr). Figure (b) illustrates the $\phi$ component of NACTs ($\tau^{\phi}_{ij}$) computed along the defined contour. Figures (c) and (d) represent the ADT angle ($\theta_{ij}$) and diagonal elements of the ADT matrix (\(A_{ii}\)), respectively. These quantities are computed using CAS(18e,12o) with AVQZ basis set along a circular contour of radius \(q = 0.1\) Bohr.}
	\end{figure}

	\noindent
	FIG.~\ref{fig:Cs_1234_adiapes} presents the PECs of O$_3$ as a function of $\phi$ around the \textit{C}$_{s}$ point (\(r = 3.85\) and \(R = 1.96\) Bohr), where the accidental \textit{C}$_{s}$ CIs between 1–2 and 3–4 electronic states are highlighted in FIG.~\ref{fig:Cs_1234_tau} by the corresponding functional forms of NACTs ($\tau_{ij}^\phi$). The presence of accidental CIs is further evidenced by the ADT angles and matrix elements illustrated in FIG.~\ref{fig:Cs_1234_angle} and FIG.~\ref{fig:Cs_1234_amat}, respectively. Specifically, the ADT angles $\theta_{12}$ and $\theta_{34}$ reach $2\pi$ and the diagonal ADT matrix elements ($A_{ii}$) undergo two sign changes at the end of the closed contour and thereby, confirming the existence of two accidental \textit{C}$_{s}$ CIs between the 1–2 and 3–4 states inside the contour. On the other hand, the negligible magnitude of $\tau_{23}^\phi$ and the zero value of ADT angle $\theta_{23}$ at the end of the close contour indicate a PJT interaction between $u_2$ and $u_3$.
	
	\noindent
	A systematic investigation of CIs among the four lowest electronic states of O$_3$ at selected geometries reveals interesting findings. In the \textit{D}$_{3h}$ geometry ($r = 3.06$, $R = 2.65$, $q = 0.25$ Bohr), a symmetry-driven CI is identified between $u_2$ and $u_3$, as indicated by the ADT angle $\theta_{23}$ attaining $\pi$ and a sign change in $A_{22}$ and $A_{33}$ at the end of closed contour. At one of the \textit{C}$_{2v}$ geometry ($r = 4.1$, $R = 3.4$, $q = 0.1$ Bohr), an accidental CI is identified between $u_2$ and $u_3$, as evidenced by the $\pi$ phase change in the ADT angle $\theta_{23}$ and the corresponding sign change in $A_{22}$ and $A_{33}$. Another \textit{C}$_{2v}$ configuration ($r = 3.85$, $R = 1.96$, $q = 0.1$ Bohr) displays accidental CIs between $u_1$–$u_2$ and $u_3$–$u_4$, as confirmed by phase changes of $\pi$ in the corresponding ADT angles and associated sign changes in the diagonal ADT matrix elements at the end of the closed contour. Apart from \textit{D}$_{3h}$ or \textit{C}$_{2v}$ geometries, \textit{C}$_s$ CIs are also present between 1-2, 2-3 and 3-4 electronic states. In summary, the present calculation on O$_3$ has the similar findings with other calculations~\cite{Alijah2018} on the location of symmetry driven (\textit{D}$_{3h}$) CI between the states, u$_2$ and u$_3$ as well as the accidental (\textit{C}$_{2v}$ and \textit{C}$_{s}$) CIs between all the adjacent electronic states (1-2, 2-3 and 3-4).
	
	\section*{IV. Hyperspherical Coordinates: Curl Condition and Diabatic PES Matrix}
	
	\noindent
	Since the hyperspherical coordinates treat both the reactant and the product channels even handedly, we intend to construct multi-state diabatic Hamiltonian for the O$_3$ system in this coordinate system so that it will be convenient to perform dynamical calculation. While constructing such Hamiltonian, we calculate adiabatic PESs using 7S-SA-MCSCF method followed by ic-MRCI(Q) approach with the CAS(18e, 12o) and AVQZ basis set as well as evaluate the NACTs by employing CP-MCSCF methodology with the same level of CAS and basis set. Once the \textit{ab initio} adiabatic PESs and NACTs are in hand over a grid of hyperangles ($\theta$ and $\phi$) for each fixed value of hyperradius $\rho$, the ADT equations are solved at each point of the configuration space to obtain the ADT angles (see Section S3 of SM) and diabatic Hamiltonian (see Section S4 of SM). In this coordinate system, the hyperradius $\rho$ characterizes the overall size of the triangle formed by the three atoms, while the hyperangles $\theta$ ($0 \le \theta \le 90^\circ$) and $\phi$ ($0 \le \phi \le 360^\circ$) define the shape and orientation of the triangle. The relations between the inter-particle distances ($R_1,R_2$ and $R_3$) of the three atoms and the hyperspherical coordinates are given by:
	
	\begin{subequations}
		\begin{eqnarray}
		R_1 &=& \dfrac{\rho}{\sqrt{2}}d_3\left[1+\sin\theta\cos(\phi+\epsilon_3)\right]^{1/2}, \\
		R_2 &=& \dfrac{\rho}{\sqrt{2}}d_1\left(1+\sin\theta\cos\phi\right)^{1/2}, \\
		R_3 &=& \dfrac{\rho}{\sqrt{2}}d_2\left[1+\sin\theta\cos(\phi-\epsilon_2)\right]^{1/2},
		\end{eqnarray}
		\label{eq:hyper}
	\end{subequations}	
	
	\noindent
	where $d_i = \sqrt{m_i(m_j+m_k)/\mu M}, \quad M = \sum_i^3 m_i, \quad \mu = \sqrt{(m_1m_2m_3/M)}, \quad	\epsilon_2 = 2\tan^{-1}(m_3/\mu)$ and $\epsilon_3 = 2\tan^{-1}(m_2/\mu)$. For O$_3$, $m_1 = m_2 = m_3 = m_{O} = 15.9994$ amu, $M = 47.9982$ amu, $\mu = 9.2373$ amu, $d_1 = d_2 = d_3 = 1.0746$ and $\epsilon_2 = \epsilon_3 = 119.9998^{\circ}$.
	
	\vspace{0.2cm}
	
\begin{figure}[!htp]
		\centering
		\includegraphics[width=0.7\linewidth]{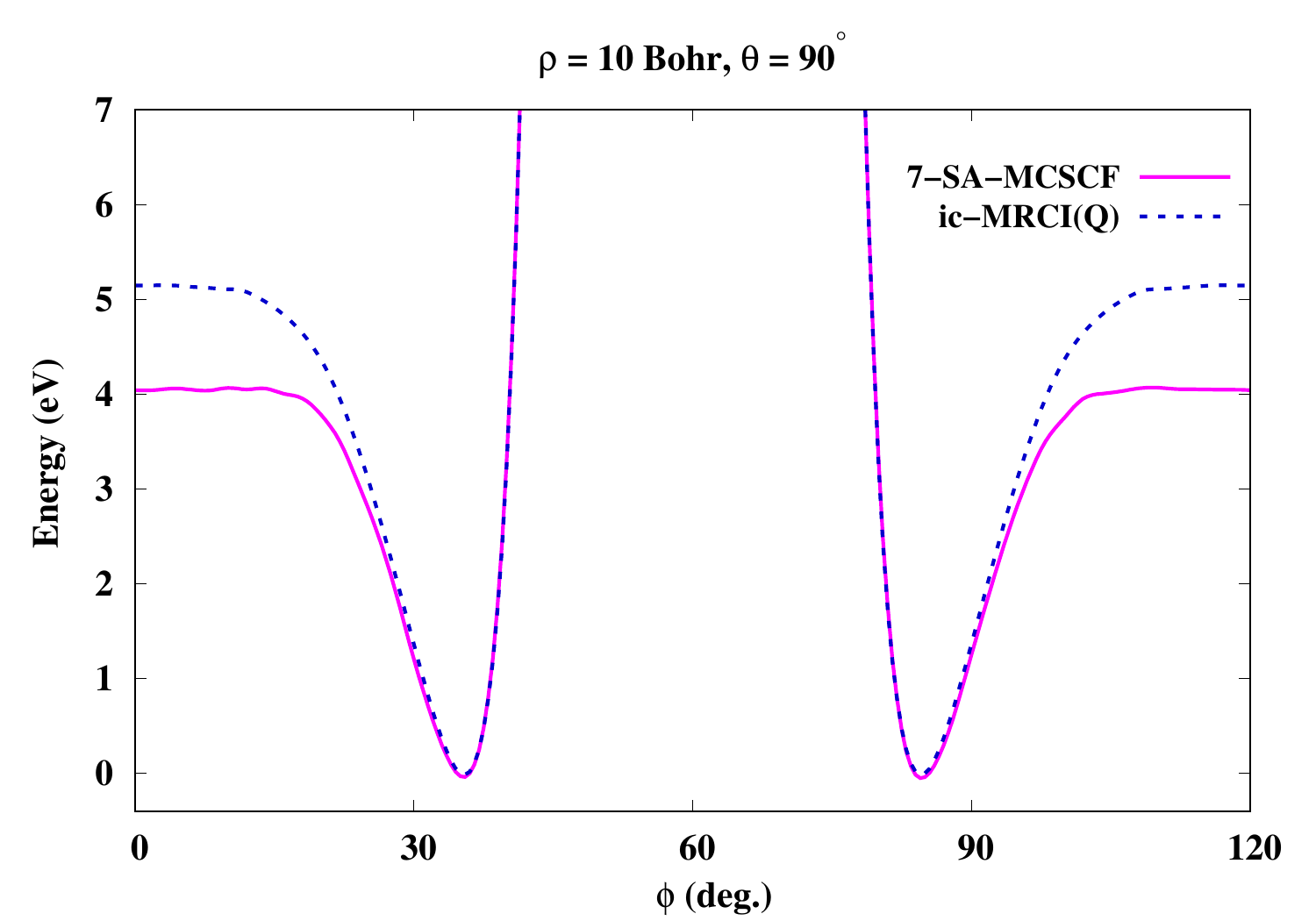}
		\caption{Ground adiabatic PEC of O$_3$ representing the dissociation of O$_2$ + O $\rightarrow$ O + O + O at $\rho$ = 10 Bohr and $\theta$ = 90$^\circ$, computed using CAS(18e,12o) and AVQZ basis set}
		\label{fig:rho_10}
	\end{figure}

	\noindent	
	At first, we calculate the dissociation energy of O$_2$ to depict the accuracy of the PEC at the asymptotic value of fixed $\rho = 10$ Bohr and $\theta = 90^\circ$, and show the ground state PEC as a function of hyperangle $\phi$ in Figure~\ref{fig:rho_10} by using 7S-SA-MCSCF followed by ic-MRCI(Q) methodology with the same CAS and basis set.
	
	\vspace{0.2cm}
	
	\vspace{0.2cm}
	
\begin{figure}[!htp]
		\centering
		\centering
		\hspace*{-3cm}
		\begin{subfigure}{0.4\linewidth}
			\centering
			\begin{overpic}[width=1.3\linewidth]{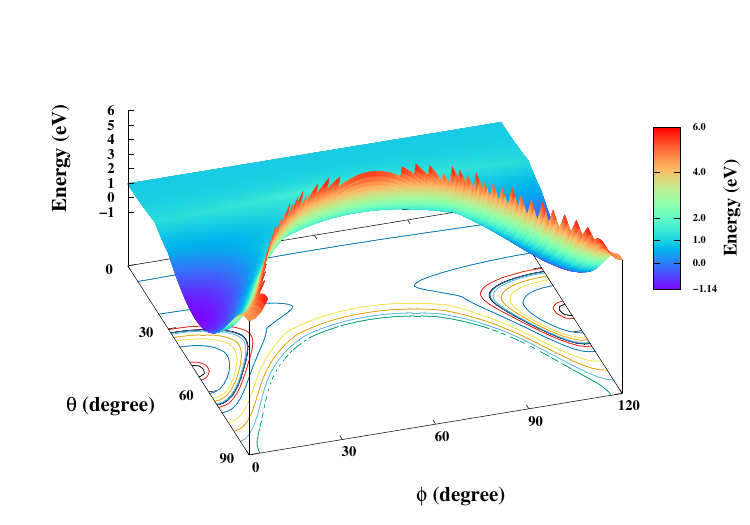}
				\put(1.5,70){\textbf{(a)} \hspace{4cm} u\textsubscript{1}} 
			\end{overpic}
			\phantomcaption
			\label{fig:u1_hyper}
		\end{subfigure}
		\hspace{2.5cm}
		\begin{subfigure}{0.4\linewidth}
			\centering
			\begin{overpic}[width=1.3\linewidth]{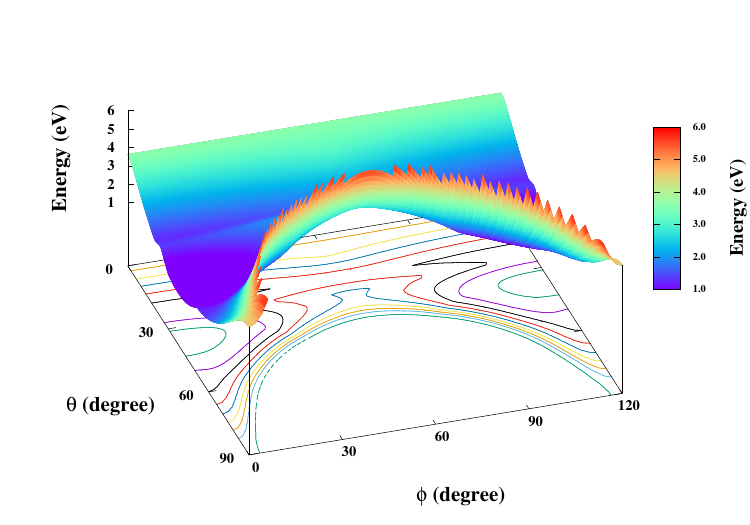}
				\put(1.5,70){\textbf{(b)} \hspace{4cm} u\textsubscript{2}}
			\end{overpic}
			\phantomcaption
			\label{fig:u2_hyper}
		\end{subfigure}
		\hspace*{-3cm}
		\begin{subfigure}{0.4\linewidth}
			\centering
			\begin{overpic}[width=1.3\linewidth]{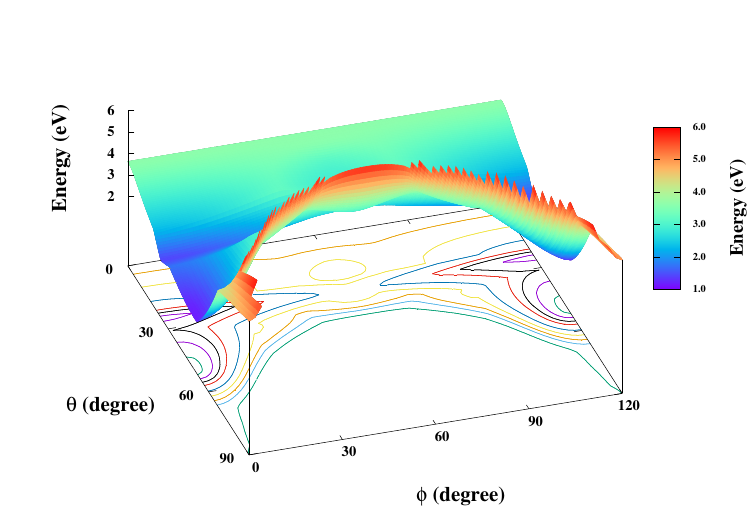}
				\put(1.5,70){\textbf{(c)} \hspace{4cm} u\textsubscript{3}}
			\end{overpic}
			\phantomcaption
			\label{fig:u3_hyper}
		\end{subfigure}
		\hspace{2.5cm}
		\begin{subfigure}{0.4\linewidth}
			\centering
			\begin{overpic}[width=1.3\linewidth]{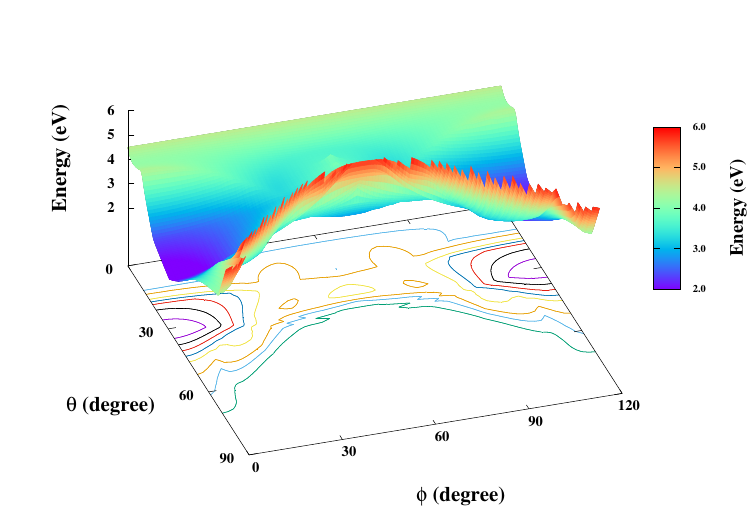}
				\put(1.5,70){\textbf{(d)} \hspace{4cm} u\textsubscript{4}}
			\end{overpic}
			\phantomcaption
			\label{fig:u4_hyper}
		\end{subfigure}			
		\caption{Four low lying adiabatic PESs of O$_3$ [Figures (a) to (d)] at $\rho = 4$ Bohr over $\theta$--$\phi$ space computed using 7S-SA-MCSCF followed by ic-MRCI(Q) with a CAS(18e,12o) and AVQZ basis set.}
		\label{fig:rho_4}
	\end{figure}

\begin{figure}[!htp]
		\centering
		\hspace*{-2cm}
		\begin{subfigure}{0.4\linewidth}
			\centering
			\begin{overpic}[width=1.3\linewidth]{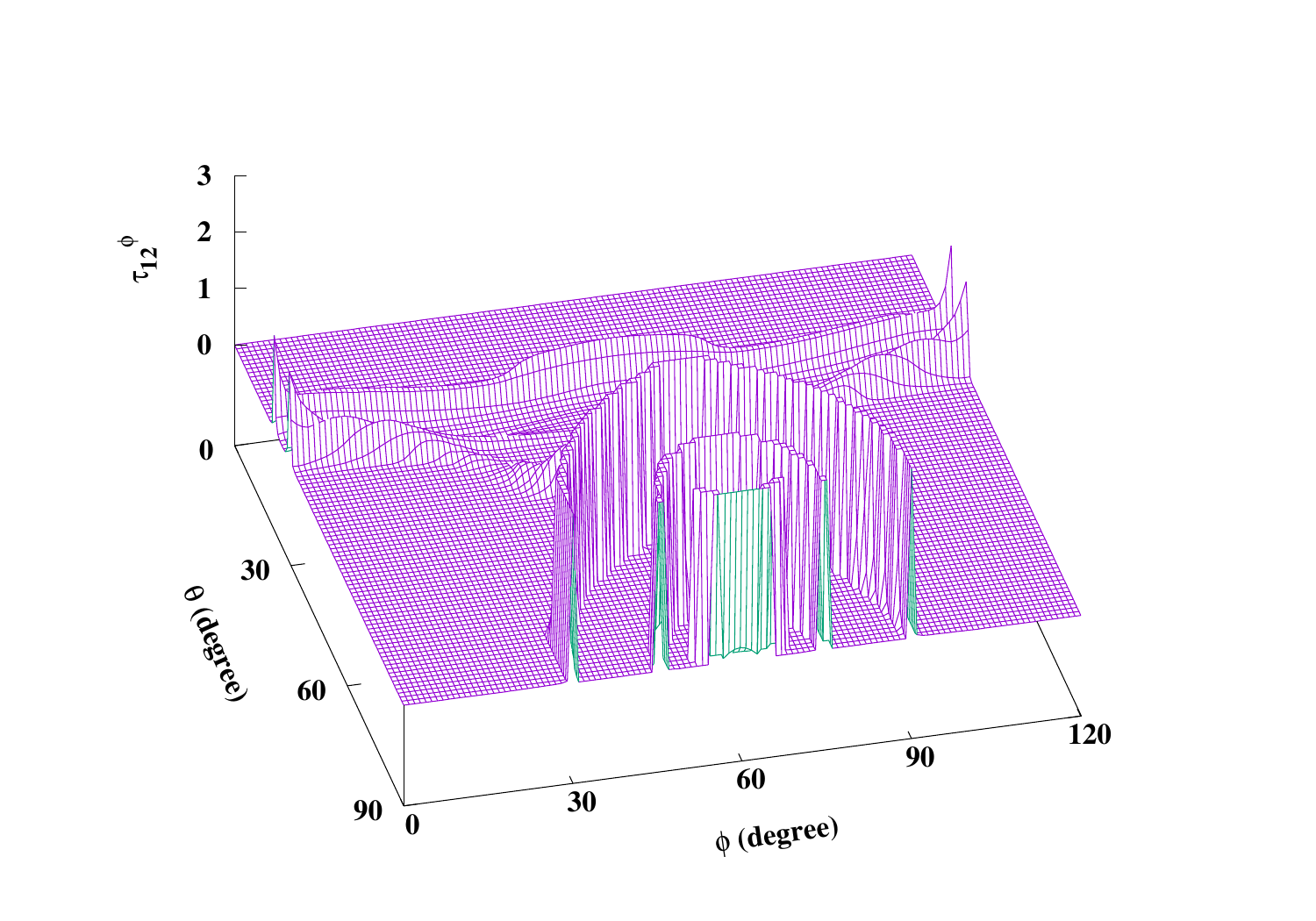}
				\put(0,70){\textbf{(a)}}
			\end{overpic}
			\phantomcaption
			\label{fig:taup_12}
		\end{subfigure}
		\hspace{2.5cm}
		\begin{subfigure}{0.4\linewidth}
			\centering
			\begin{overpic}[width=1.3\linewidth]{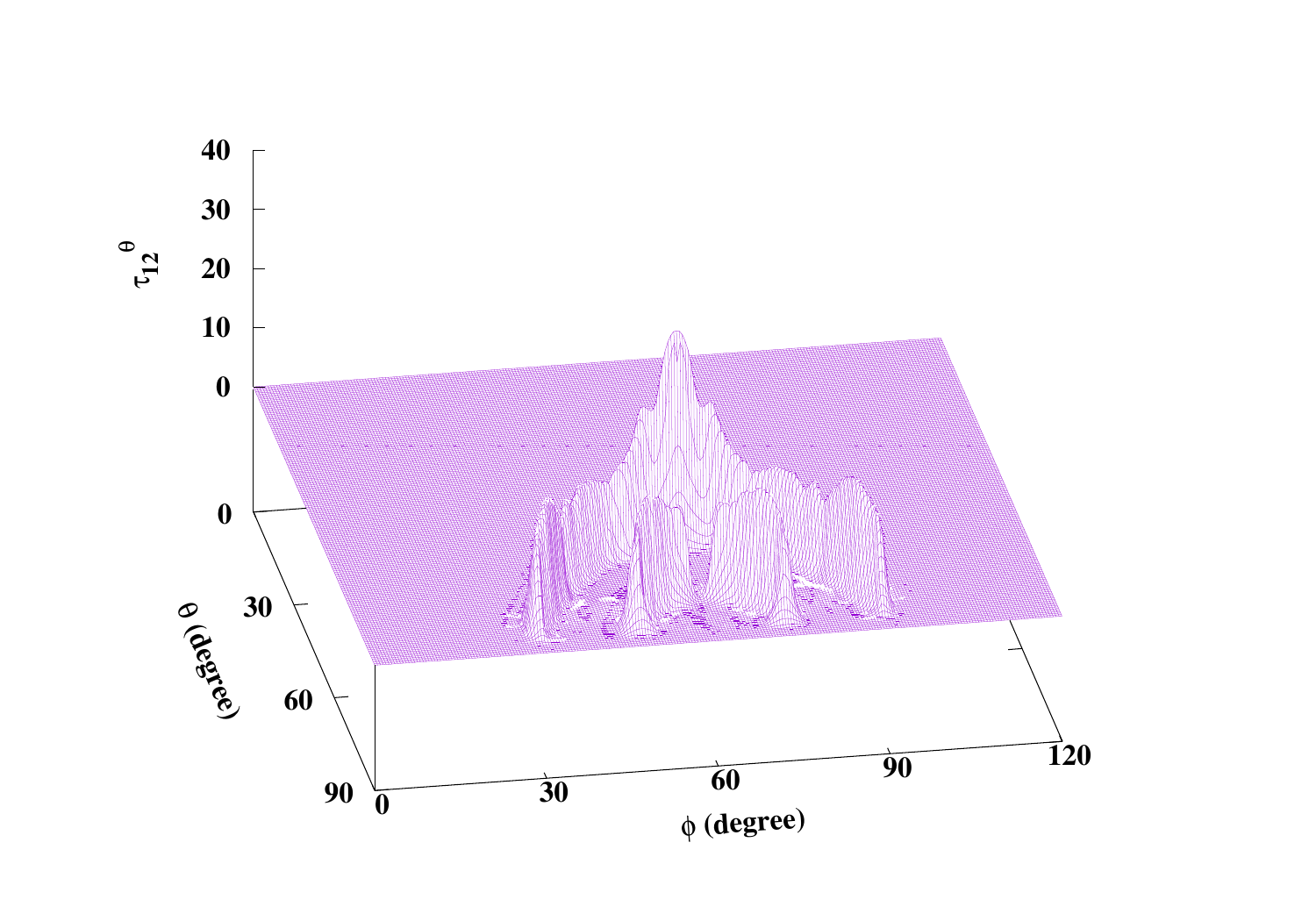}
				\put(1.5,70){\textbf{(b)}}
			\end{overpic}
			\phantomcaption
			\label{fig:taut_12}
		\end{subfigure}
		\hspace*{-2cm}
		\begin{subfigure}{0.4\linewidth}
			\centering
			\begin{overpic}[width=1.3\linewidth]{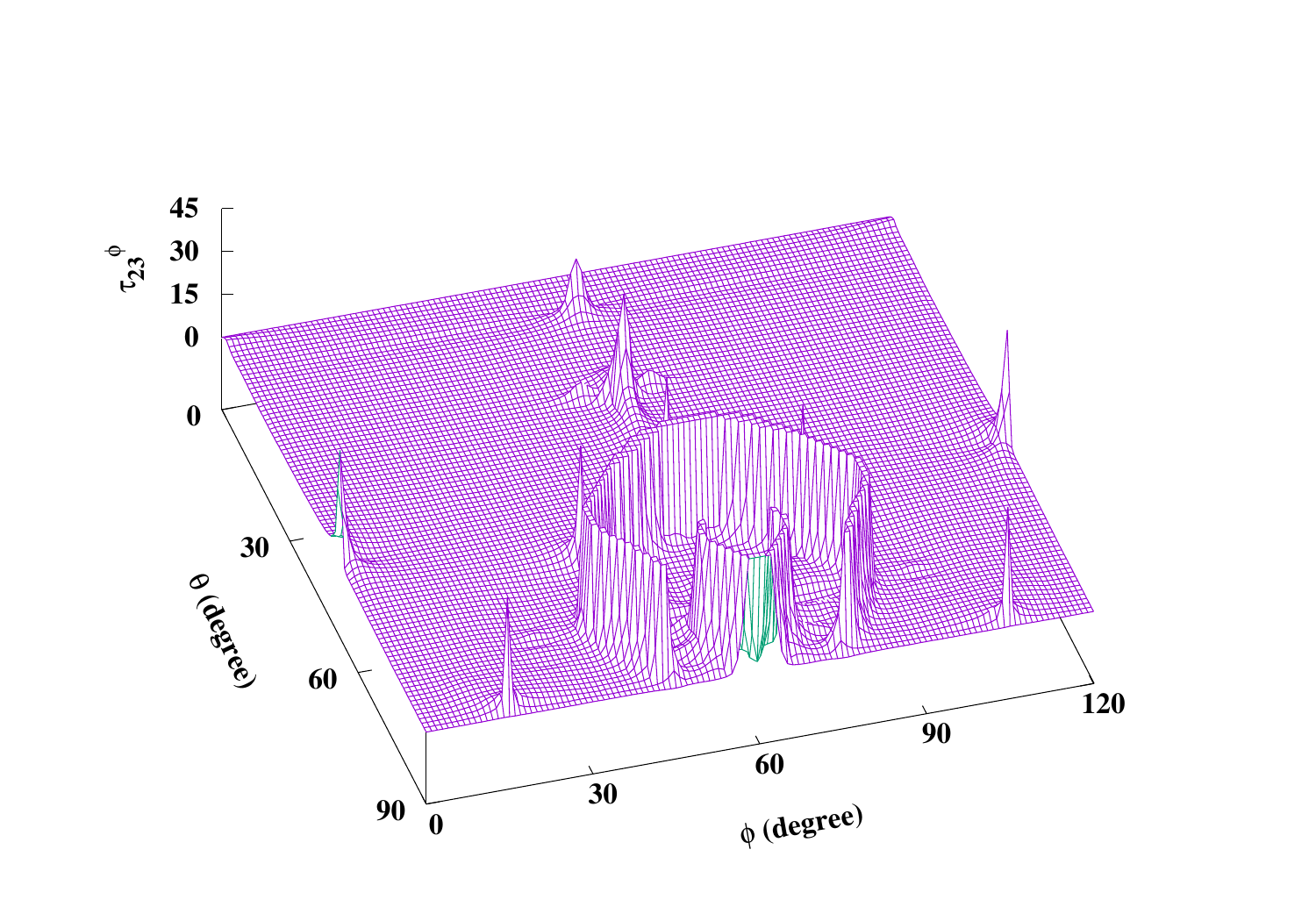}
				\put(1.5,70){\textbf{(c)}}
			\end{overpic}
			\phantomcaption
			\label{fig:taup_23}
		\end{subfigure}
		\hspace{2.5cm}
		\begin{subfigure}{0.4\linewidth}
			\centering
			\begin{overpic}[width=1.3\linewidth]{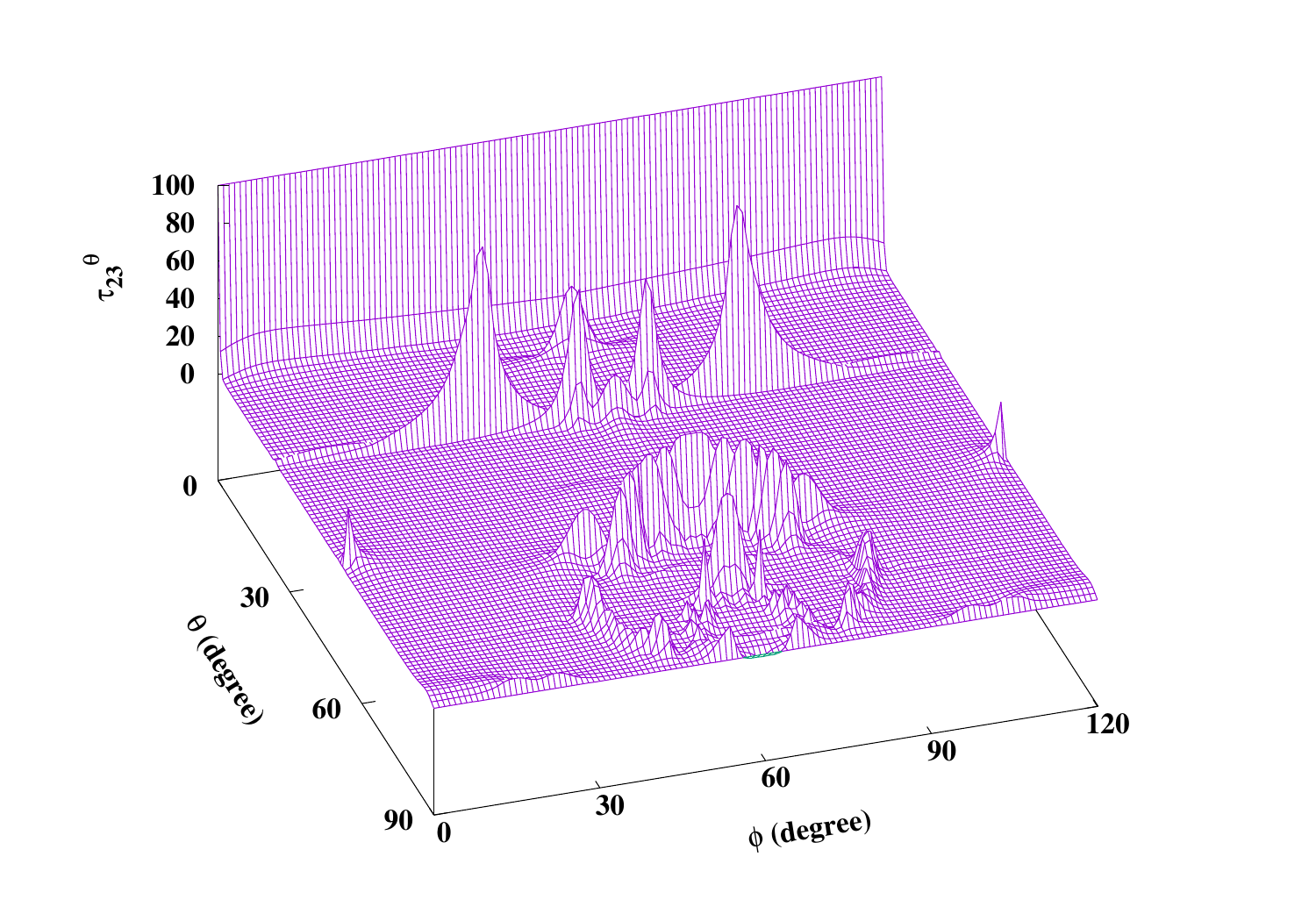}
				\put(1.5,70){\textbf{(d)}}
			\end{overpic}
			\phantomcaption
			\label{fig:taut_23}
		\end{subfigure}
		\hspace*{-2cm}
		\begin{subfigure}{0.4\linewidth}
			\centering
			\begin{overpic}[width=1.3\linewidth]{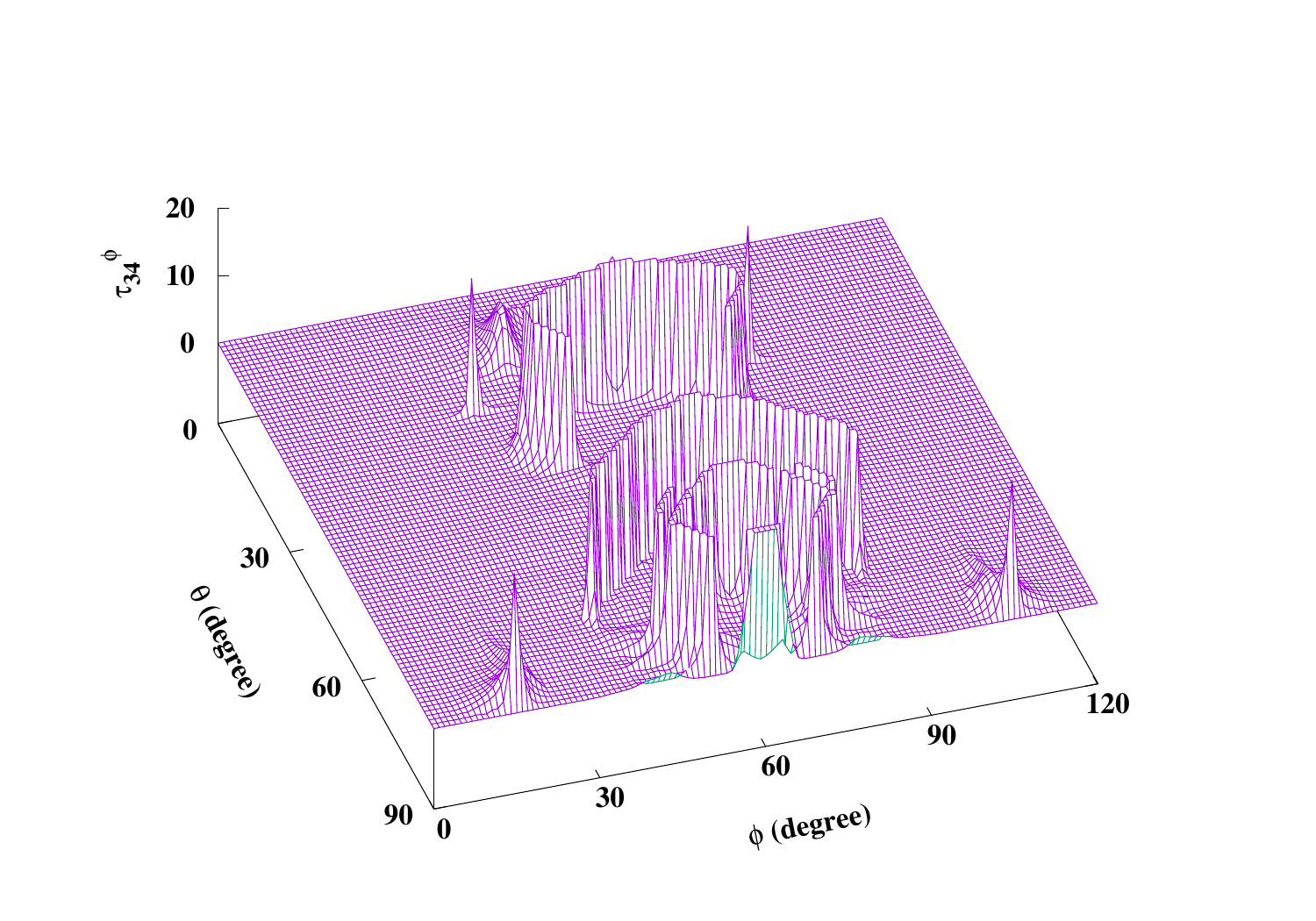}
				\put(1.5,70){\textbf{(e)}}
			\end{overpic}
			\phantomcaption
			\label{fig:taup_34}
		\end{subfigure}
		\hspace{2.5cm}
		\begin{subfigure}{0.4\linewidth}
			\centering
			\begin{overpic}[width=1.3\linewidth]{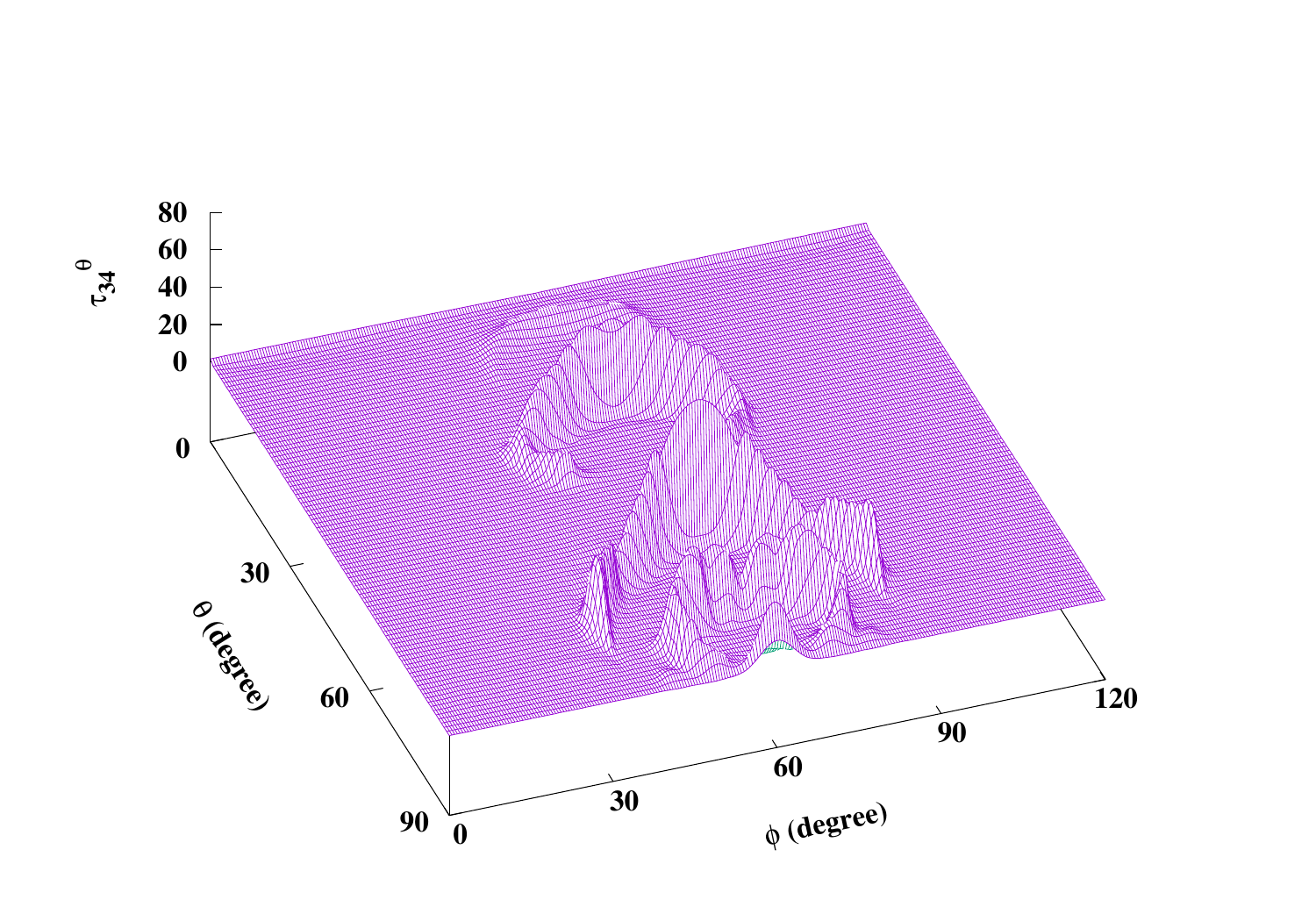}
				\put(1.5,70){\textbf{(f)}}
			\end{overpic}
			\phantomcaption
			\label{fig:taut_34}
		\end{subfigure}	
		\caption{Hyperspherical $\phi$ and $\theta$ components of NACTs of O$_3$ using the CP-MCSCF methodology over $ \theta $-$ \phi $ configuration space for a fixed value of $ \rho $ = 4.0 Bohr.}
		\label{fig:nact_rho4}
	\end{figure}

	\noindent
	At $\rho$ = 4.0 Bohr, \textit{ab initio} adiabatic PESs and NACTs are computed over a $\theta$–$\phi$ grid ranging from $0 \leq \theta \leq 90^\circ$ and $0 \leq \phi \leq 360^\circ$, using 91 $\times$ 361 grid points as displayed Figure \ref{fig:rho_4} and \ref{fig:nact_rho4}, respectively to construct the diabatic PES matrix. Since the adiabatic PESs and NACTs of the triatomic system, O$_3$ have the inherent symmetry along the hyperangle $\phi$ and those quantities show repetition for every after $\phi$ = 120$^\circ$ at fixed $\rho$ and $\theta$, it is possible to reduce the computational cost by one-third.

	\vspace*{1cm}

	\subsection*{IV.A. Existence of four-state SHS: Curl Condition}
	
	\noindent
	While locating the symmetry-driven and accidental CIs as well as constructing the diabatic surfaces of O$_3$ system, it is necessary to explore the numerical ``accuracy'' of \textit{ab initio} calculated NACTs and then, to find out the existence of a SHS. In the process of solving ADT equations (see Section S3 of SM) by plugging the \textit{ab initio} calculated NACTs, we numerically calculate the mathematical curl ($Z_{\theta \phi}^{ij}$) and the ADT curl ($C_{\theta\phi}^{ij}$), and determine their equality at each point of the configuration space (CS) for a given sub-space. The detail formulation of those equations can be found elsewhere~\cite{SHS2023}. In Figure S8 of SM, we present the cuts of the mathematical curl ($Z_{\theta\phi}^{ij}$) and the ADT curl ($C_{\theta \phi}^{ij}$) as well as the Curl condition ($F_{\theta \phi}^{ij}$) as  a function of hyperspherical angle $ \phi $ for different $ \theta $s, namely, 50$^\circ$, 70$^\circ$ and 90$^\circ$ at fixed $ \rho $ = 4 Bohr. Note that those profiles are obtained by employing the NACTs between the lowest four electronic states of O$_3$. Although the numerical values of $Z_{\theta \phi}^{ij}$ and $C_{\theta \phi}^{ij}$ are substantially large, in particular, around the CIs, those are practically identical and the difference $F_{\theta \phi}^{ij}$ is almost zero, which confirms the validity of curl condition for the lowest four electronic states of O$_3$ forming the SHS.

	\subsection*{IV.B. Diabatic PES Matrix}
\begin{figure}[!htp]
		\centering
		\hspace*{-2cm}
		\begin{subfigure}{0.4\linewidth}
			\centering
			\begin{overpic}[width=1.4\linewidth]{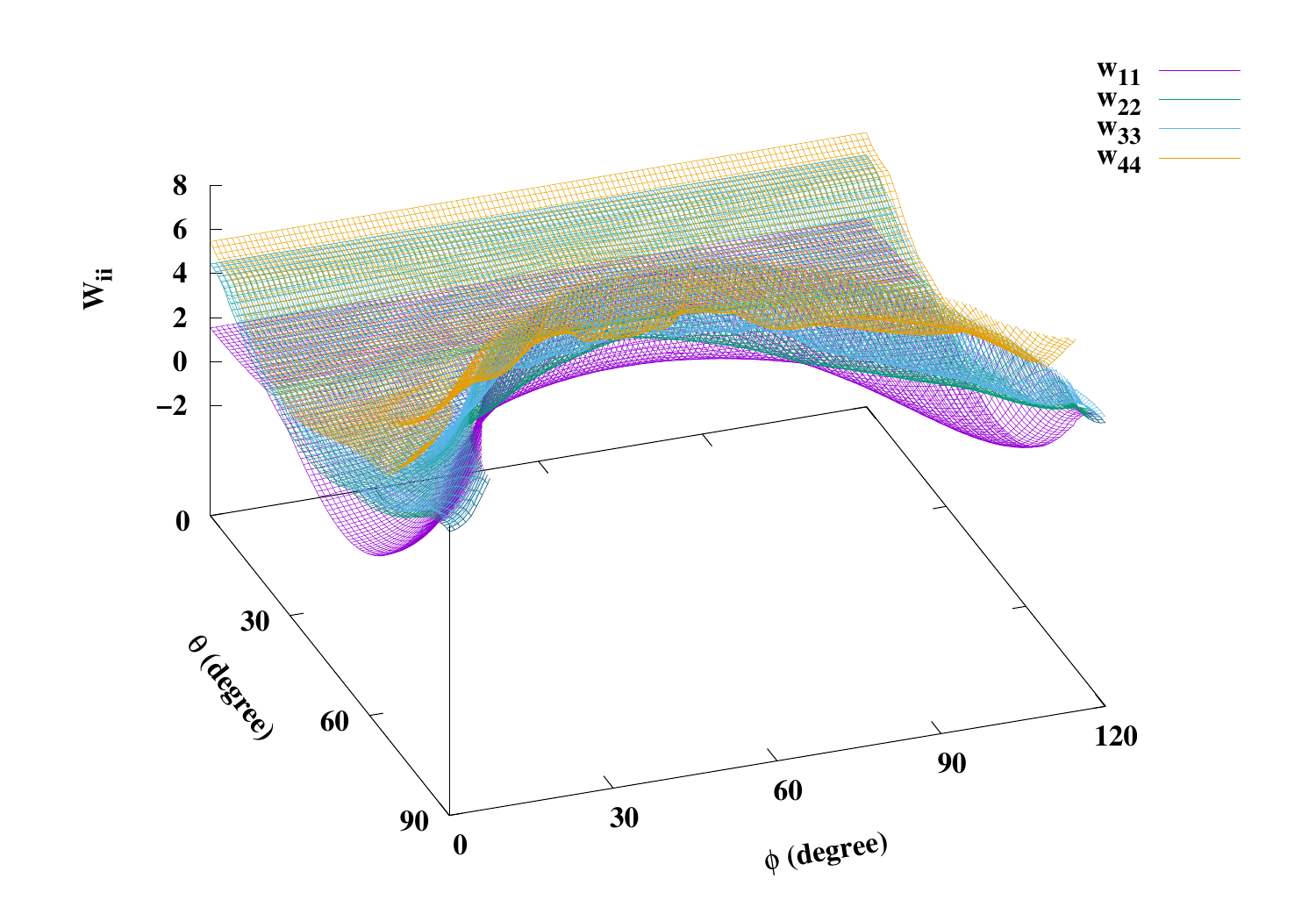}
				\put(0,70){\textbf{(a)}}
			\end{overpic}
			\phantomcaption
			\label{fig:w11}
		\end{subfigure}
		\hspace{2.5cm}
		\begin{subfigure}{0.4\linewidth}
			\centering
			\begin{overpic}[width=1.4\linewidth]{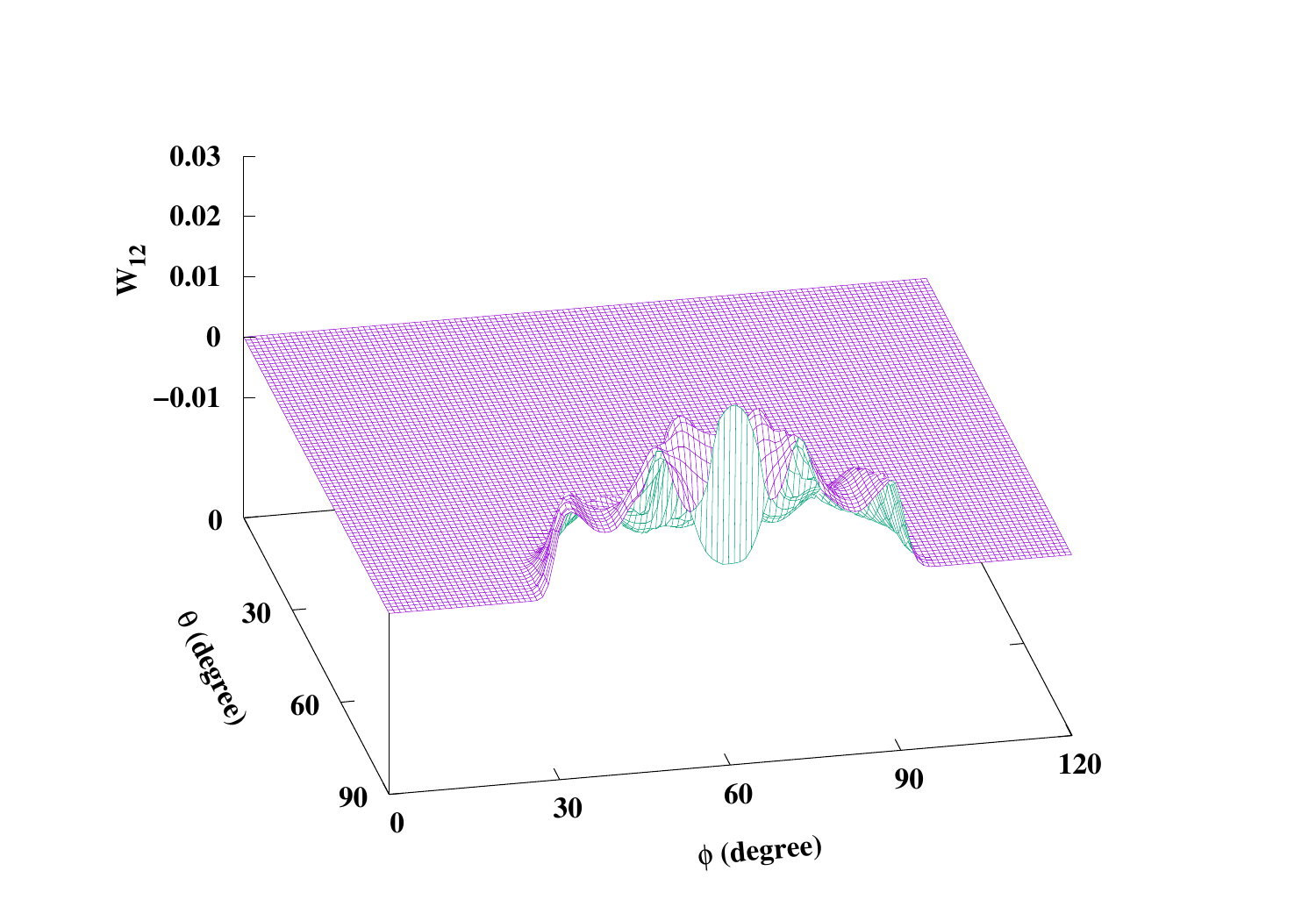}
				\put(1.5,70){\textbf{(b)}}
			\end{overpic}
			\phantomcaption
			\label{fig:w22}
		\end{subfigure}
		\hspace*{-2cm}
		\begin{subfigure}{0.4\linewidth}
			\centering
			\begin{overpic}[width=1.4\linewidth]{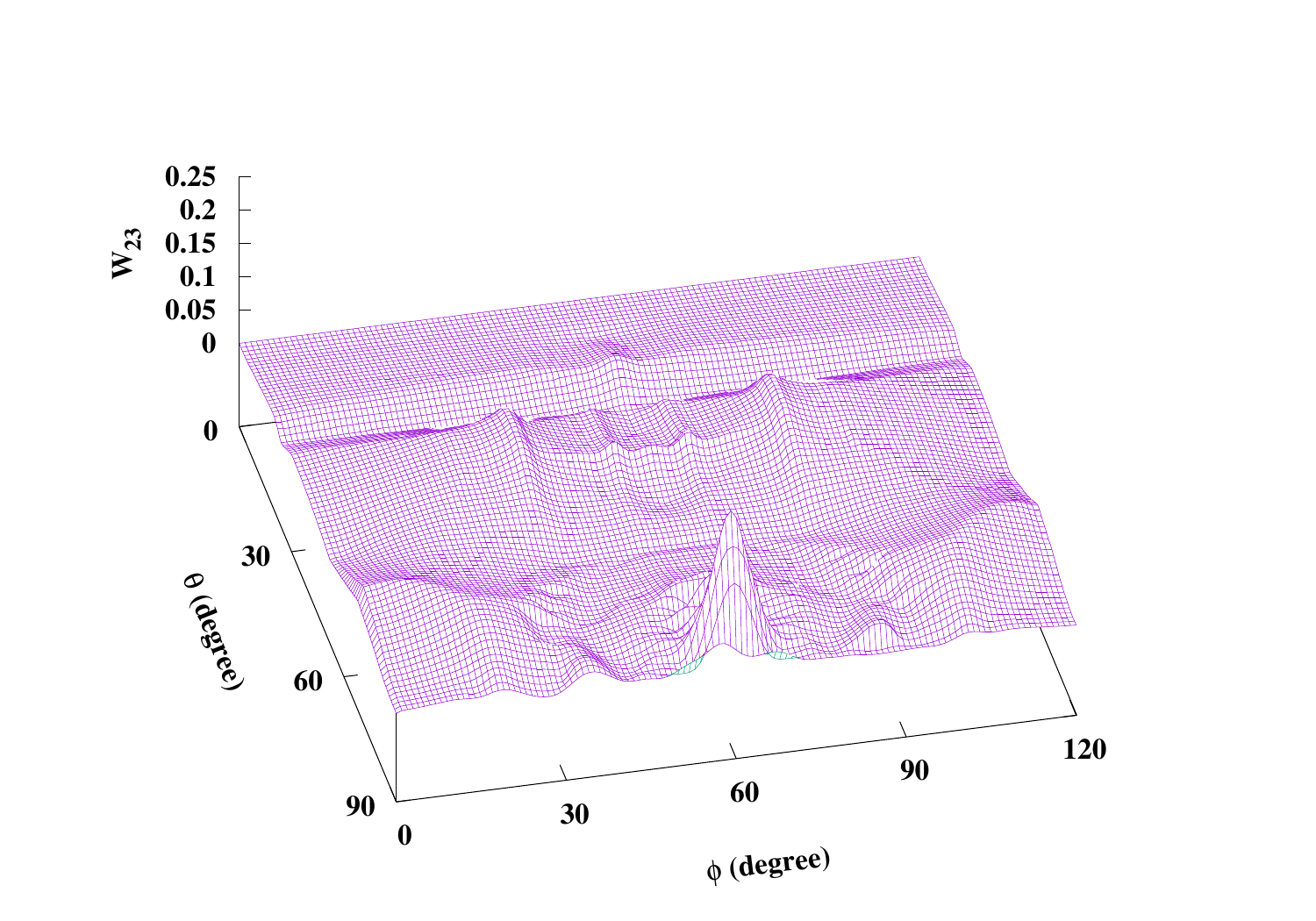}
				\put(1.5,70){\textbf{(c)}}
			\end{overpic}
			\phantomcaption
			\label{fig:w12}
		\end{subfigure}
		\hspace{2.5cm}
	   	\begin{subfigure}{0.4\linewidth}
	    	\centering
	    	\begin{overpic}[width=1.4\linewidth]{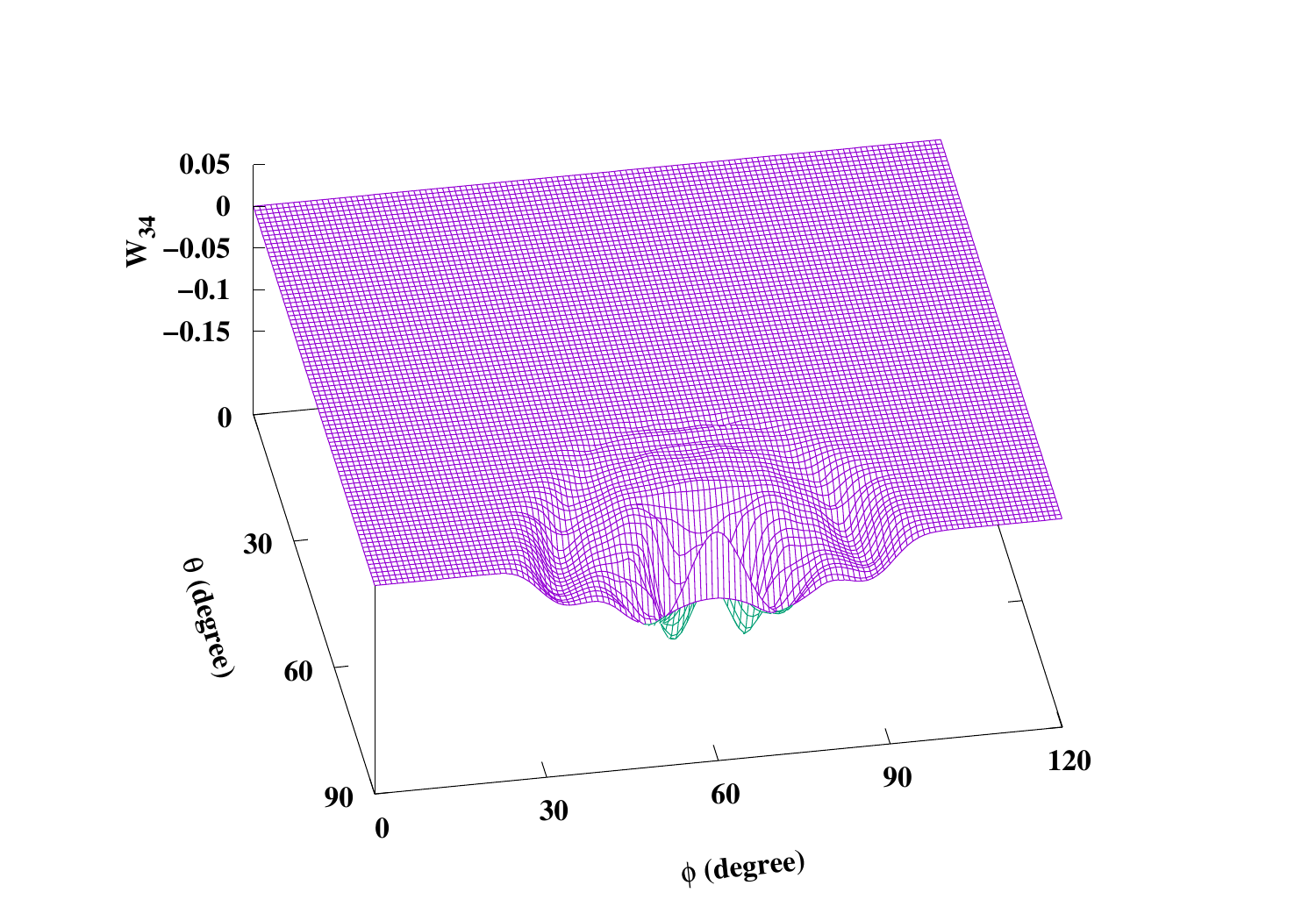}
	    		\put(1.5,70){\textbf{(d)}}
	    	\end{overpic}
	    	\phantomcaption
	    	\label{fig:w23}
	    \end{subfigure}		
		\caption{Figures (a) presents the diabatic PESs W$_{ii}$, while Figures (b)-(d) show the diabatic couplings W$_{12}$, W$_{23}$ and W$_{34}$, respectively, over the $\theta$–$\phi$ space at $\rho = 4$ Bohr.}
		\label{fig:diaPES_rho4}
	\end{figure}

	\noindent
	Once the \textit{ab initio} adiabatic PESs and NACTs are in hand, we explore the existence of SHS for the low-lying four singlet electronic states O$_3$. While constructing the diabatic PESs in hyperspherical coordinates, we first calculate the ADT angles ($\Theta_{ij}$) also known as mixing angles between the four electronic states by integrating the corresponding differential equations (see Section S3 of SM) over the 2D $\theta$–$\phi$ grid for fixed $\rho$ = 4 Bohr. With those calculated ADT angles, we construct the diabatic PES matrix (see Section S4 of SM). Figure \ref{fig:diaPES_rho4} depict diabatic PESs ($W_{ii}$s) as well as the coupling ($W_{ij}$s) for $\rho$ = 4 Bohr over the $\theta$–$\phi$ grid. It is evident from the presented profiles of diabatic potential matrix elements that the calculated diabatic PESs and the coupling elements over the $\theta$-$\phi$ space appear to be single-valued, continuous, smooth and symmetric. 

	\vspace{0.2cm}
	
	\newpage
	\noindent
	In this process of construction, we consider two distinct rectangular integration paths, denoted as \textbf{P1} and \textbf{P2}. In path \textbf{P1}, the differential equations are first solved along the $\theta$ grid with a positive increment from $\theta = 0^\circ$ to $90^\circ$ at a fixed $\phi = 0^\circ$. Subsequently, for each $\theta$ value, integration is carried out along the $\phi$ grid from $\phi = 0^\circ$ to $360^\circ$. In contrast, path \textbf{P2} begins by integrating along the $\phi$ grid from $\phi = 0^\circ$ to $360^\circ$ at fixed $\theta = 0^\circ$ and then, continues along the $\theta$ grid for each $\phi$ value from $\theta = 0^\circ$ to $90^\circ$. These integration schemes are illustrated in Figure S1 of the section S5 in SM. These two paths lead to different solutions of the ADT equations, resulting in path-dependent diabatic PES matrices. However, such matrices are related by another orthogonal matrix, \cite{Mukherjee2019_IRPC_BBO,Dutta2020_PCCP_perspective,Mantu2023} ensuring that all physical observables derived from them remain invariant \cite{Mukherjee2013_JPCA_117}. Since path \textbf{P1} facilitates the identification of closed contours, it is particularly convenient for locating CIs. On the other hand, to achieve greater numerical accuracy and to exploit the underlying symmetry of the system, path \textbf{P2} is employed for constructing the diabatic PES matrices only upto $\phi$ = 120$^\circ$ for each grid of $\rho$-$\theta$ space and thereby, can reduce the numerical effort (computational cost) by one third.

	\vspace{0.2cm}

	\section*{CONCLUSION}
	The ``accuracy'' of the molecular parameters as well as adiabatic PESs is one of the most important aspect for constructing the diabatic Hamiltonian. On the other hand, the location of CIs as well as the ``accuracy'' of associated NACTs are equally important, if not more, to determine the lifetime of the species from the excited electronic states. Moreover, the kinetic process for this chemical reaction will depend on the ``accurate'' adiabatic PESs as well as NACTs for different/between electronic states.
	
	\vspace{0.2cm}
	
	\noindent
	 While exploring the details on the process of constructing diabatic Hamiltonian, it is important to follow the mechanism of the chemical reaction from the BBO perspective of structural (electronic) aspects of O$_3$: (a) location and nature of symmetry driven and accidental Jahn-Teller CIs to understand radiationless energy transfer; (b) the ``accuracy'' of non-adiabatic coupling elements through quantization of NACTs and their topological implications on the reaction attributes; (c) existence of SHS by examining the validaty of Curl Condition for low energy collision process, which varies from one reactive process to another; (d) construction of multi-state diabatic Hamiltonian matrix, which is, at present, four (N = 4) for O$_3$ case. Once such diabatic Hamiltonians are constructed involving even with singularities of NACTs, reaction attributes will be calculated and compared with experimental data as well as other theoretical results.
	
	\vspace{0.2cm}
	
	\noindent
	In this work, we present lowest four singlet adiabatic PESs ($\tilde{X}^1A'$, $1^1A''$, $1^1A'$ and $2^1A''$) of ozone as functions of Jacobi coordinates \{$r$, $R$\} for $\gamma$ = 0$^\circ$, 45$^\circ$ and 90$^\circ$, computed using 7S-SA-MCSCF method followed by ic-MRCI(Q) with a full valence CAS(18e,12o) and AVQZ basis set, where the geometries and energetics at \textit{C}$_{2v}$ and \textit{D}$_{3h}$ configurations as well as the O$_2$ + O channel and high energy O + O + O asymptote are depicted with emphasis. Convergence of geometrical parameters, energetics and locations of minimum energy separation are also described with increasing basis set size. The NACTs are calculated employing 7S-SA-MCSCF method followed by CP-MCSCF approach along carefully chosen circular contours. A symmetry-driven CI involving the 2-3 electronic states at the \textit{D}$_{3h}$ geometry is characterized, along with the accidental \textit{C}$_{2v}$ and \textit{C}$_{s}$ ones between all the adjacent states (1-2, 2-3 and 3-4). The quantization of NACTs along the circular contours confirms the presence of CIs. Finally, for the purpose of future reactive scattering calculations on O$_3$ system and its isotopic variants, the adiabatic PESs [ic-MRCI(Q)], NACTs (CP-MCSCF), and the diabatic PES matrix are constructed using the same active space and basis set over the $\theta$–$\phi$ space at a specific $\rho$ (= 4 Bohr), which is a 2D cross section of the global 3D surface ($\rho$, $\theta$, $\phi$). Although the current model primarily includes low-lying singlet states, we have also investigated the effect of higher spin multiplicities by incorporating three triplet and three quintet states along with three singlets in the SA-MCSCF calculations. However, this extended state-averaging does not yield any improvement in the accuracy of the adiabatic PES compared to the singlet-only approach.
	
	\vspace{0.2cm}
	
	\noindent
	In our future study, we aim to explicitly incorporate SO interactions, which are crucial for capturing intersystem crossing effects and achieving a more complete representation of the nonadiabatic dynamics of ozone. Overall, the methodologies and insights presented in this study lay a basic foundation for the construction of global diabatic PES matrix for the application of BBO based scattering dynamics in ozone.

	\section*{SUPPLEMENTARY MATERIAL}
	See the SM for explicit expressions of state specific ADT equations, diabatic potential matrix elements, path dependence of diabatic PES matrices, loation of CIs, curl condition and additional figures on adiabatic PESs.
	
	\section*{ACKNOWLEDGMENTS}
	Both Avik Guchait and Gourhari Jana contributed equally. Avik Guchait acknowledges CSIR (File No. 09/0080(22904)/2025-EMR-I) for the research	fellowship. Koushik Naskar thanks to IACS. Satyam Ravi thanks VIT Bhopal for the seed grant and SERB (File No. SRG/2023/001624) for research funding. Gourhari Jana and Satrajit Adhikari acknowledge the SERB (Project No. CRG/2023/000611) for
	research funding. Satrajit Adhikari also acknowledges IACS for the supercomputing facility.
	
	\section*{AUTHOR DECLARATIONS}
	\subsection*{Conflict of Interest}
	The authors have no conflicts to disclose
	
	\section*{DATA AVAILABILITY}
	The data that supports the findings of this study are available within the article and its supplementary material.
	
	\clearpage
	
	\renewcommand{\thetable}{S\arabic{table}}
	\renewcommand{\thefigure}{S\arabic{figure}}
	\renewcommand{\thesection}{S\arabic{section}}
	\renewcommand{\theequation}{S\arabic{equation}}
	\renewcommand{\thepage}{S\arabic{page}}
	
	\setcounter{table}{0}
	\setcounter{figure}{0}
	\setcounter{equation}{0}
	\setcounter{section}{0}
	\setcounter{page}{1}
	
	\begin{center}
		\textbf{{\Large Supplementary Material}}
	\end{center}
	
	\section{Progresses on Adiabatic Potential Energy Surfaces of Ozone}
	\noindent
	In the early 1980s, the first computational work~\cite{wilson1981theoretical} on ozone was focused on its electronic structure and potential energy surfaces (PES). An analytical PES functions\cite{varandas1982} of ozone was derived by Varandas \textit{et. al.} for the ground state by minimizing differences between observed and variationally calculated vibrational spectra, yielding accurate equilibrium force fields on which quasiclassical trajectory (QCT) calculation\cite{varandas1982dynamics} had beed carried out. Later, using the Double Many-Body Expansion (DMBE), a simple and globally valid analytical PES for O$_3$ was developed\cite{varandas1988} by fitting with experimental scattering and dissociation data. Rama Krishna \textit{et al.}~\cite{krishna1987theoretical} explored the dissociation energy of ozone using methods like Configuration Interaction and multi-configurational self-consistent field (MCSCF). This investigation highlighted the critical role of electronic correlation and paved the way for more refined models on the complex electronic and nuclear interactions of ozone. In 1991, Xantheas \textit{et al.}\cite{xantheas1991potential} carried out full valence space MCSCF calculations on O$_3$ and revealed its critical features in ground-state adiabatic PES, including open and ring minima, and symmetry driven intersection seams. Yamashita \textit{et al.}~\cite{yamashita1992new} designed the initial wave packets for UV photodissociation studies rather than exploring dissociation in the ground electronic state. Subsequently, in trajectory calculations, the PES was empirically modified to remove the relatively high dissociation barrier. In 1997, Woywod \textit{et al.}\cite{Zhu2006} constructed the three low-lying 1\,$^1$A$^\prime$, 1\,$^1$A$^{\prime\prime}$ and 2\,$^1$A$^{\prime\prime}$ states of ozone and the corresponding transition-dipole-moment surfaces using CASSCF and CASPT2 methods, which are relevant for calculating Chappuis band. Their work also highlighted the importance of the $^1$A$^{\prime\prime}$ - 2\,$^1$A$^{\prime\prime}$ conical intersection in ozone photodissociation dynamics. Other studies, including those by Banichevich \textit{et al.} \cite{banichevich1993potential} and Borowski \textit{et al.}~\cite{borowski1995theoretical}, provided partial PESs, while Xie \textit{et al.}~\cite{xie2000accurate} presented an accurate PES for near-equilibrium geometries. Tyuterev \textit{et al.} \cite{tyuterev1999} developed an effective ground-state potential energy function for ozone in the vicinity of the \textit{C}$_{2v}$ equilibrium configuration by least-squares fitting of extensive high-resolution experimental spectroscopic data. The potential was subsequently refined using an exact kinetic energy operator\cite{tyuterev2000variational}, achieving near-spectroscopic accuracy for both vibrational and rotational energy levels and exhibiting physically correct long-range behavior. Later on, Banichevich \textit{et al.}~\cite{banichevich1993potential} explored PESs for ground and excited states of ozone using MRCI methods to study dissociation pathways and photodissociation processes. Based on this study, Atchity and Ruedenberg (1997)\cite{atchity1997global} introduced global PESs for the lowest two $^1A^\prime$ states, offering critical insights into the dissociation dynamics of ozone. Though the electronic structure of low-lying states of ozone were investigated by numerous researchers, only a couple of global PES for ozone were attempted using \textit{ab initio} methods over limited region\cite{yamashita1992new,siebert2001spectroscopy}.
	
	\section{Strategy of ab initio calculations for adiabatic PESs and NACTs of Ozone (O$_3$)}
	
	\noindent
	Since we intend to develop multi-surface global diabatic Hamiltonian to explore the effect of electron-nuclear coupling through conical intersections between the surfaces on reactive cross-section and rate constant, it appears to be a necessity to calculate adiabatic potential energy surfaces as well as non-adiabatic coupling terms at the highest possible level of ``accuracy'' in \textit{ab initio} calculation with the following motivation: 
	
	\begin{itemize}
		\item All molecular as well as geometric parameters are in closest agreement with the latest experimental as well as other \textit{ab initio} calculation(s).
		
		\item Locations and energetics of symmetry-driven as well as accidental degeneracies are reproduced within numerical ``accuracy'', if such informations are available by other theoretical calculations.
		
		\item The quantization of \textit{ab initio} calculated NACTs encapsulating $n\ (=1,2,...)$ number of conical intersections is verified by the magnitude, $n\pi$ (Jahn-Teller), $n.2\pi$ (Renner-Teller) and $0$ (Psedo Jahn-Teller)
		
		\item The existence of sub-Hilbert space through the equality of mathematical ($Z_{\theta\phi}^{ij}$) and ADT ($C_{\theta\phi}^{ij}$) curl using \textit{ab initio} calculated NACTs over the entire configuration space.
	\end{itemize}
	
	\noindent
	Once the \textit{ab initio} adiabatic PESs and NACTs are available over the entire configuration space and those quantities satisfy the above conditions within numerical ``accuracy'', we construct the global diabatic Hamiltonian for scattering calculation to obtain cross-section and rate constants. As for examples, we have constructed so far the following diabatic surfaces of triatomic species and perform reaction dynamics:
	
	\begin{itemize}
		\item F+H$_2$: CAS(9e,10o), basis set: aug-cc-pVQZ, method: three state MRCI+Q
		
		\vspace{-0.2cm}
		
		Ref: \textit{J. Chem. Phys. \textbf{153}, 174301, 2021}; \textit{J. Phys. Chem. A \textbf{128} (8) 1438–1456, 2024}
		
		\item H$_3^+$: CAS(2e,10o), basis set: cc-pV5Z, method: three state MRCI 
		
		\vspace{-0.2cm}
		
		Ref: \textit{J. Chem. Phys. \textbf{141}, 204306, 2014}; \textit{J. Chem. Phys. \textbf{147} (7),  074105, 2017}; \textit{Phys. Chem. A \textbf{125}, (3) 731–74, 2021}
		
		\item HeH$_2^+$: CAS(3e,8o), basis set: aug-cc-pVQZ, method: four state ic-MRCI 
		
		\vspace{-0.2cm}
		
		Ref:\textit{J. Phys. Chem. A \textbf{127}, 3832, 2023}; \textit{J. Phys. Chem. A \textbf{128} (8), 1438–1456, 2024}
		
		\vspace{-0.2cm}
		
		\item H$_3$: CAS(3e,16o), basis set: aug-cc-pVQZ, method: three state MRCI
		
		\vspace{-0.2cm}
		
		Ref:\textit{J. Phys. Chem. A \textbf{129}, 6312, 2025}
	\end{itemize}
	
	\noindent
	The associated diabatic Hamiltonians [in hyperspherical coordinates ($\rho$, $\theta$ and $\phi$) inter-convertible to Jacobi coordinates ($r, R$ and $\gamma$)] are published as subroutine (see the references above). In all the cases, the fitting of the diabatic matrix elements are avoided using general analytic functions rather locally fitted numerically for all required geometries in the dynamical calculation to attain higher accuracy. Since the fitting functions for ADT angles (or diabatic couplings) are not well known or even not an established procedure yet, we calculate astronomically large number of \textit{ab initio} points to get accurate local fitting  in all the above cases. Moreover, calculated cross section and rate constants using those diabatic surfaces for all the above systems show closest agreement with experimentally ones in comparison with results obtained from other existing diabatic PESs (see the above references on scattering calculations). Actually, the inclusion of NACTs in the construction of multi-state diabatic Hamiltonian is the single most important quantity after the ``accuracy'' of the adiabatic PESs to reproduce experimental data through scattering calculation.
	
	\vspace{0.2cm}
	
	\noindent
	Like the other systems (F+H$_2$, H$_3^+$, HeH$_2^+$ and H$_3$), in case of ozone (O$_3$), we are supposed to carry out nearly 1.5 million \textit{ab initio} calculations at the ic-MRCI(Q) level for the low-lying four singlet adiabatic PESs and nearly two million \textit{ab initio} points for NACT calculations as functions of hyperspherical coordiates ($2\le\rho\le20$ Bohr, $0\le\theta\le\pi/2$, $0\le\phi\le2\pi$). Indeed, we get the advantage of inherent symmetry of the species (O$_3$) that reduces the actual number of \textit{ab initio} calculations. The other important aspect is the choice of the CAS(18e,12o) rather than the CAS(24e,15o). Although the latter [CAS(24e,15o)] can improves the dissociation energy of B1 [O$_3$ $\rightarrow$ O$_2$ + O] process, but it overestimates the dissociation energy of B2 process [O$_2$ $\rightarrow$ 2O]. On the other hand, the calculation of the global adiabatic and NACTs [followed by the construction of diabatic PES matrix] with the choice of CAS(24e,15o) at the ic-MRCI(Q) level could be virtually impossible due to the computational cost (see Table S1). Therefore, the following choice on methods and CAS/basis for calculating adiabatic PESs and NACTs at each \textit{ab initio} point:
	
	\begin{itemize}
		\item \textbf{Adiabatic PESs:} Using CAS(18e,12o) and aug-cc-pVQZ basis, State Average MCSCF including seven low-lying singlet states for variational convergence followed by ic-MRCI incorporating Davidson correction to obtain the lowest four global adiabatic PESs.
		
		\item \textbf{NACTs:} Using CAS(18e,12o) and aug-cc-pVQZ basis, State Average MCSCF including seven low-lying singlet states for variational convergence followed by CP-MCSCF to calculate fifty four (54) nuclear gradients between the lowest four adiabatic PESs.
	\end{itemize}
	
	\noindent
	is the optimal one for reaching reasonable ``accuracy'' of the molecular parameters as well as for different processes (B1, B2, ..) and the coupling among the electronic states. Since our purpose is to construct the four state diabatic PES matrix and thereafter, to calculate cross section/rate constant for comparison with experimental results, the present level of calculation is itself a humongous task that could be recognized. Incidently, for the entire configuration space ($2\le\rho\le20$ Bohr, $0\le\theta\le\pi/2$, $0\le\phi\le2\pi$), calculations for the four (4) adiabatic PESs and six (6) NACTs at the 7S-SA-MCSCF and CP-MCSCF level, respectively, are already performed, but the ic-MRCI(Q) calculation for those adiabatic PESs are going on.
	
	\begin{table}[h!]
		\caption{Average computational time (in hours) and scratch memory usage (in GB) at the ic-MRCI(Q) level in the vicinity of different geometries of O$_3$ using CAS(18e,12o) and CAS(24e,15o) with increasing basis set size (AVQZ $\rightarrow$ AV6Z).}
		\label{tab:comp_cost}
		\renewcommand{\arraystretch}{1.0}
		\begin{center}
			\begin{tabular}{lcccccc}
				\toprule
				\textbf{Geometry}
				& \multicolumn{3}{c}{\textbf{CAS(18e,12o)}} 
				& \multicolumn{3}{c}{\textbf{CAS(24e,15o)}} \\
				\cmidrule(lr){2-4} \cmidrule(lr){5-7}
				
				& \textbf{AVQZ} & \textbf{AV5Z} & \textbf{AV6Z} 
				& \textbf{AVQZ} & \textbf{AV5Z} & \textbf{AV6Z} \\
				\midrule
				
				& \multicolumn{6}{c}{\textit{\textbf{Time (hours)}}} \\
				C$_{2v}$ minimum      & 10.8 & 19.8 & 37.0 & 38.2 & 80.8 & 118.9 \\
				D$_{3h}$ minimum      &  8.2 & 11.3 & 38.9 & 41.2 & 80.4 & 115.2 \\
				O$_2$ + O channel     & 10.4 & 13.0 & 30.9 & 21.4 & 38.4 & 133.0 \\
				O + O + O asymptote   &  8.3 & 16.7 & 20.2 & 18.9 & 36.2 & 106.1 \\
				
				\addlinespace
				& \multicolumn{6}{c}{\textit{\textbf{Memory (GB)}}} \\
				C$_{2v}$ minimum      & 9.2 & 18.1 & 35.6 & 29.9 & 55.0 &	95.9 \\
				D$_{3h}$ minimum      & 8.6 & 18.1 & 35.6 & 29.9 & 55.0 &	95.9 \\
				O$_2$ + O channel     & 8.6 &	18.0 & 35.4	& 29.8 & 55.0 & 95.8 \\
				O + O + O asymptote   & 8.5 &	18.1 & 35.4	& 29.8 & 55.0 &	95.7 \\
				
				\bottomrule
			\end{tabular}
		\end{center}
	\end{table}
	
	\newpage
	\section{Adiabatic to Diabatic Transformation (ADT) equations for four (4) electronic state Sub-Hilbert space}
	\noindent
	In case of four (4) electronic state sub-Hilbert space, six (6) elementary rotation matrices can be multiplied in 6! ways to obtain the model ADT matrix where numerical solutions of any set of ADT equations originating from different permutation of elementary rotation matrix appear to be same.~\cite{ADT_JCTC_16} In 
	the present work, the following permutation is adapted:
	
	\begin{eqnarray*}
		\textbf{A} &=& \textbf{A}^{12}(\Theta_{12}).\textbf{A}^{13}(\Theta_{13}).\textbf{A}^{23}(\Theta_{23}).\textbf{A}^{14}(\Theta_{14})
		.\textbf{A}^{24}(\Theta_{24}).\textbf{A}^{34}(\Theta_{34}),
		\label{Eq:ADTmat}
	\end{eqnarray*}
	where $\Theta$s are the ADT angles. The ADT equations are as follows:
	
	\begin{small}
		\begin{subequations}
			\begin{eqnarray}
			\vec{\nabla}\Theta_{12}&=& -\vec{\tau}_{12}-\sin\Theta_{12}\tan\Theta_{13}\vec{\tau}_{13}-\cos\Theta_{12}\tan\Theta_{13}
			\vec{\tau}_{23}-\sin\Theta_{12}\sec\Theta_{13}\tan\Theta_{14}\vec{\tau}_{14}\nonumber\\
			&&-\cos\Theta_{12}\sec\Theta_{13}\tan\Theta_{14}
			\vec{\tau}_{24}\\
			\vec{\nabla}\Theta_{13}&=& -\cos\Theta_{12}\vec{\tau}_{13}+\sin\Theta_{12}\vec{\tau}_{23}-\cos\Theta_{12}\sin\Theta_{13}
			\tan\Theta_{14}\vec{\tau}_{14}+\sin\Theta_{12}\sin\Theta_{13}\tan\Theta_{14}\vec{\tau}_{24}\nonumber\\
			&&-\cos\Theta_{13}\tan\Theta_{14}\vec{\tau}_{34}\\
			\vec{\nabla}\Theta_{23}&=&-\cos\Theta_{13}[\vec{\tau}_{13}\sin\Theta_{12}\sec^2\Theta_{13}+\cos\Theta_{23}\sec\Theta_{14}(\vec{\tau}_{34}-\vec{\tau}_{24}\sin\Theta_{12}\tan\Theta_{13})\tan\Theta_{24}\nonumber \\
			&&+\sin\Theta_{12}\sec\Theta_{13}\tan\Theta_{13}\vec{\tau}_{14}\tan\Theta_{14}+\sin\Theta_{23}\vec{\tau}_{14}\sec\Theta_{14}\tan\Theta_{24}
			\nonumber\\
			&&+\cos\Theta_{12}\lbrace\vec{\tau}_{23}\sec^2\Theta_{13}+\tan\Theta_{13}\cos\Theta_{23}\vec{\tau}_{14}\sec\Theta_{14}
			\tan\Theta_{24}+\sec\Theta_{13}(\tan\Theta_{13}\vec{\tau}_{24}
			\tan\Theta_{14}\nonumber \\
			&&+\sin\Theta_{23}\vec{\tau}_{24}\sec\Theta_{14}
			\tan\Theta_{24})\rbrace]\\
			\vec{\nabla}\Theta_{14}&=&-\cos\Theta_{12}\cos\Theta_{13}\vec{\tau}_{14}+\sin\Theta_{12}\cos\Theta_{13}\vec{\tau}_{24}
			+\sin\Theta_{13}\vec{\tau}_{34}\\
			\vec{\nabla}\Theta_{24}&=&\sin\Theta_{23}(-\sin\Theta_{12}\sin\Theta_{13}\vec{\tau}_{24}\sec\Theta_{14}+
			\cos\Theta_{13}\vec{\tau}_{34}\sec\Theta_{14})-
			\sin\Theta_{12}\cos\Theta_{23}\vec{\tau}_{14}\sec\Theta_{14}\nonumber \\
			&&+\cos\Theta_{12}(\sin\Theta_{13}
			\sin\Theta_{23}\vec{\tau}_{14}\sec\Theta_{14}-\cos\Theta_{23}\vec{\tau}_{24}\sec\Theta_{14}) \\
			\vec{\nabla}\Theta_{34}&=&\sin\Theta_{12}\lbrace-\sin\Theta_{23}\vec{\tau}_{14}\sec\Theta_{14}\sec\Theta_{24}+\sin\Theta_{13}\cos\Theta_{23}\vec{\tau}_{24}\sec\Theta_{14}\sec\Theta_{24}\rbrace\nonumber\\
			&&+\cos\Theta_{12}[-\sec\Theta_{24}\lbrace\sin\Theta_{13}\cos\Theta_{23}\vec{\tau}_{14}\sec\Theta_{14}+\sin\Theta_{23}
			\vec{\tau}_{24}\sec\Theta_{14}\rbrace]\nonumber\\
			&&-\cos\Theta_{13}\cos\Theta_{23}\vec{\tau}_{34}\sec\Theta_{14}\sec\Theta_{24}
			\end{eqnarray}
		\end{subequations}
	\end{small}

	\section{Diabatic potential matrix elements for four (4) electronic state Sub-Hilbert space}
	\begin{small}
		\begin{subequations}
			\begin{eqnarray}
			W_{11} &=&
			u_{1} \cos^{2}{(\Theta_{12} )} \cos^{2}{(\Theta_{13} )} \cos^{2}{(\Theta_{14} )} + u_{2} \sin^{2}{(\Theta_{12} )} \cos^{2}{(\Theta_{13} )} \cos^{2}{(\Theta_{14} )} \nonumber \\
			&&+ u_{3} \sin^{2}{(\Theta_{13} )} \cos^{2}{(\Theta_{14} )} + u_{4} \sin^{2}{(\Theta_{14} )}\\
			\nonumber\\
			W_{22} &=&
			u_{1} [\{\sin{(\Theta_{12} )} \cos{(\Theta_{23} )} - \sin{(\Theta_{13} )} \sin{(\Theta_{23} )} \cos{(\Theta_{12} )}\} \cos{(\Theta_{24} )} - \sin{(\Theta_{14} )} \sin{(\Theta_{24} )} \cos{(\Theta_{12} )} \cos{(\Theta_{13} )}]^{2} \nonumber\\
			&&+ u_{2} [\{\sin{(\Theta_{12} )} \sin{(\Theta_{13} )} \sin{(\Theta_{23} )} + \cos{(\Theta_{12} )} \cos{(\Theta_{23} )}\} \cos{(\Theta_{24} )} + \sin{(\Theta_{12} )} \sin{(\Theta_{14} )} \sin{(\Theta_{24} )} \cos{(\Theta_{13} )}]^{2}\nonumber \\
			&&+ u_{3} \{\sin{(\Theta_{13} )} \sin{(\Theta_{14} )} \sin{(\Theta_{24} )} - \sin{(\Theta_{23} )} \cos{(\Theta_{13} )} \cos{(\Theta_{24} )}\}^{2} \nonumber\\
			&&+ u_{4} \sin^{2}{(\Theta_{24} )} \cos^{2}{(\Theta_{14} )}\\
			\nonumber\\
			W_{33} &=&
			u_{1} \big[- [\{\sin{(\Theta_{12} )} \cos{(\Theta_{23} )} - \sin{(\Theta_{13} )} \sin{(\Theta_{23} )} \cos{(\Theta_{12} )}\} \sin{(\Theta_{24} )}\nonumber \\
			&& + \sin{(\Theta_{14} )} \cos{(\Theta_{12} )} \cos{(\Theta_{13} )} \cos{(\Theta_{24} )}] \sin{(\Theta_{34} )}  \nonumber\\
			&&+ (\sin{(\Theta_{12} )} \sin{(\Theta_{23} )} + \sin{(\Theta_{13} )} \cos{(\Theta_{12} )} \cos{(\Theta_{23} )}) \cos{(\Theta_{34} )}\big]^{2} \nonumber \\
			&&+ u_{2} \big[- [\{\sin{(\Theta_{12} )} \sin{(\Theta_{13} )} \sin{(\Theta_{23} )} + \cos{(\Theta_{12} )} \cos{(\Theta_{23} )}\} \sin{(\Theta_{24} )} \nonumber\\
			&&- \sin{(\Theta_{12} )} \sin{(\Theta_{14} )} \cos{(\Theta_{13} )} \cos{(\Theta_{24} )}] \sin{(\Theta_{34} )} \nonumber \\
			&&+ \{- \sin{(\Theta_{12} )} \sin{(\Theta_{13} )} \cos{(\Theta_{23} )} + \sin{(\Theta_{23} )} \cos{(\Theta_{12} )}\} \cos{(\Theta_{34} )}\big]^{2} \nonumber \\
			&&+ u_{3} [- \{- \sin{(\Theta_{13} )} \sin{(\Theta_{14} )} \cos{(\Theta_{24} )} - \sin{(\Theta_{23} )} \sin{(\Theta_{24} )} \cos{(\Theta_{13} )}\}\sin{(\Theta_{34} )}  \nonumber\\
			&&+ \cos{(\Theta_{13} )} \cos{(\Theta_{23} )} \cos{(\Theta_{34} )}]^{2} + u_{4} \sin^{2}{(\Theta_{34} )} \cos^{2}{(\Theta_{14} )} \cos^{2}{(\Theta_{24} )}\\
			\nonumber\\
			W_{44} &=&
			u_{1} \big[[\{\sin{(\Theta_{12} )} \cos{(\Theta_{23} )} - \sin{(\Theta_{13} )} \sin{(\Theta_{23} )} \cos{(\Theta_{12} )}\} \sin{(\Theta_{24} )}\nonumber\\
			&&+ \sin{(\Theta_{14} )} \cos{(\Theta_{12} )} \cos{(\Theta_{13} )} \cos{(\Theta_{24} )}] \cos{(\Theta_{34} )}\nonumber \\
			&&+ (\sin{(\Theta_{12} )} \sin{(\Theta_{23} )} + \sin{(\Theta_{13} )} \cos{(\Theta_{12} )} \cos{(\Theta_{23} )}) \sin{(\Theta_{34} )}\big]^{2} + u_{2} \big[[\{\sin{(\Theta_{12} )} \sin{(\Theta_{13} )} \sin{(\Theta_{23} )} \nonumber\\
			&&+ \cos{(\Theta_{12} )} \cos{(\Theta_{23} )}\} \sin{(\Theta_{24} )} - \sin{(\Theta_{12} )} \sin{(\Theta_{14} )} \cos{(\Theta_{13} )} \cos{(\Theta_{24} )}] \cos{(\Theta_{34} )} \nonumber\\
			&&+ (- \sin{(\Theta_{12} )} \sin{(\Theta_{13} )} \cos{(\Theta_{23} )} + \sin{(\Theta_{23} )} \cos{(\Theta_{12} )}) \sin{(\Theta_{34} )}\big]^{2} + u_{3} [\{- \sin{(\Theta_{13} )} \sin{(\Theta_{14} )} \cos{(\Theta_{24} )}\nonumber \\
			&&- \sin{(\Theta_{23} )} \sin{(\Theta_{24} )} \cos{(\Theta_{13} )}\} \cos{(\Theta_{34} )} + \sin{(\Theta_{34} )} \cos{(\Theta_{13} )} \cos{(\Theta_{23} )}]^{2}\nonumber \\
			&&+ u_{4} \cos^{2}{(\Theta_{14} )} \cos^{2}{(\Theta_{24} )} \cos^{2}{(\Theta_{34} )}
			\end{eqnarray}
		\end{subequations}
	\end{small}
	
	\begin{small}
		\begin{subequations}
			\begin{eqnarray}
			W_{12} &=&
			u_{1} [\{\sin{(\Theta_{12} )} \cos{(\Theta_{23} )} - \sin{(\Theta_{13} )} \sin{(\Theta_{23} )} \cos{(\Theta_{12} )}\} \cos{(\Theta_{24} )}\nonumber\\
			&&-\sin{(\Theta_{14} )} \sin{(\Theta_{24} )} \cos{(\Theta_{12} )} \cos{(\Theta_{13} )}] \cos{(\Theta_{12} )} \cos{(\Theta_{13} )} \cos{(\Theta_{14} )}\nonumber\\
			&& - u_{2} [\{\sin{(\Theta_{12} )} \sin{(\Theta_{13} )} \sin{(\Theta_{23} )} + \cos{(\Theta_{12} )} \cos{(\Theta_{23} )}\} \cos{(\Theta_{24} )}\nonumber\\
			&& + \sin{(\Theta_{12} )} \sin{(\Theta_{14} )} \sin{(\Theta_{24} )} \cos{(\Theta_{13} )}] \sin{(\Theta_{12} )} \cos{(\Theta_{13} )} \cos{(\Theta_{14} )}
			- u_{3} \{\sin{(\Theta_{13} )} \sin{(\Theta_{14} )} \sin{(\Theta_{24} )}\nonumber \\
			&&- \sin{(\Theta_{23} )} \cos{(\Theta_{13} )} \cos{(\Theta_{24} )}\} \sin{(\Theta_{13} )} \cos{(\Theta_{14} )} + u_{4} \sin{(\Theta_{14} )} \sin{(\Theta_{24} )} \cos{(\Theta_{14} )}\\
			\nonumber\\
			W_{13} &=&
			u_{1} \big[- [\{\sin{(\Theta_{12} )} \cos{(\Theta_{23} )} - \sin{(\Theta_{13} )} \sin{(\Theta_{23} )} \cos{(\Theta_{12} )}\} \sin{(\Theta_{24} )}\nonumber\\
			&&+ \sin{(\Theta_{14} )} \cos{(\Theta_{12} )} \cos{(\Theta_{13} )} \cos{(\Theta_{24} )}] \sin{(\Theta_{34} )}\nonumber \\
			&&+ \{\sin{(\Theta_{12} )} \sin{(\Theta_{23} )} + \sin{(\Theta_{13} )} \cos{(\Theta_{12} )} \cos{(\Theta_{23} )}\} \cos{(\Theta_{34} )}\big] \cos{(\Theta_{12} )} \cos{(\Theta_{13} )} \cos{(\Theta_{14} )}\nonumber \\
			&&- u_{2} \big[- [\{\sin{(\Theta_{12} )} \sin{(\Theta_{13} )} \sin{(\Theta_{23} )} + \cos{(\Theta_{12} )} \cos{(\Theta_{23} )}\} \sin{(\Theta_{24} )}\nonumber \\
			&&- \sin{(\Theta_{12} )} \sin{(\Theta_{14} )} \cos{(\Theta_{13} )} \cos{(\Theta_{24} )}] \sin{(\Theta_{34} )} + \{- \sin{(\Theta_{12} )} \sin{(\Theta_{13} )} \cos{(\Theta_{23} )}\nonumber\\
			&& + \sin{(\Theta_{23} )} \cos{(\Theta_{12} )}\} \cos{(\Theta_{34} )}\big] \sin{(\Theta_{12} )} \cos{(\Theta_{13} )} \cos{(\Theta_{14} )}\nonumber \\
			&&- u_{3} [- \{- \sin{(\Theta_{13} )} \sin{(\Theta_{14} )} \cos{(\Theta_{24} )} - \sin{(\Theta_{23} )} \sin{(\Theta_{24} )} \cos{(\Theta_{13} )}\} \sin{(\Theta_{34} )}\nonumber \\
			&&+ \cos{(\Theta_{13} )} \cos{(\Theta_{23} )} \cos{(\Theta_{34} )}] \sin{(\Theta_{13} )} \cos{(\Theta_{14} )} \nonumber\\
			&&+ u_{4} \sin{(\Theta_{14} )} \sin{(\Theta_{34} )} \cos{(\Theta_{14} )} \cos{(\Theta_{24} )}
			\end{eqnarray}
		\end{subequations}
	\end{small}
	
	\begin{small}
		\begin{subequations}
			\begin{eqnarray}
			W_{23} &=&
			u_{1} \big[- [\{\sin{(\Theta_{12} )} \cos{(\Theta_{23} )} - \sin{(\Theta_{13} )} \sin{(\Theta_{23} )} \cos{(\Theta_{12} )}\} \sin{(\Theta_{24} )}
			\nonumber\\
			&&+ \sin{(\Theta_{14} )} \cos{(\Theta_{12} )} \cos{(\Theta_{13} )} \cos{(\Theta_{24} )}] \sin{(\Theta_{34} )}\nonumber \\
			&&+ \{\sin{(\Theta_{12} )} \sin{(\Theta_{23} )} + \sin{(\Theta_{13} )} \cos{(\Theta_{12} )} \cos{(\Theta_{23} )}\} \cos{(\Theta_{34} )}\big] [\{\sin{(\Theta_{12} )} \cos{(\Theta_{23} )}\nonumber \\
			&&- \sin{(\Theta_{13} )} \sin{(\Theta_{23} )} \cos{(\Theta_{12} )}\} \cos{(\Theta_{24} )} - \sin{(\Theta_{14} )} \sin{(\Theta_{24} )} \cos{(\Theta_{12} )} \cos{(\Theta_{13} )}]\nonumber\\
			&& + u_{2} \big[- [\{\sin{(\Theta_{12} )} \sin{(\Theta_{13} )} \sin{(\Theta_{23} )} + \cos{(\Theta_{12} )} \cos{(\Theta_{23} )}\} \sin{(\Theta_{24} )}\nonumber \\
			&&- \sin{(\Theta_{12} )} \sin{(\Theta_{14} )} \cos{(\Theta_{13} )} \cos{(\Theta_{24} )}] \sin{(\Theta_{34} )} + (- \sin{(\Theta_{12} )} \sin{(\Theta_{13} )} \cos{(\Theta_{23} )}\nonumber \\
			&&+ \sin{(\Theta_{23} )} \cos{(\Theta_{12} )}) \cos{(\Theta_{34} )}\big] [\{\sin{(\Theta_{12} )} \sin{(\Theta_{13} )} \sin{(\Theta_{23} )} + \cos{(\Theta_{12} )} \cos{(\Theta_{23} )}\} \cos{(\Theta_{24} )} \nonumber\\
			&&+ \sin{(\Theta_{12} )} \sin{(\Theta_{14} )} \sin{(\Theta_{24} )} \cos{(\Theta_{13} )}]\nonumber\\
			&&+ u_{3} [- \{- \sin{(\Theta_{13} )} \sin{(\Theta_{14} )} \cos{(\Theta_{24} )} - \sin{(\Theta_{23} )} \sin{(\Theta_{24} )} \cos{(\Theta_{13} )}\} \sin{(\Theta_{34} )}\nonumber \\
			&&+ \cos{(\Theta_{13} )} \cos{(\Theta_{23} )} \cos{(\Theta_{34} )}] \{\sin{(\Theta_{13} )} \sin{(\Theta_{14} )} \sin{(\Theta_{24} )} - \sin{(\Theta_{23} )} \cos{(\Theta_{13} )} \cos{(\Theta_{24} )}\}\nonumber\\
			&& + u_{4} \sin{(\Theta_{24} )} \sin{(\Theta_{34} )} \cos^{2}{(\Theta_{14} )} \cos{(\Theta_{24} )}\\
			\nonumber\\
			W_{14} &=&
			u_{1} \big[[\{\sin{(\Theta_{12} )} \cos{(\Theta_{23} )} - \sin{(\Theta_{13} )} \sin{(\Theta_{23} )} \cos{(\Theta_{12} )}\} \sin{(\Theta_{24} )}\nonumber\\
			&& + \sin{(\Theta_{14} )} \cos{(\Theta_{12} )} \cos{(\Theta_{13} )} \cos{(\Theta_{24} )}] \cos{(\Theta_{34} )} + \{\sin{(\Theta_{12} )} \sin{(\Theta_{23} )} \nonumber\\
			&& + \sin{(\Theta_{13} )} \cos{(\Theta_{12} )} \cos{(\Theta_{23} )}\} \sin{(\Theta_{34} )}\big] \cos{(\Theta_{12} )} \cos{(\Theta_{13} )} \cos{(\Theta_{14} )}\nonumber\\
			&&  - u_{2} \big[[\{\sin{(\Theta_{12} )} \sin{(\Theta_{13} )} \sin{(\Theta_{23} )} + \cos{(\Theta_{12} )} \cos{(\Theta_{23} )}\} \sin{(\Theta_{24} )} \nonumber\\
			&& - \sin{(\Theta_{12} )} \sin{(\Theta_{14} )} \cos{(\Theta_{13} )} \cos{(\Theta_{24} )}] \cos{(\Theta_{34} )} + \{- \sin{(\Theta_{12} )} \sin{(\Theta_{13} )} \cos{(\Theta_{23} )}\nonumber\\
			&&   + \sin{(\Theta_{23} )} \cos{(\Theta_{12} )}\} \sin{(\Theta_{34} )}\big] \sin{(\Theta_{12} )} \cos{(\Theta_{13} )} \cos{(\Theta_{14} )} - u_{3} [\{- \sin{(\Theta_{13} )} \sin{(\Theta_{14} )} \cos{(\Theta_{24} )} \nonumber\\
			&&  - \sin{(\Theta_{23} )} \sin{(\Theta_{24} )} \cos{(\Theta_{13} )}\} \cos{(\Theta_{34} )} + \sin{(\Theta_{34} )} \cos{(\Theta_{13} )} \cos{(\Theta_{23} )}] \sin{(\Theta_{13} )} \cos{(\Theta_{14} )}\nonumber \\
			&&   - u_{4} \sin{(\Theta_{14} )} \cos{(\Theta_{14} )} \cos{(\Theta_{24} )} \cos{(\Theta_{34} )}
			\end{eqnarray}
		\end{subequations}
	\end{small}
	
	\begin{small}
		\begin{subequations}
			\begin{eqnarray}
			W_{24} &=&
			u_{1} \big[[\{\sin{(\Theta_{12} )} \cos{(\Theta_{23} )} - \sin{(\Theta_{13} )} \sin{(\Theta_{23} )} \cos{(\Theta_{12} )}\} \sin{(\Theta_{24} )}  \nonumber\\
			&&+ \sin{(\Theta_{14} )} \cos{(\Theta_{12} )} \cos{(\Theta_{13} )} \cos{(\Theta_{24} )}] \cos{(\Theta_{34} )} + \{\sin{(\Theta_{12} )} \sin{(\Theta_{23} )}  \nonumber\\
			&&+ \sin{(\Theta_{13} )} \cos{(\Theta_{12} )} \cos{(\Theta_{23} )}\} \sin{(\Theta_{34} )}\big] [\{\sin{(\Theta_{12} )} \cos{(\Theta_{23} )} - \sin{(\Theta_{13} )} \sin{(\Theta_{23} )} \cos{(\Theta_{12} )}\} \cos{(\Theta_{24} )} \nonumber \\
			&&- \sin{(\Theta_{14} )} \sin{(\Theta_{24} )} \cos{(\Theta_{12} )} \cos{(\Theta_{13} )}] + u_{2} \big[[\{\sin{(\Theta_{12} )} \sin{(\Theta_{13} )} \sin{(\Theta_{23} )} + \cos{(\Theta_{12} )} \cos{(\Theta_{23} )}\} \sin{(\Theta_{24} )} \nonumber \\
			&&- \sin{(\Theta_{12} )} \sin{(\Theta_{14} )} \cos{(\Theta_{13} )} \cos{(\Theta_{24} )}] \cos{(\Theta_{34} )} + \{- \sin{(\Theta_{12} )} \sin{(\Theta_{13} )} \cos{(\Theta_{23} )} \nonumber\\
			&& + \sin{(\Theta_{23} )} \cos{(\Theta_{12} )}\} \sin{(\Theta_{34} )}\big] [\{\sin{(\Theta_{12} )} \sin{(\Theta_{13} )} \sin{(\Theta_{23} )} + \cos{(\Theta_{12} )} \cos{(\Theta_{23} )}\} \cos{(\Theta_{24} )}  \nonumber\\
			&&+ \sin{(\Theta_{12} )} \sin{(\Theta_{14} )} \sin{(\Theta_{24} )} \cos{(\Theta_{13} )}] \nonumber\\
			&&+ u_{3} [\{- \sin{(\Theta_{13} )} \sin{(\Theta_{14} )} \cos{(\Theta_{24} )} - \sin{(\Theta_{23} )} \sin{(\Theta_{24} )} \cos{(\Theta_{13} )}\} \cos{(\Theta_{34} )} \nonumber \\
			&&+ \sin{(\Theta_{34} )} \cos{(\Theta_{13} )} \cos{(\Theta_{23} )}] \{\sin{(\Theta_{13} )} \sin{(\Theta_{14} )} \sin{(\Theta_{24} )}  \nonumber\\
			&&- \sin{(\Theta_{23} )} \cos{(\Theta_{13} )} \cos{(\Theta_{24} )}\} - u_{4} \sin{(\Theta_{24} )} \cos^{2}{(\Theta_{14} )} \cos{(\Theta_{24} )} \cos{(\Theta_{34} )}\\
			\nonumber\\
			W_{34} &=&
			u_{1} \big[- [\{\sin{(\Theta_{12} )} \cos{(\Theta_{23} )} - \sin{(\Theta_{13} )} \sin{(\Theta_{23} )} \cos{(\Theta_{12} )}\} \sin{(\Theta_{24} )} \nonumber\\
			&&+ \sin{(\Theta_{14} )} \cos{(\Theta_{12} )} \cos{(\Theta_{13} )} \cos{(\Theta_{24} )}] \sin{(\Theta_{34} )} \nonumber \\
			&& + \{\sin{(\Theta_{12} )} \sin{(\Theta_{23} )} + \sin{(\Theta_{13} )} \cos{(\Theta_{12} )} \cos{(\Theta_{23} )}\} \cos{(\Theta_{34} )}\big] \big[[\{\sin{(\Theta_{12} )} \cos{(\Theta_{23} )}  \nonumber\\
			&&- \sin{(\Theta_{13} )} \sin{(\Theta_{23} )} \cos{(\Theta_{12} )}\} \sin{(\Theta_{24} )} + \sin{(\Theta_{14} )} \cos{(\Theta_{12} )} \cos{(\Theta_{13} )} \cos{(\Theta_{24} )}] \cos{(\Theta_{34} )} \nonumber\\
			&& + \{\sin{(\Theta_{12} )} \sin{(\Theta_{23} )} + \sin{(\Theta_{13} )} \cos{(\Theta_{12} )} \cos{(\Theta_{23} )}\} \sin{(\Theta_{34} )}\big] + u_{2} \big[- [\{\sin{(\Theta_{12} )} \sin{(\Theta_{13} )} \sin{(\Theta_{23} )} \nonumber \\
			&&+ \cos{(\Theta_{12} )} \cos{(\Theta_{23} )}\} \sin{(\Theta_{24} )} - \sin{(\Theta_{12} )} \sin{(\Theta_{14} )} \cos{(\Theta_{13} )} \cos{(\Theta_{24} )}] \sin{(\Theta_{34} )} \nonumber \\
			&&+ \{- \sin{(\Theta_{12} )} \sin{(\Theta_{13} )} \cos{(\Theta_{23} )} + \sin{(\Theta_{23} )} \cos{(\Theta_{12} )}\} \cos{(\Theta_{34} )}\big] \big[[\{\sin{(\Theta_{12} )} \sin{(\Theta_{13} )} \sin{(\Theta_{23} )} \nonumber \\
			&&+ \cos{(\Theta_{12} )} \cos{(\Theta_{23} )}\} \sin{(\Theta_{24} )} - \sin{(\Theta_{12} )} \sin{(\Theta_{14} )} \cos{(\Theta_{13} )} \cos{(\Theta_{24} )}] \cos{(\Theta_{34} )} \nonumber\\
			&& + \{- \sin{(\Theta_{12} )} \sin{(\Theta_{13} )} \cos{(\Theta_{23} )} + \sin{(\Theta_{23} )} \cos{(\Theta_{12} )}\} \sin{(\Theta_{34} )}\big] + u_{3} [- \{- \sin{(\Theta_{13} )} \sin{(\Theta_{14} )} \cos{(\Theta_{24} )} \nonumber \\
			&&- \sin{(\Theta_{23} )} \sin{(\Theta_{24} )} \cos{(\Theta_{13} )}\} \sin{(\Theta_{34} )} + \cos{(\Theta_{13} )} \cos{(\Theta_{23} )} \cos{(\Theta_{34} )}] [\{- \sin{(\Theta_{13} )} \sin{(\Theta_{14} )} \cos{(\Theta_{24} )}  \nonumber\\
			&&- \sin{(\Theta_{23} )} \sin{(\Theta_{24} )} \cos{(\Theta_{13} )}\} \cos{(\Theta_{34} )} + \sin{(\Theta_{34} )} \cos{(\Theta_{13} )} \cos{(\Theta_{23} )}] \nonumber \\
			&&- u_{4} \sin{(\Theta_{34} )} \cos^{2}{(\Theta_{14} )} \cos^{2}{(\Theta_{24} )} \cos{(\Theta_{34} )}
			\end{eqnarray}
		\end{subequations}
	\end{small}	
	\clearpage
	
	\section{Diabatic PES matrices constructed along different paths of integration}	
	\begin{figure}[!htp]
		\centering
		\includegraphics[width=\linewidth]{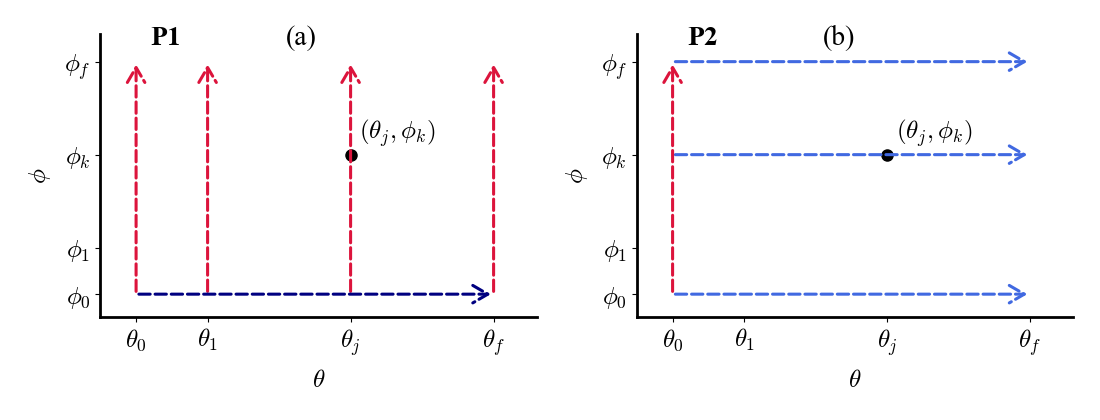}
		\caption{The 2D rectangular paths along which the ADT equations are solved.}
		\label{fig:adtpath}
	\end{figure}
	
	\noindent
	Let us choose two different 2D contour integration of ADT equations along $\theta$ and $\phi$ for a fixed $\rho$ as shown in Figures~\ref{fig:adtpath}(a) and \ref{fig:adtpath}(b) and such ADT matrices are termed as $A_1$ and $A_2$, where those matrices are orthonormal by construction at any point in configuration space:
	
	\begin{subequations}
		\begin{eqnarray}
		\mathbf{A_{1}{^\dagger}} (\theta, \phi) \, \mathbf{A_{1}} (\theta, \phi) = \mathbf{I} \\
		\mathbf{A_{2}{^\dagger}} (\theta, \phi) \, \mathbf{A_{2}} (\theta, \phi) = \mathbf{I} 
		\end{eqnarray} \label{eq:1}
	\end{subequations}
	
	\noindent
	Furthermore, we consider the following product matrix:
	
	\begin{eqnarray}
	\mathbf{B} (\theta, \phi) = \mathbf{A_{1}{^\dagger}} (\theta, \phi) \mathbf{A_{2}} (\theta, \phi),
	\label{eq:2}  
	\end{eqnarray}
	
	\noindent
	which also appears as an orthogonal matrix:
	
	\begin{eqnarray}
	\mathbf{B{^\dagger}} \, \mathbf{B} &=& \left(\mathbf{A_{1}{^\dagger}} \, \mathbf{A_{2}}\right)^{\dagger} \, \left(\mathbf{A_{1}{^\dagger}} \, \mathbf{A_{2}}\right) \nonumber \\
	&=& \mathbf{A_{2}{^\dagger}} \left(\mathbf{A_{1}} \, \mathbf{A_{1}{^\dagger}}\right) \, \mathbf{A_{2}} \nonumber \\ 
	&=& \mathbf{I}
	\end{eqnarray}\label{eq:3}
	
	\noindent
	From Equation~\ref{eq:2}, one can get the following relation:
	
	\begin{eqnarray}
	\mathbf{A_{1}}  \, \mathbf{B}  = \mathbf{A_2}, 
	\end{eqnarray} \label{eq:4}
	
	\noindent
	which indicates that the ADT matrices belonging to two different paths are related through another orthogonal matrix (\textbf{B}).
	
	\vspace{0.5cm}
	
	\noindent
	On the other hand, the diabatic PESs matrices along two different paths of integration takes the following form:
	
	\begin{subequations}
		\begin{eqnarray}
		\mathbf{W_1} &=& \mathbf{A_{1}{^\dagger}} \mathbf{U} \mathbf{A_1} \\ 
		\mathbf{W_2} &=& \mathbf{A_{2}{^\dagger}} \mathbf{U} \mathbf{A_2}, \qquad \text{$\mathbf{U}$ is adiabatic PESs matrix (diagonal)} 
		\end{eqnarray}\label{eq:5}
	\end{subequations}
	
	\noindent
	which leads to
	
	\begin{subequations}
		\begin{eqnarray}
		\mathbf{A_1} \, \mathbf{W_1}\, \mathbf{A_{1}{^\dagger}} =  \mathbf{U}  \\
		\mathbf{A_2} \, \mathbf{W_2}\, \mathbf{A_{2}{^\dagger}} = \mathbf{U} 
		\end{eqnarray} \label{eq:6}
	\end{subequations}
	
	\noindent
	Comparing Equation~\ref{eq:6}(a) and \ref{eq:6}(b), we obtain:
	
	\begin{eqnarray}
	\mathbf{A_1} \, \mathbf{W_1}\, \mathbf{A_{1}{^\dagger}} &=& \mathbf{A_2} \, \mathbf{W_2}\, \mathbf{A_{2}{^\dagger}} \nonumber \\
	\mathbf{A_{1}{^\dagger}} \, \mathbf{A_1} \, \mathbf{W_1}\, \mathbf{A_{1}{^\dagger}} \, \mathbf{A_1}  &=& \mathbf{A_{1}{^\dagger}} \, \mathbf{A_2} \, \mathbf{W_2}\, \mathbf{A_{2}{^\dagger}} \, \mathbf{A_1} \nonumber \\ 
	\mathbf{W_1} &=& \mathbf{B} \, \mathbf{W_2} \, \mathbf{B}^{\dagger}
	\label{eq:8}
	\end{eqnarray}
	
	\noindent
	Therefore, the diabatic PESs matrices along two different paths are related through another orthogonal matrix (B) and thus, calculated observables should be path independent in theoretical consideration.
	
	\pagebreak
	
	\section{7S-SA-MCSCF level Adiabatic PESs using CAS(18e,12o) and AVQZ basis}
	
	\begin{figure}[!htp]
		\centering
		\includegraphics[width=\linewidth]{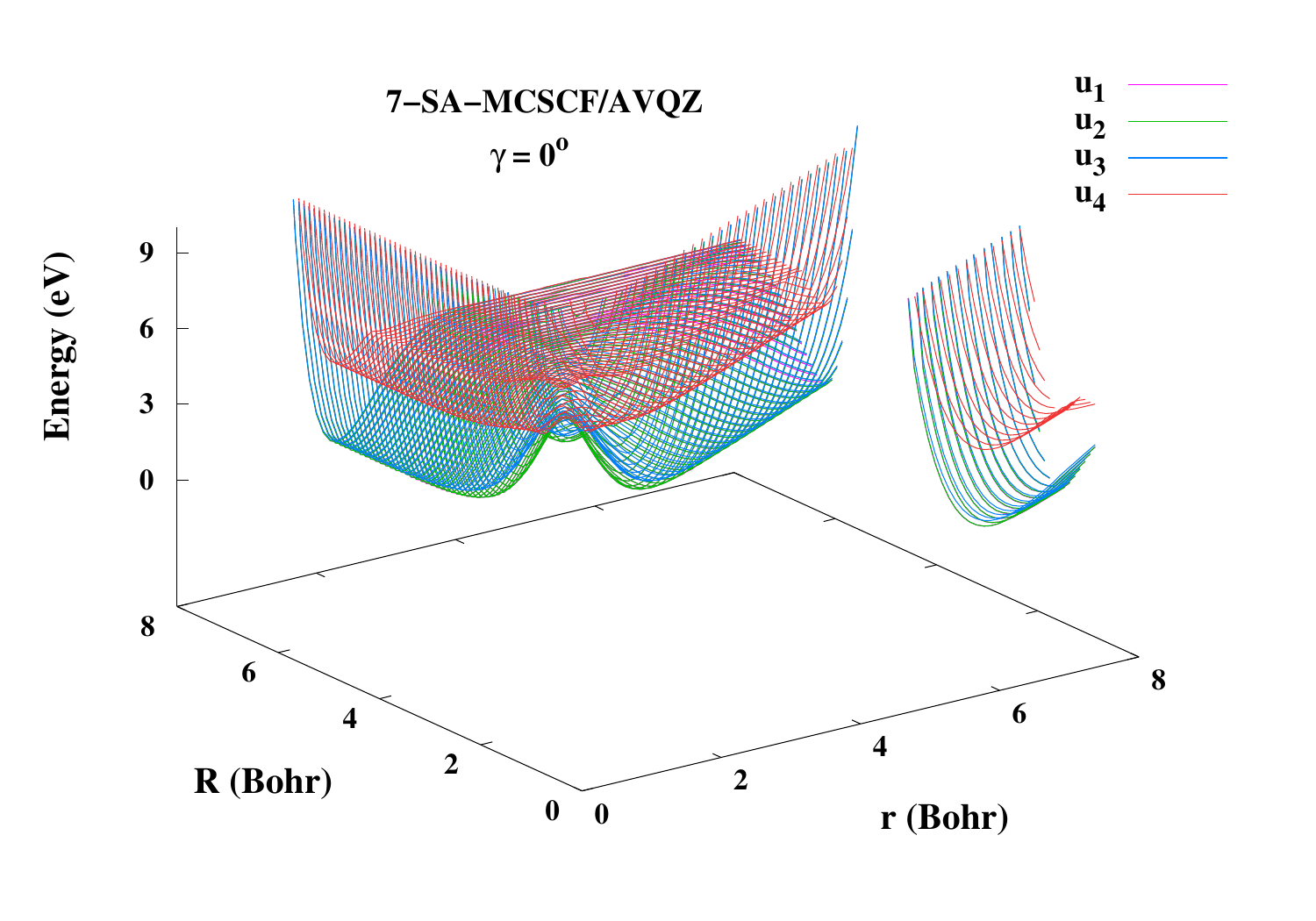}
		\caption{Lowest four adiabatic PESs of O$_3$ at $\gamma = 0^\circ$ using 7S-SA-MCSCF method with CAS(18e,12o) and the AVQZ basis set.}
		\label{fig:gamma_0_4_PES}
	\end{figure}
	
	\vspace{0.2cm}
	
	\begin{figure}[!htp]
		\centering
		\includegraphics[width=\linewidth]{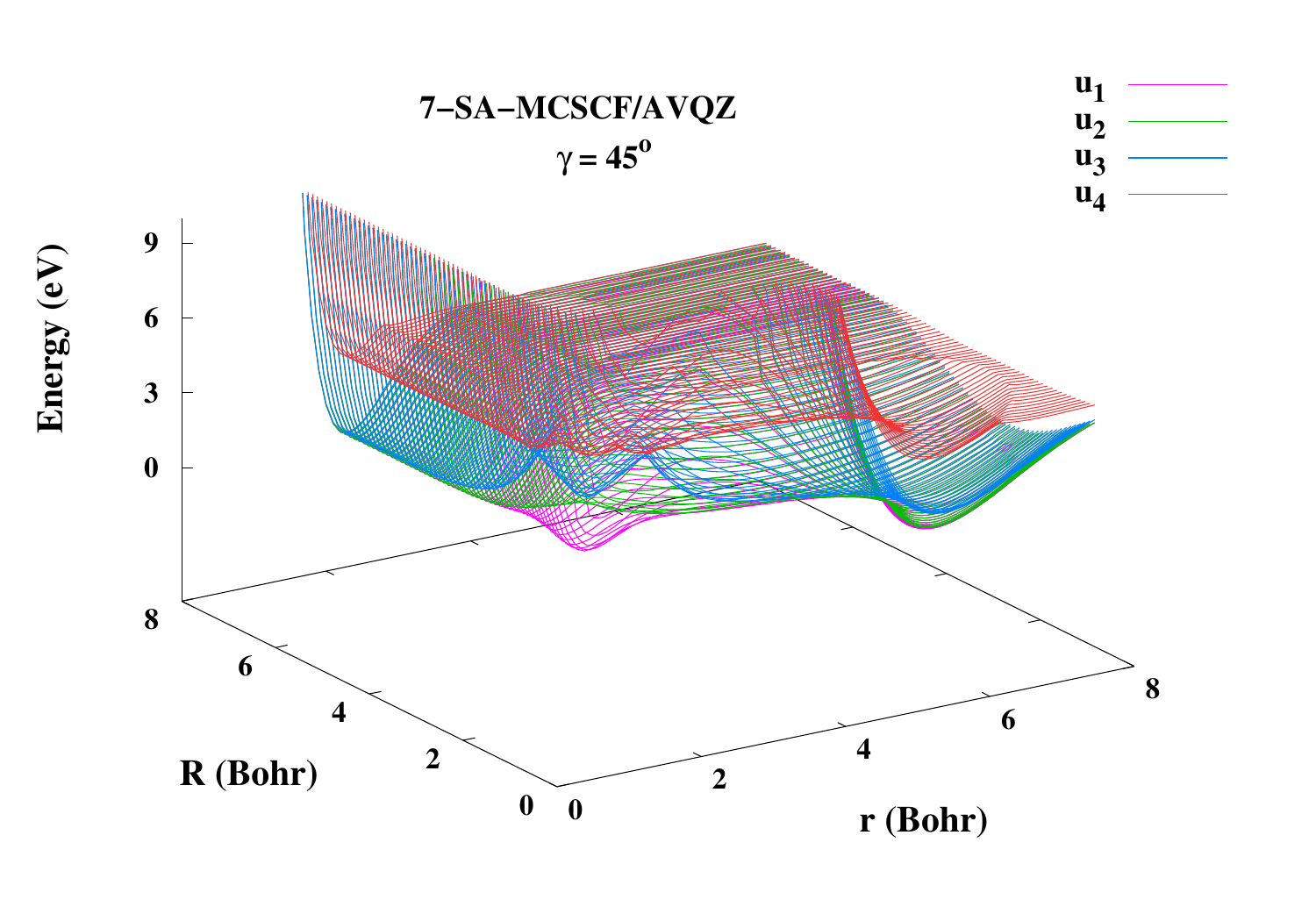}
		\caption{Lowest four adiabatic PESs of O$_3$ at $\gamma = 45^\circ$ using 7S-SA-MCSCF method with CAS(18e,12o) and the AVQZ basis set.}
		\label{fig:gamma_45_4_PES}
	\end{figure}

	\begin{figure}[!htp]
		\centering
		\includegraphics[width=\linewidth]{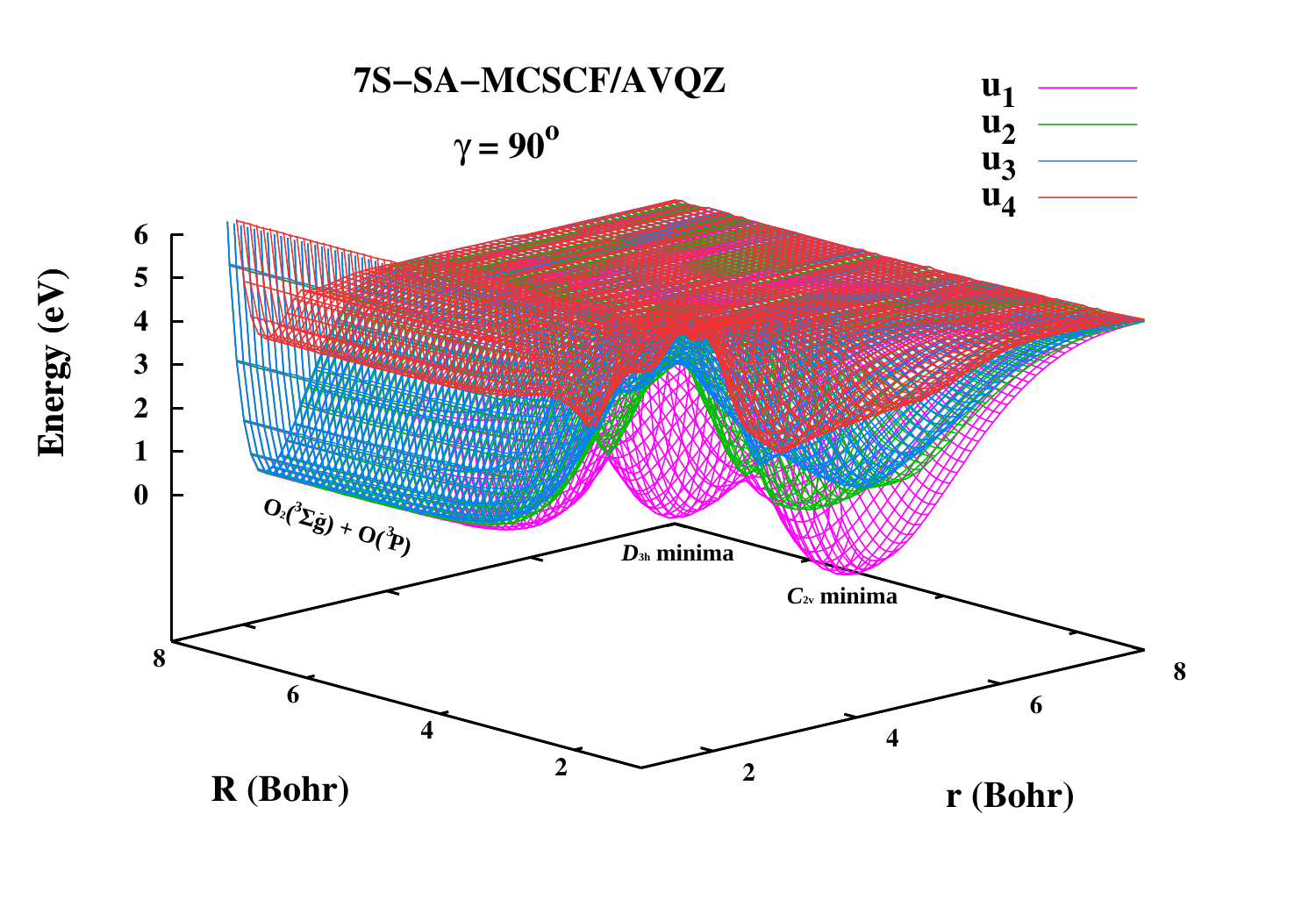}
		\caption{Lowest four adiabatic PESs of O$_3$ at $\gamma = 90^\circ$ using 7S-SA-MCSCF method with CAS(18e,12o) and the AVQZ basis set.}
		\label{fig:gamma_90_4_PES}
	\end{figure}
	
	\pagebreak
	\section{Location of CIs}
	\vspace{0.5cm}
	\begin{figure}[!htp]
		\centering
		\hspace*{-1cm}
		\begin{subfigure}{0.4\linewidth}
			\centering
			\begin{overpic}[width=1.4\linewidth]{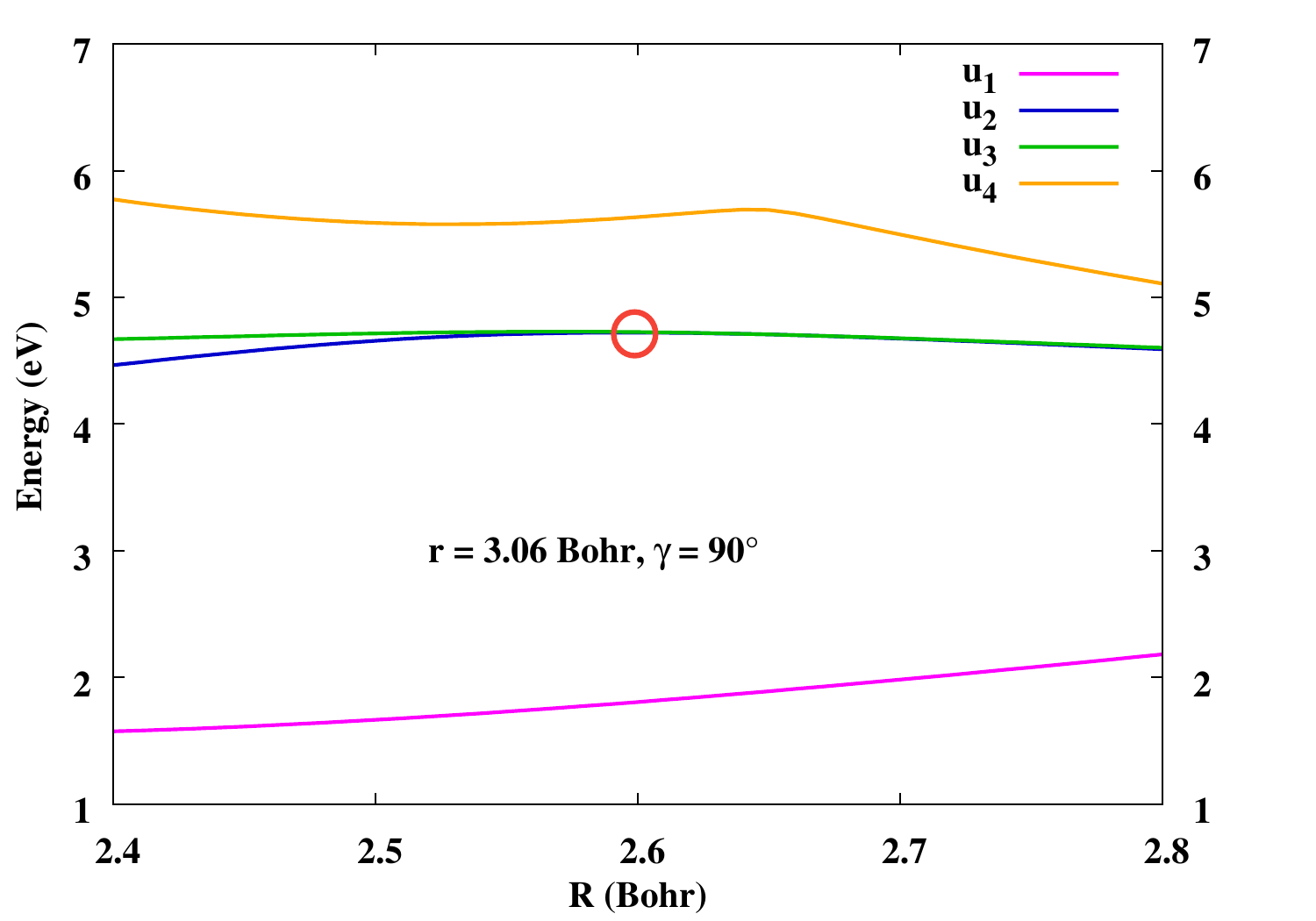}
				\put(5,75){\textbf{(a)}} 
			\end{overpic}
			\phantomcaption
			\label{fig:C2v_1234}
		\end{subfigure}
		\hspace{2.5cm}
		\begin{subfigure}{0.4\linewidth}
			\centering
			\begin{overpic}[width=1.4\linewidth]{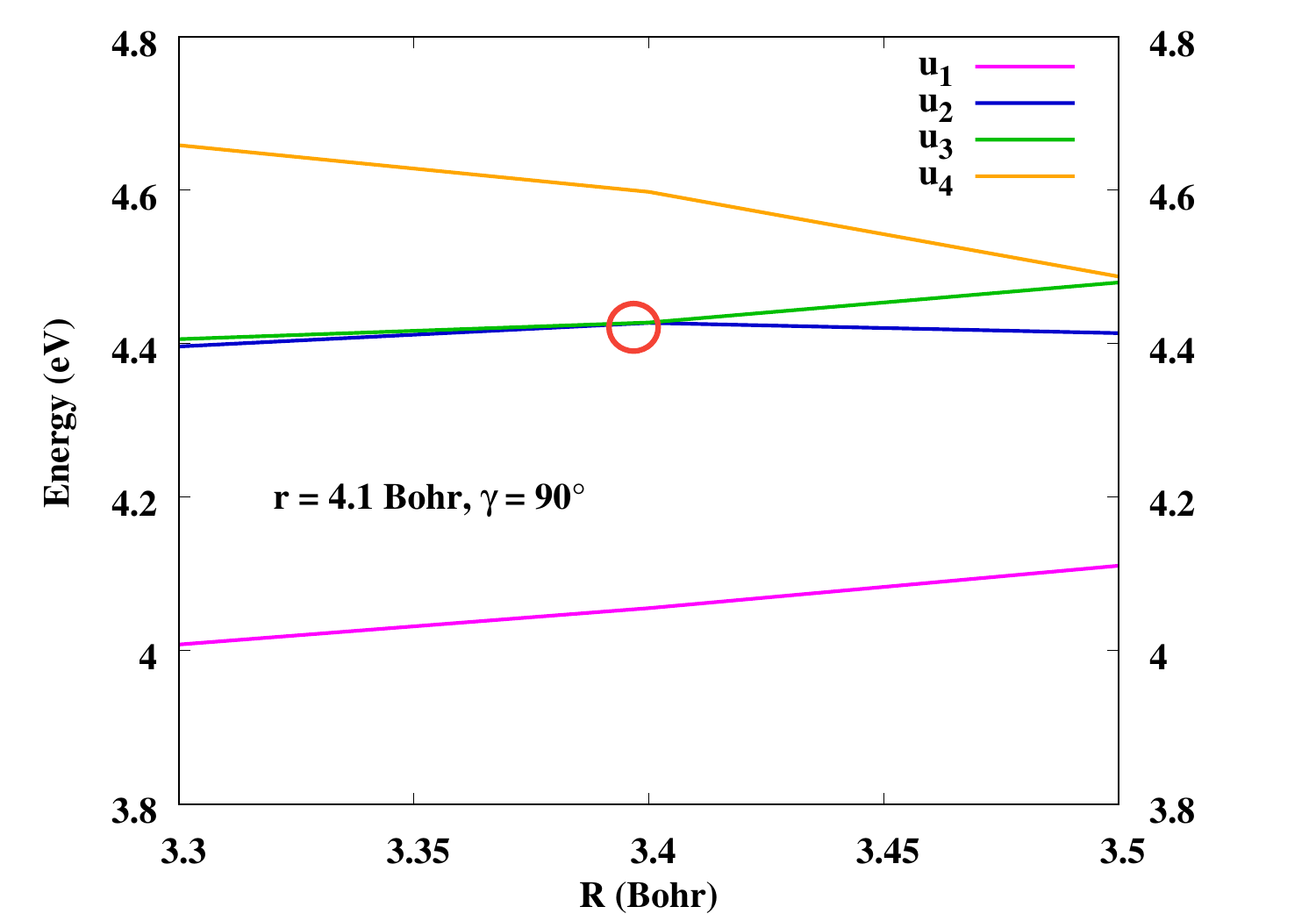}
				\put(5,75){\textbf{(b)}} 
			\end{overpic}
			\phantomcaption
			\label{fig:D3h_23}
		\end{subfigure}
		\vspace{0.5cm}
		\begin{subfigure}{0.4\linewidth}
			\centering
			\begin{overpic}[width=1.4\linewidth]{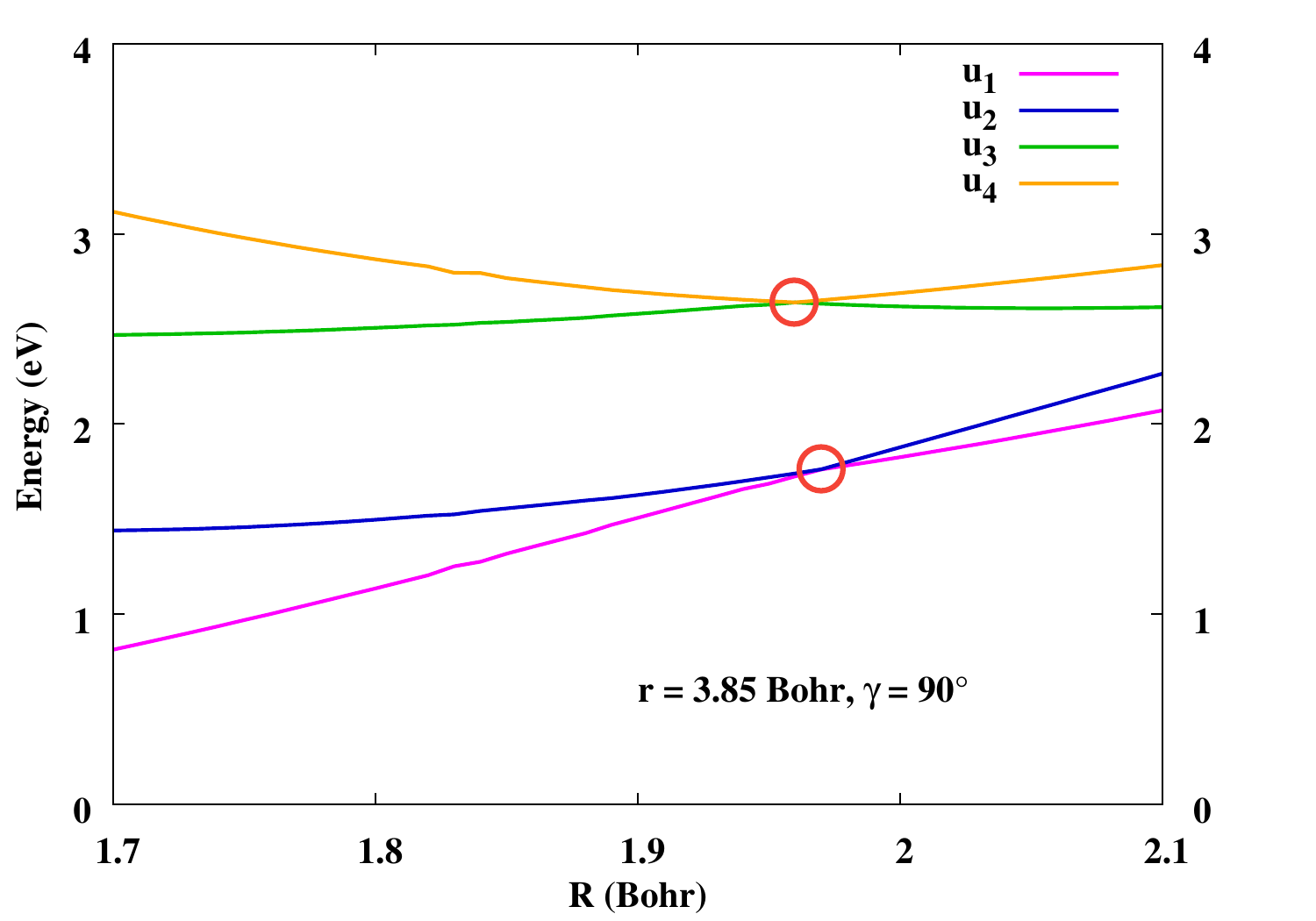}
				\put(0,70){\textbf{(c)}}
			\end{overpic}
			\phantomcaption
			\label{fig:C2v_23}
		\end{subfigure}		
		\label{fig:1DPECs}
		\caption{The adiabatic PECs at different $R$ as a function of $r$ at $\gamma$ = 90$^\circ$ are depicted as function of $R$ for various values of $r$ to explore the location of \textit{C}$_{2v}$ and \textit{D}$_{3h}$ CIs.}
	\end{figure}
	
	\pagebreak	
	
	\section{Curl Condition}
	\begin{figure}[!htp]
		\centering
		\includegraphics[width=\linewidth]{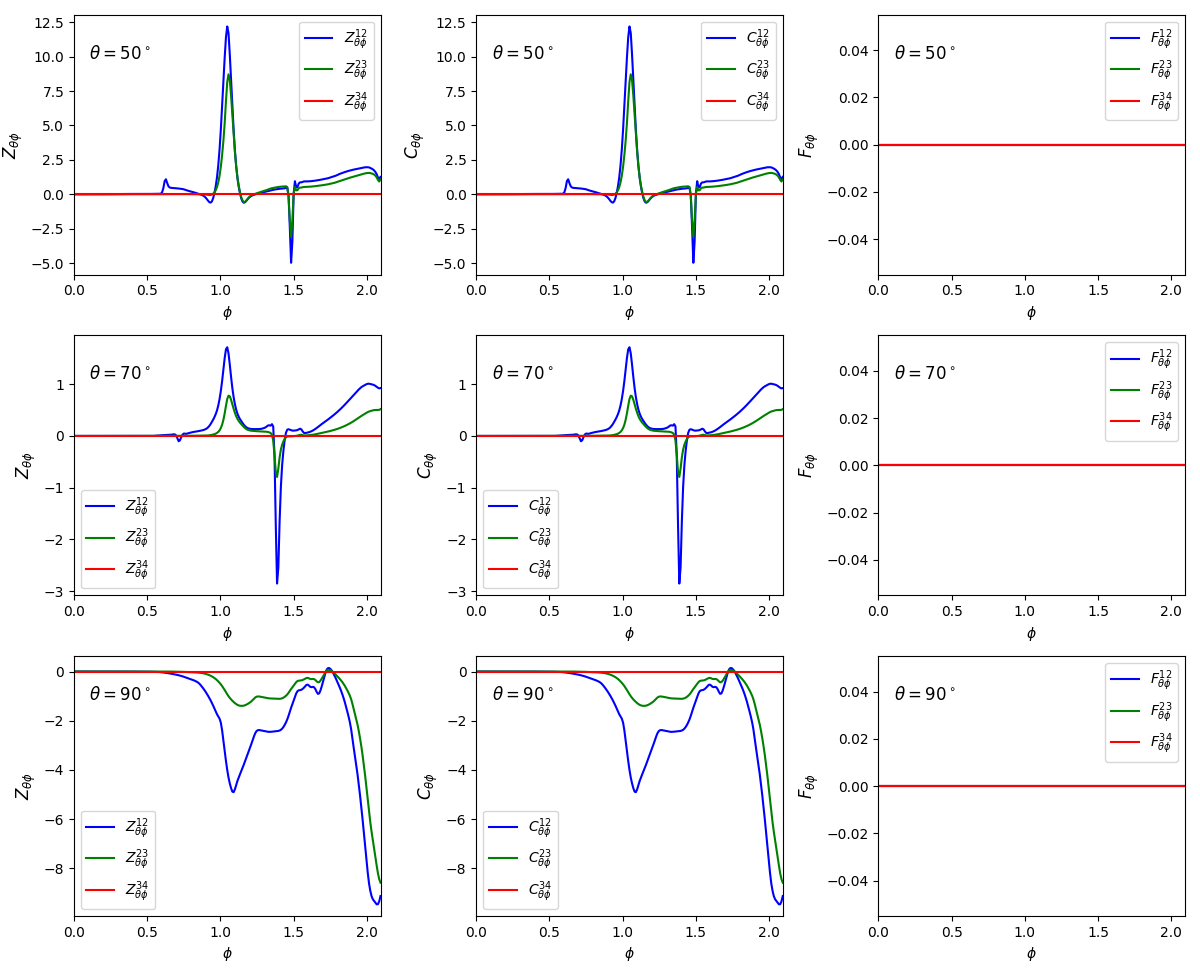}
		\caption{Mathematical curl ($Z_{\theta\phi}^{ij}$) and the ADT curl ($C_{\theta\phi}^{ij}$) as well as Curl Condition ($F_{\theta\phi}^{ij}$) are presented as a function of $\phi$ for fixed $\rho$ = 4 Bohr at different $\theta$, namely, 50$^\circ$, 70$^\circ$ and 90$^\circ$}
		\label{fig:curl}
	\end{figure}
	\pagebreak
	
	\bibliographystyle{aipnum4-1}
	\bibliography{ref} 
	
\end{document}